\documentclass[11pt,letterpaper]{article}
\pdfoutput=1
\usepackage{jheppub}
\usepackage[dvipsnames]{xcolor}
\usepackage{graphicx}

\usepackage{verbatim}
\usepackage{amsmath}
\usepackage{amssymb}
\usepackage{url}
\usepackage{bbold}
\usepackage{xspace}
\usepackage{tabularx}
\usepackage{makecell}
\usepackage[makeroom]{cancel}
\usepackage{multirow}
\usepackage{color}
\usepackage{subfig}
\usepackage{rotating}
\usepackage{cancel}

\def\beq{\begin{equation}}
	\def\eeq#1{\label{#1}\end{equation}}
\def\eeqn{\end{equation}}
\def\beqa{\begin{eqnarray}}
\def\eeqa#1{\label{#1}\end{eqnarray}}
\def\eeqan{\end{eqnarray}}
\def\CR{\nonumber \\ }
\def\leqn#1{(\ref{#1})}
\def\bea{\begin{eqnarray}}
\def\eea{\end{eqnarray}}

\def\mev{\mathrm{\ MeV}}

\renewcommand{\d}{\mathrm{d}}

\newcommand{\centeron}[2]{{\setbox0=\hbox{#1}\setbox1=\hbox{#2}\ifdim
\wd1>\wd0\kern.5\wd1\kern-.5\wd0\fi \copy0
\kern-.5\wd0\kern-.5\wd1\copy1\ifdim\wd0>\wd1
\kern.5\wd0\kern-.5\wd1\fi}}
\newcommand{\ltap}{\>\centeron{\raise.35ex\hbox{$<$}}
{\lower.65ex\hbox{$\sim$}}\>}
\newcommand{\gtap}{\>\centeron{\raise.35ex\hbox{$>$}}
{\lower.65ex\hbox{$\sim$}}\>}
\newcommand{\gsim}{\mathrel{\gtap}}
\newcommand{\lsim}{\mathrel{\ltap}}

\def\O{{\cal O}}
\def\gap{M_{\rm gap}}
\def\rhoc{\rho_{\scriptstyle_{\rm CFT}}}
\def\Pc{P_{\scriptstyle_{\rm CFT}}}
\def\rhos{\rho_{\scriptstyle_{\rm SM}}}
\def\Ps{P_{\scriptstyle_{\rm SM}}}
\def\l{\lambda_{\scriptstyle_{\rm CFT}}}
\def\L{\Lambda_{\scriptstyle_{\rm CFT}}}
\def\Oc{{\cal O}_{\scriptstyle_{\rm CFT}}}
\def\Os{{\cal O}_{\scriptstyle_{\rm SM}}}
\def\Tc{T_{\scriptstyle_{\rm D}}}
\def\Ts{T_{\scriptstyle_{\rm SM}}}
\def\mdm{m_{\scriptstyle_{\rm DM}}}
\def\ds{d_{\scriptstyle_{\rm SM}}}

\newcommand{\CFT}{\mathrm{CFT}}

\begin{document}
\title{Dark Matter from a Conformal Dark Sector}

\author[a,b,c]{Sungwoo Hong,}
\author[a,d]{Gowri Kurup}
\author[a]{and Maxim Perelstein}

\affiliation[a]{Department of Physics, LEPP, Cornell University, Ithaca, NY 14853, USA}
\affiliation[b]{Department of Physics, The University of Chicago, Chicago, IL 60637 , USA }
\affiliation[c]{Argonne National Laboratory, Lemont, IL 60439, USA}
\affiliation[d]{Rudolf Peierls Centre for Theoretical Physics, University of Oxford, Parks Rd, Oxford OX1 3PJ, United Kingdom}

\date{22 February,  2023}                                                                                                                                                                                                                                         

\abstract{
We consider theories in which a dark sector is described by a Conformal Field Theory (CFT) over a broad range of energy scales. A coupling of the dark sector to the Standard Model breaks conformal invariance. While weak at high energies, the breaking grows in the infrared, and at a certain energy scale the theory enters a confined (hadronic) phase. One of the hadronic excitations can play the role of dark matter. We study a ``Conformal Freeze-In" cosmological scenario, in which the dark sector is populated through its interactions with the SM at temperatures when it is conformal. In this scenario, the dark matter relic density is determined by the CFT data, such as the dimension of the CFT operator coupled to the Standard Model. We show that this simple and highly predictive model of dark matter is phenomenologically viable. The observed relic density is reproduced for a variety of SM operators (``portals") coupled to the CFT, and the resulting models are consistent with observational constraints. The mass of the COFI dark matter candidate is predicted to be in the keV-MeV range.   
}



\maketitle

\section{Introduction}

The microscopic nature of dark matter (DM) is one of the most pressing issues in fundamental physics, as no known elementary particle has the right properties to make up DM. An interesting possibility is that DM particles are part of a ``dark sector", a set of fields that are uncharged under the Standard Model (SM) gauge group~\cite{Dienes:2022zbh, Essig:2013lka}. A dark sector may contain its own gauge interactions and matter fields, and may indeed have a level of complexity and structure similar to or exceeding the SM. Such dark sectors are very natural from a theoretical point of view, and in fact are ubiquitous in string theory constructions incorporating the SM. 

As there are very few theoretical constraints on the nature of the dark sector, it is important to explore a wide range of possibilities that may lead to viable DM candidates. In this paper, we will study the scenario where the dark sector possesses {\it conformal symmetry}. Conformal field theories (CFT's) are generic in the landscape of quantum field theories, arising whenever renormalization group evolution has a non-trivial attractive fixed point~\cite{CFT:BigBook, Ginsparg:1988ui, Rychkov:2016iqz}. Moreover, while CFT's are generally strongly-coupled and cannot be studied via perturbative techniques, the conformal symmetry is often sufficient to make non-trivial physical predictions in these theories. In practice, this will allow us to construct models of dark matter in which observables such as relic density are both calculable and differ parametrically from the prediction of any perturbative model of the dark sector. In fact, in many cases the only input needed from the CFT side is the two-point function of the CFT operator coupled to the SM, which is completely determined by the dimension of this operator and conformal invariance.

Suppose that a conformally-invariant dark sector exists, and some energy is injected into this sector in the early universe. Conformal symmetry implies that in the expanding universe, the energy density of the dark sector will scale as $\rho_{\rm dark} \propto a^{-4}$, where $a$ is the scale factor. This scaling is that of radiation, not non-relativistic matter, leading to an immediate objection to the idea of dark matter made out of a CFT. However, very generically, we can expect the dark sector to interact, at some level, with the non-conformal sector containing SM.\footnote{Here, we consider non-gravitational coupling of the SM to the CFT. Models with gravitational coupling of the two sectors were studied in~\cite{Redi:2020ffc}.} These interactions necessarily lead to breaking of the conformal symmetry in the dark sector. While the SM-CFT coupling may be perturbatively small in the UV, it grows with decreasing energy if the interaction involves a relevant operator (dimension$<4$) in the CFT. Eventually, the conformal symmetry is completely broken at an IR scale $\gap$. Below this scale the theory enters a ``hadronic" phase, with ordinary massive particle excitations in the spectrum. These particles can play the role of Cold Dark Matter (CDM). While the DM today consists of ``normal" particles in this scenario, it is possible that the processes that are responsible for populating the dark sector (thus fixing the relic density of the DM) occurred when the dark sector was in the conformal regime.  

If the SM-CFT coupling is sufficiently strong for the two sectors to come to thermodynamic equilibrium in the early universe, a rough estimate shows that the observed relic density of DM requires $\gap\sim 10-100$~eV. This scenario would lead to hot dark matter, ruled out by observations of large-scale structure. (It is possible to avoid this conclusion if the DM can effectively annihilate to the SM in the hadronic phase, but in that case, the relic density would be completely determined by the ordinary particle physics of the hadronic phase, not the CFT.) We will therefore focus on the case when the CFT does {\it not} come into thermal equilibrium with the SM due to weakness of the coupling between the two sectors. We assume that the CFT sector is not populated by inflaton decays, since otherwise the DM relic density becomes just an initial condition with no physical origin. (A model in which such ``asymmetric reheating" is realized naturally is discussed in Ref.~\cite{Chiu:2022bni}.) 
The interactions with the SM then provide the main mechanism for populating the CFT sector in the early universe. Such a non-thermal production mechanism in the case of ordinary particles is known as ``freeze-in". The scenario studied in this paper can then be described as ``conformal freeze-in (COFI)", the term that was first introduced in Ref.~\cite{Hong:2019nwd}, where we considered a specific realization of this scenario. In this paper, we present a systematic study of the COFI mechanism, including several possible SM operators, or ``portals", that can couple to the CFT dark sector, as well as effects of operator mixing. We also include an updated analysis of astrophysical constraints from stellar cooling and other sources. We find that the COFI scenario is very generic and can occur for any of the portals we consider, and in many cases the resulting DM candidate is phenomenologically viable.     

The rest of the paper is organized as follows. In Section~\ref{sec:gen}, we describe the model of the dark sector and its interactions with the SM underlying our scenario. This includes the discussion of the CFT phase, the hadronic phase that emerges at low energies after the conformal invariance is broken, and a possible UV completion of the CFT by a gauge theory with a strongly-interacting Banks-Zaks fixed point. In Section~\ref{sec:Ops}, we describe the cosmological evolution of the dark sector in the COFI scenario, and calculate the dark matter relic density. The figures in this section provide a snapshot of the parameter space in various COFI models containing a viable dark matter candidate, along with observational and theoretical constraints on these models. The derivation of these constraints is presented in Section~\ref{sec:pheno}. Finally, we summarize and conclude in Section~\ref{sec:out}. Technical details of calculations of relic density and stellar cooling rates are contained in the appendix.     

\section{Theoretical Framework}
\label{sec:gen}

We consider a theory in which a Dark Sector ({\it i.e.} a set of fields with no direct charges under SM gauge groups) is described by a Conformal Field Theory (CFT) across a broad range of energy scales, between the ``gap scale" $\gap$ in the infrared (IR), and the ultraviolet (UV) cutoff $\Lambda_{\rm UV}\gg\gap$. We discuss the theory in the CFT window and its interactions with the Standard Model (SM) in Section~\ref{subsec:CFT}. We describe the mechanism that generates the gap scale and the physics at and below that scale in Section~\ref{subsec:IR}. For completeness, we  outline a possible UV completion above $\Lambda_{\rm UV}$ in Section~\ref{subsec:UV}, although that theory is not directly relevant for the discussion of dark matter. 

\subsection{Conformal Dark Sector}
\label{subsec:CFT}

At energy scales between $\gap$ and $\Lambda_{\rm UV}$, the Dark Sector is described by a CFT. We assume that the CFT contains an operator $\Oc$ with a scaling dimension $d<4$, {\it i.e.} a relevant operator. Generically the CFT is strongly coupled, and $d$ need not be integer. Further, we assume that $\Oc$ is charged under a global symmetry ${\cal G}$ (for example a discrete $\mathbb{Z}_2$), which forbids a Lagrangian term of the form $c\, \Oc$. Standard Model (SM) fields are not charged under ${\cal G}$.

We consider a coupling between the SM and the dark CFT of the form
\beq
{\cal L}_{\rm int} = \frac{\l}{\L^{D-4}} \,\Os \Oc\,.
\eeq{eqn:interaction}
where $\Os$ is an operator made out of SM fields. Here $\l$ is a dimensionless constant, while $\L$ is a mass scale. Further, 
\beq
D = d + \ds\,,
\eeq{Ddef}
where $\ds$ is the scaling dimension of $\Os$. The interaction term~\leqn{eqn:interaction} explicitly breaks both conformal symmetry (since the SM is not conformal), and the global symmetry ${\cal G}$. We consider the regime where this interaction is small enough to consider this breaking perturbatively, and work to leading order in the interaction strength. 

Since the dark sector does not carry SM gauge charges, $\Os$ must be gauge-invariant, but there are {\it a priori} no other restrictions on this operator. For simplicity, we assume that {\it at tree level}, there is a single SM operator interacting with the CFT via Eq.~\leqn{eqn:interaction}. (Of course, couplings between $\Oc$ and other SM operators will generically be induced by quantum corrections, as discussed below.) To illustrate the range of possibilities, we consider several possible portal operators $\Os$, which couple the CFT to quark, lepton, and gauge sectors of the SM.  We can classify these operators into two types: {\it Type-I operators} that acquire a non-zero vacuum expectation value (VEV) in the IR, and {\it Type-II operators} that do not. We consider three Standard Model operators in the class of type-I operators: 

\begin{itemize}
	\item Higgs portal, $\mathbf{H^\dag H}$, \ 
	
	\item Quark portal, $\mathbf{HQ_L^\dag q_R}$, \ and \
	
	\item Gluon portal, $\mathbf{G^{\mu\nu}G_{\mu\nu}}$. 
\end{itemize}
The Higgs portal operator gets a VEV at the weak scale, while the quark and gluon portals get VEVs at the QCD confinement scale. Further, we consider three examples of type-II operators: 
\begin{itemize}
	\item Lepton portal $\mathbf{H L^\dagger \ell_R}$, \
	
	\item Weak-gauge portal $\mathbf{W^{\mu\nu}W_{\mu\nu}}$, \ and \ 
	
	\item Hypercharge-gauge portal $\mathbf{B^{\mu\nu}B_{\mu\nu}}$. 
\end{itemize}
All our examples involve relevant or marginal SM operators, which are expected to be dominant at low energies. Also, all operators we consider are Lorentz scalars. For an example of a non-scalar portal, namely $\Os=B_{\mu\nu}$ which results in a composite dark photon, see Ref.~\cite{COFI_DP}. 

In the case of quark and lepton portals, the flavor structure of the CFT coupling to the SM needs to be specified. We will assume that the portal operators are flavor-diagonal in the SM mass eigenbasis. The coupling to CFT can then be written in this basis as
\beq
{\cal L}_{\rm int} = \frac{\l}{\L^{D-4}} \, \Oc\, \cdot \left( \sum_i \kappa_i {\cal O}_{\rm Yuk}^i \right)\,,
\eeq{Flavor}
where the sum runs over the six flavors of SM quarks or three flavors of charged leptons, and ${\cal O}_{\rm Yuk}^i$ is the SM Yukawa operator for each flavor. The constants $\kappa_i$ encode the flavor dependence of the CFT-SM interactions. Specifically, we will consider three cases:

\begin{itemize}
	
	\item Minimal Flavor Violation (MFV), with entries proportional to SM Yukawas: $\kappa_i=y_i$, $i=1\ldots 6$ for quarks and $1\ldots 3$ for charged leptons.  
	
	\item Democratic, with all entries the same: $\kappa_i=1$.
	
	\item First-Generation Only: $\kappa_i=1$ for the first-generation quarks or electrons, and 0 for the second and third generations. 
	
\end{itemize}

\subsection{CFT Breaking in the Infrared}
\label{subsec:IR}

Since the Dark Sector CFT contains a relevant operator $\Oc$, the generic expectation is that the conformal symmetry is broken in the infrared (IR). Specifically, if the Lagrangian contains a term
\beq
{\cal L} \,=\, c \, \Oc\,,
\eeq{deform}
where $c$ is a constant of mass dimension $4-d>0$, the conformal symmetry is broken at the ``gap" mass scale
\beq
\gap \sim c^{1/(4-d)}\,.
\eeq{gap}
Here and below, we make use of Naive Dimensional Analysis (NDA) to estimate various quantities of interest up to order-one factors. In most cases, more precise analytic results are not available due to the strongly-coupled nature of the underlying theory. NDA estimates will be sufficient to establish the basic features of the dark matter model and establish its viability. At energy scales below $\gap$, the theory is no longer conformal. In this subsection, we will first estimate the gap scale for each of the six SM portals, and then describe the physics at low energies below $\gap$. 

\subsubsection{Estimates of the Gap Scale}

Global symmetry ${\cal G}$ forbids the deformation~\leqn{deform} within the CFT itself, and the infrared breaking of the CFT is entirely due to its interaction with the SM, Eq.~\leqn{eqn:interaction}. For each portal operator $\Os$, there are several distinct contributions to $\gap$, with the NDA estimates for each of them summarized in Table~\ref{tab:gaps}. Below, we will discuss each of these contributions.  

For type-I operators, a non-zero VEV directly leads to an effective Lagrangian of the form~\leqn{deform}, with
a coefficient
\beq
c = \frac{\l}{\L^{D-4}} \langle \Os \rangle.
\eeq{TypeIdef}
An estimate of the corresponding contribution to the gap scale $\gap$ for each of the three type-I portals is listed in the first column of Table~\ref{tab:gaps}. We refer to this contribution as ``tree-level". Note that since these are NDA-level estimates, all QCD condensates are simply taken to be $\Lambda_{\rm QCD}$ to the appropriate power.  

\begin{figure}[t]
	\begin{center}
		\includegraphics[width=14cm]{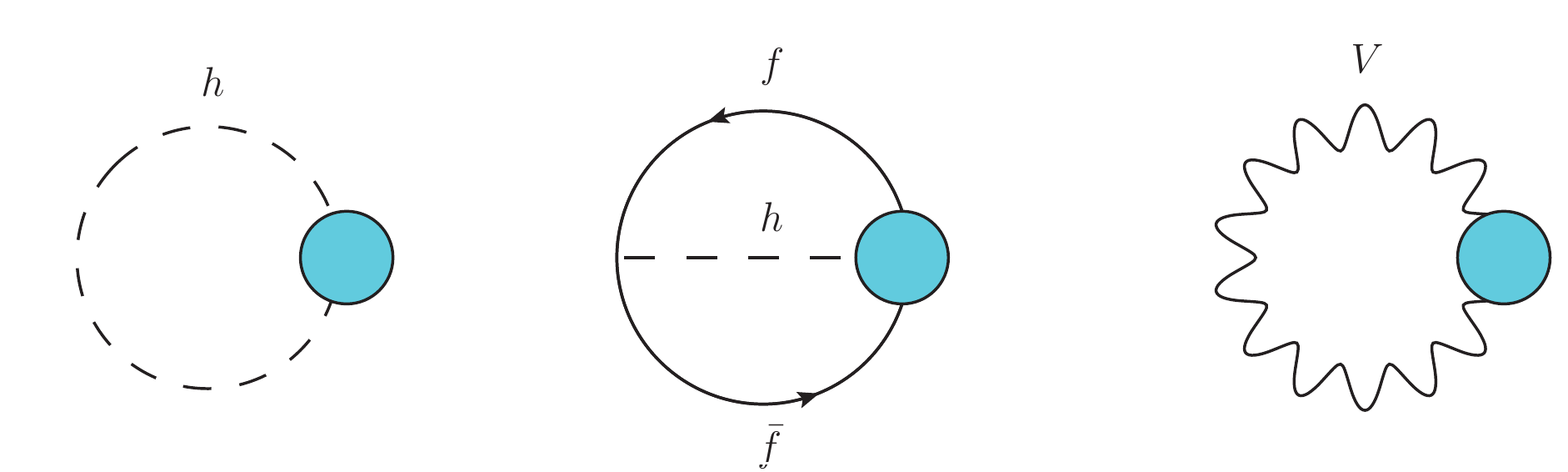}
		\caption{Contributions to conformal symmetry breaking via ``radiative direct" diagrams, in the Higgs, quark/lepton and gluon/weak boson portals respectively. Blue circles indicate CFT operator insertions.}
		\label{fig:raddirect}
	\end{center}
\end{figure}

For both type-I and type-II operators, the deformation~\leqn{deform} is induced by quantum corrections. For example, the leading contributions of this type for Higgs,  quark/lepton and gluon/weak boson/hypercharge boson portals are illustrated in Fig.~\ref{fig:raddirect}. We refer to these contributions as ``radiative direct".  The Feynman diagrams that contribute are generally UV-divergent, and the NDA estimates of their contributions are proportional to powers of the scale $\Lambda_{\rm SM}$ which serves as the UV cutoff of the SM loops. The LHC constraints generally imply $\Lambda_{\rm SM} \gsim 1$~TeV. Note that if $\Lambda_{\rm SM} \gg 4\pi v$, the observed weak scale requires strong fine-tuning. A similar fine-tuning may or may not occur in the SM loop contributions to~\leqn{deform}, and the gap scale in this scenario is strongly model-dependent.  For concreteness, we will use $\Lambda_{\rm SM} =2\pi v \sim 1.5$~TeV in the estimates of this paper. The NDA estimates of this contribution to $\gap$ for each portal are collected in the second column of 
Table~\ref{tab:gaps}.

\begin{figure}[t!]
	\begin{center}
		\includegraphics[width=15cm]{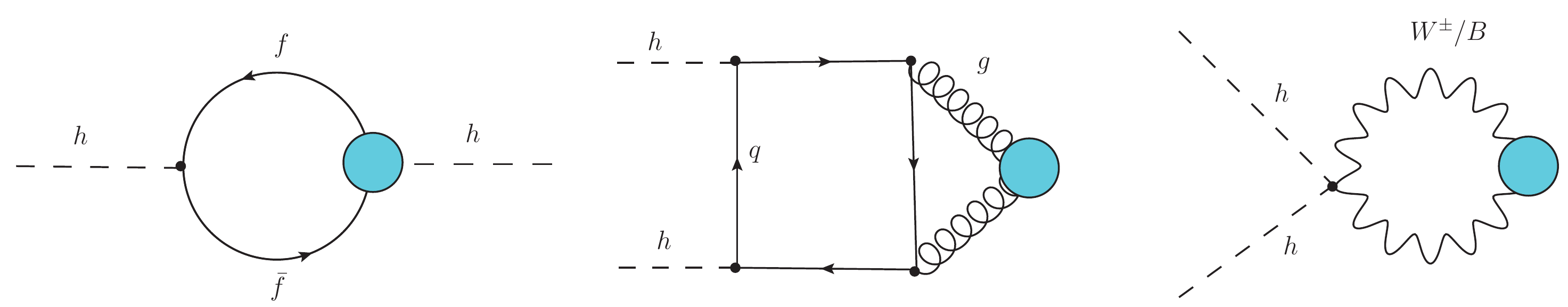}
		\caption{Diagrams that contribute to conformal symmetry breaking via mixing with the Higgs, in the quark/lepton portal, the gluon portal and the electroweak boson portal respectively. Blue circles indicate CFT operator insertions. }
		\label{fig:higgsmix}
	\end{center}
\end{figure}

Quantum corrections in the SM also introduce mixing among the SM operators. In effect, for each choice of the portal operator in Eq.~\leqn{eqn:interaction}, interactions of $\Oc$ with all other gauge-invariant SM operators are induced, with loop-suppressed coefficients. In particular, a coupling of the CFT to the Higgs portal operator is always generated. The leading contributions to this coupling for lepton, quark, gluon, weak and hypercharge portals are illustrated in Fig.~\ref{fig:higgsmix}. Below the weak scale, this coupling induces the deformation~\leqn{deform}. We refer to this mechanism as ``radiative mixing". The NDA estimates of the corresponding contribution to the gap scale for each portal are summarized in the third column of Table~\ref{tab:gaps}. Mixing with the other two type-I operators is also generically present, but their effect is subdominant since $\Lambda_{\rm QCD}\ll v$.    

\begin{figure}[t!]
	\begin{center}
		\includegraphics[width=14cm]{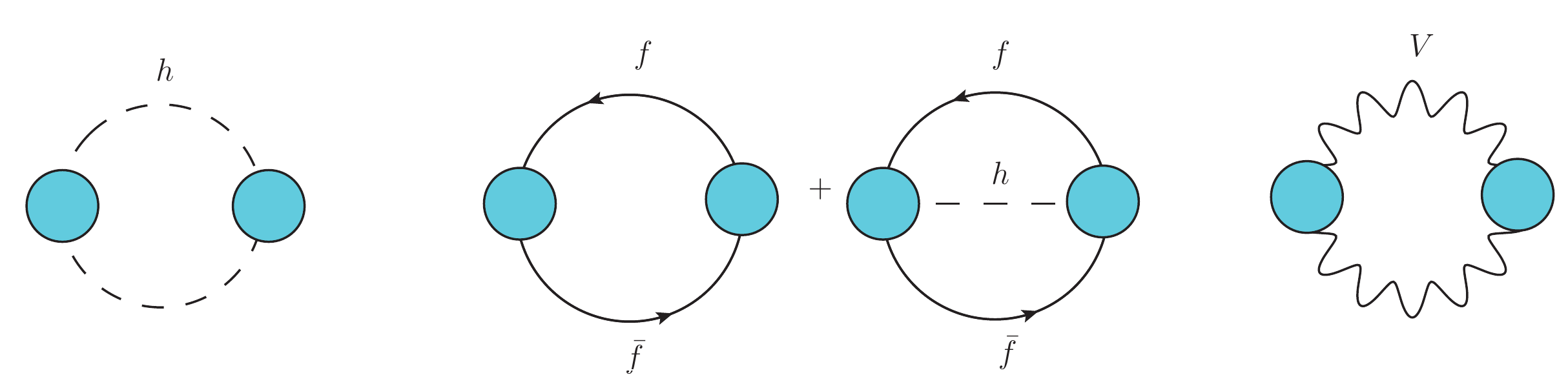}
		\caption{Diagrams that contribute to conformal symmetry breaking via generation of $\Oc^2$, in the Higgs portal, quark/lepton portal, and the gluon/weak boson portal respectively. Blue circles indicate CFT operator insertions. }
		\label{fig:oc2}
	\end{center}
\end{figure}

Another potential source of radiative breaking of conformal symmetry is the deformation
\beq
{\cal L} \,=\, c^\prime \, \Oc^2\,,
\eeq{deform2}
which can also be generated through SM loops. For example, the relevant diagrams for each portal are shown in  Fig.~\ref{fig:oc2}. If $\Oc^2$ is a relevant operator (which in the large-$N$ limit corresponds to $\Oc$ having $d \lesssim 2$), this leads to IR breaking of the conformal symmetry and generation of the gap scale. The NDA estimates of the resulting contribution to the gap scale are listed in the last column of Table~\ref{tab:gaps}. 

\begin{sidewaystable}[ph!]
	\begin{center}
		\begin{tabular}{ |c||c|c|c|c| }
			\hline
			&&&&\\[-0.7em]
			$\Os$ & $\Oc$ from             & Loop-induced & Loop-induced                          & $\Oc^2$ \\
			&  $\left< \Os \right>$  &  $\Oc$       &  $\Oc$ from $\left< H^\dagger H \right>$ &
			\\[0.4em]\hline &&&& \\[-0.7em] 
			$H^\dagger H$  &  $ \left( \frac{\l}{\L^{d-2}} v^2 \right)^\frac{1}{4-d}$ & 
			$\left(\frac{\l}{\L^{d-2}} \, \frac{\Lambda_{\scriptstyle_{\rm SM}}^2}{16\pi^2}\right)^\frac{1}{4-d}$ &
			--- &  $\left( \frac{\l}{\L^{d-2}} \frac{v}{m_h} \right)^\frac{1}{2-d}$
			\\[1em]\hline &&&&\\[-0.7em]  
			$H Q_L^\dagger q_R$  &  $ \left( \frac{\l}{\L^{d}} v \Lambda_{\scriptstyle_{\rm QCD}}^3 \right)^\frac{1}{4-d}$ & $\left( \frac{\l}{\L^{d}}\frac{ (\sum_i \kappa_i y_{q_i}) \, v^2 \Lambda_{\scriptstyle_{\rm SM}}^2}{(16\pi^2)^2} \right)^\frac{1}{4-d}$ &
			$\left( \frac{\l}{\L^{d}}\frac{ (\sum_i \kappa_i y_{q_i}) \, v^2 \Lambda_{\scriptstyle_{\rm SM}}^2}{16\pi^2} \right)^\frac{1}{4-d}$ &
			$\left( \frac{\l}{\L^d} \sqrt{ \frac{\Lambda_{\scriptstyle_{\rm SM}}^4}{(16\pi^2)^2} + \frac{v^4}{16\pi^2}}\right)^\frac{1}{2-d}$
			\\[1em]\hline &&&&\\[-0.7em]  
			$G^{\mu\nu}G_{\mu\nu}$  &  $\left( \frac{\l}{\L^{d}} \Lambda_{\scriptstyle_{\rm QCD}}^4 \right)^\frac{1}{4-d}$ & 
			$\left(\frac{\l}{\L^{d}} \, \frac{\Lambda_{\scriptstyle_{\rm SM}}^4}{16\pi^2}\right)^\frac{1}{4-d}$&
			$\left( \frac{\l}{\L^{d}}\frac{(\sum_i y^2_{q_i})\,\alpha_{\rm \scriptstyle s}\, v^2 \Lambda_{\scriptstyle_{\rm SM}}^2}{64\pi^3} \right)^\frac{1}{4-d}$ &
			$\left(\frac{\l}{\L^{d}} \frac{\Lambda_{\scriptstyle_{\rm SM}}^2}{4\pi} \right)^\frac{1}{2-d}$
			\\[1em]\hline &&&&\\[-0.7em] 
			$H L^\dagger \ell_R$ &  ---  & $\left( \frac{\l}{\L^{d}}\frac{(\sum_i \kappa_i y_{\ell_i}) \, v^2 \Lambda_{\scriptstyle_{\rm SM}}^2}{(16\pi^2)^2} \right)^\frac{1}{4-d}$ &
			$\left( \frac{\l}{\L^{d}}\frac{(\sum_i \kappa_i y_{\ell_i}) \, v^2 \Lambda_{\scriptstyle_{\rm SM}}^2}{16\pi^2} \right)^\frac{1}{4-d}$ & $\left( \frac{\l}{\L^d}  \sqrt{ \frac{\Lambda_{\scriptstyle_{\rm SM}}^4}{(16\pi^2)^2} + \frac{v^4}{16\pi^2}}\right)^\frac{1}{2-d}$
			\\[1em]\hline &&&&\\[-0.7em] 
			$W^{\mu\nu}W_{\mu\nu}$ &  ---  & 
			$\left(\frac{\l}{\L^{d}} \, \frac{\Lambda_{\scriptstyle_{\rm SM}}^4}{16\pi^2}\right)^\frac{1}{4-d}$&
			$\left( \frac{\l}{\L^{d}}\frac{g^2}{4} \frac{v^2 \Lambda_{\scriptstyle_{\rm SM}}^2}{16\pi^2} \right)^\frac{1}{4-d}$ & $\left(\frac{\l}{\L^{d}} \frac{\Lambda_{\scriptstyle_{\rm SM}}^2}{4\pi} \right)^\frac{1}{2-d}$
			\\[1em]\hline &&&&\\[-0.7em] 
			$B^{\mu\nu}B_{\mu\nu}$ &  ---  & 
			$\left(\frac{\l}{\L^{d}} \, \frac{\Lambda_{\scriptstyle_{\rm SM}}^4}{16\pi^2}\right)^\frac{1}{4-d}$& 
			$\left( \frac{\l}{\L^{d}}\frac{g^{\prime\, 2}}{4} \frac{v^2 \Lambda_{\scriptstyle_{\rm SM}}^2}{16\pi^2} \right)^\frac{1}{4-d}$ & $\left(\frac{\l}{\L^{d}} \frac{\Lambda_{\scriptstyle_{\rm SM}}^2}{4\pi} \right)^\frac{1}{2-d}$
			\\[1em]\hline
		\end{tabular}
		\caption{Standard Model operators and corresponding mass gaps generated through different sources. $\Lambda_{\scriptstyle_{\rm SM}}$ is the SM cutoff scale.} 
		\label{tab:gaps}
	\end{center}
\end{sidewaystable}

Depending on the parameters $\l$, $\L$ and $d$, each of the conformal symmetry-breaking contributions listed in Table~\ref{tab:gaps} may be dominant. We found that in the parameter space where the models successfully reproduce the observed dark matter relic density via freeze-in, $\Oc^2$ deformations are sub-leading to $\Oc$ deformations for all operators studied here. For the Higgs portal, the tree-level contribution to the gap scale dominates. For quark and lepton portals, the dominant source of conformal symmetry breaking is radiative mixing. For gauge-boson portals (gluon, weak and hypercharge), the radiative direct contribution is dominant. Note that for the quark and gluon portals, radiative contributions dominate over the tree-level one; this is primarily due to the hierarchy $v\gg \Lambda_{\rm QCD}$.

\subsubsection{Physics Below the Gap Scale}
\label{subsubsec:IRphysics}

Below the conformal symmetry breaking scale $\gap$, the dark sector is populated by particle-like excitations which are hadronic composite states of the original CFT degrees of freedom.\footnote{While a hadronic phase seems generic, another possible IR phase suggested by certain five-dimensional CFT duals is a ``gapped continuum"~\cite{Cabrer:2009we}. For a recent example of viable dark matter models with gapped continuum, see~\cite{Csaki:2021gfm,Csaki:2021xpy}.}
Predicting the spectrum of these excitations in a given CFT requires non-perturbative analysis, which is outside the scope of this paper. Instead, we will make a few simple, realistic assumptions about the properties of the low-energy theory, which will be sufficient to estimate the dark matter density and other quantities of interest up to order-one factors. 

We assume that the lightest of the CFT composite states $\chi$ is stable on cosmological time scales. This particle plays the role of dark matter. Stability may be due to a conserved global (discrete or continuous) symmetry under which $\chi$ (and possibly some other CFT composites) are charged, but SM states are all neutral. Further, as in \cite{Hong:2019nwd}, we posit that the DM particle is a pseudo-Goldstone boson (PGB) of an approximate global symmetry spontaneously broken at $\gap$. In this case, $\mdm \ll \gap$ is natural, with the DM mass dictated by the amount of explicit symmetry breaking.
This is necessary to satisfy self-interaction constraints~\cite{Markevitch:2003at, Lin:2019uvt}, as will be discussed in Section \ref{sec:pheno}. Notably, both the PGB property and a $\mathbb{Z}_2$ global symmetry are in fact realized for pions in QCD, although in that case the would-be stabilizing symmetry is anomalous leading to $\pi^0\to 2\gamma$ decay. (For other examples of models with dark pion playing the role of dark matter, see {\it e.g.}~\cite{Frigerio:2012uc,Hochberg:2014kqa}.)

Note that the ratio $r = \mdm/\gap$ is a free parameter of the theory. Phenomenologically, the value of $r$ is bounded from above by the self-interaction bound and from below by the warm dark matter constraint (since very light DM states can disrupt structure formation). 
It turns out that these considerations restrict $r$ to a parametrically narrow range, so that the theory remains highly predictive with respect to the DM mass and other relevant quantities. Fig.~\ref{fig:rplot} illustrates this for one of the models studied in this paper, while Section~\ref{sec:pheno} explains these constraints in detail.

In addition to $\chi$, the low-energy theory generically contains a set of bound states with masses $\sim \gap$. 
These states will couple to $\chi$ and mediate both DM self-interactions and its interactions with the Standard Model. We model these couplings as  
\beq
\mathcal{L} \sim g_\star \rho^\mu \left( \chi^\dagger \partial_\mu \chi + {\rm h.c.} \right)\,,
\eeq{rhoDM}
for a vector mediator $\rho^\mu$, and    
\beq
\mathcal{L} \sim \frac{g_\star}{\gap} \phi \left( \partial \chi \right)^2 ,
\eeq{phiDM}
for a scalar mediator $\phi$. The characteristic coupling can be estimated in the large-$N$ limit as
\beq
g_\star \sim \frac{4\pi}{\sqrt{N}}.
\eeq{gstar}
In a generic theory (such as QCD), both vector and scalar mesons will be present with comparable masses. 

The interactions of $\chi$ with the SM are obtained by matching the interaction Lagrangian in the CFT phase, Eq.~\leqn{eqn:interaction}, to the low-energy effective theory. Dimensional analysis and large-$N$ arguments suggest 
\beq
\Oc \longrightarrow \,\frac{\gap^{d-1}}{g_\star}\,\phi\,,
\eeq{Op_match}
while contributions from $\rho^\mu$ and $\chi$ are subdominant. 
This is seen by first noting that $\Oc$ is a scalar operator with scaling dimension $d$. Once the CFT confines, it is expected to ``interpolate'' a scalar operator made up of canonically normalized field operators of composite states. A single trace interpolation is given by the above equation where $\phi$ is a gauge invariant operator for a composite scalar. The factor $\gap^{d-1}$ is fixed by the dimensional analysis, while the factor $1/g_{\star}$ is determined by the large-$N$ counting. Explicitly, in the large-$N$ limit, $\langle \Oc \Oc \rangle \sim \frac{N}{16\pi^2} = \frac{1}{g_\star^2}$, suggesting that $\Oc \propto \frac{1}{g_\star}$. For $\rho^\mu$ or $\chi$, the interpolation relation is either that of a ``descendant'' or multi-trace. This is simply because $\Oc \sim \partial_\mu \rho^\mu$ by Lorentz invariance and $\Oc \sim (\partial \chi)
^2$ by the shift symmetry of $\chi$. This amounts to raising the effective dimension with more suppression by inverse powers of $\gap$, rendering them subdominant in the low-energy effective theory. 

\subsection{Ultraviolet Completion}
\label{subsec:UV}

There exists a natural UV completion of a dark-sector CFT considered above: SU(N) gauge theories with fixed points in the infrared \textit{a la} Banks-Zaks \cite{Caswell,BanksZaks}.\footnote{The UV theory may be any gauge theory with an interacting IR fixed point. The gauge group need not be $SU(N)$ and also we do not require the fixed point to be weakly interacting.} In the UV, an operator of this gauge theory, for example, a fermion bilinear, is coupled to the SM. At some scale $\L$, there is a fixed point and the UV gauge theory has a phase transition into the (generically strongly coupled) conformal phase. $\Oc$ is the operator in the conformal phase that corresponds to the original operator of the gauge theory. The matching for the example of a fermion bilinear operator is,
\beqa
{\cal L}_{\rm UV} = \frac{\lambda_{\rm BZ}}{M_{\rm BZ}^{\ds -1}} \O_{\rm SM} \bar{\Psi}\Psi \xrightarrow{\L}  \frac{\l}{\L^{D-4}} \,\Os \Oc\, \Rightarrow 
\l \approx \lambda_{\rm BZ} \left(\frac{\L}{M_{\rm BZ}}\right)^{\ds - 1}\,,
\eeqa{eq:BZint}
where $M_{\rm BZ}$ is the UV cutoff scale of the gauge theory, $\lambda_{\rm BZ}$ is the coupling and $\Psi$ is a fermion in the UV. We impose $\lambda_{\rm BZ} \sim \cal{O}$(1) as a naturalness condition in all the models we consider in the paper. Since $\ds > 1$ and $\L < M_{\rm BZ}$ , it is natural for $\l$ to be very small. The dark sector is never in equilibrium with the Standard Model, and dark sector energy density is produced through the freeze-in mechanism. In the next section, we will show that this mechanism can provide dark matter with the observed relic density.

\section{Cosmology and Relic Density}
\label{sec:Ops}

In this section, we outline the cosmological history of the dark sector, and estimate the resulting dark matter relic density for the six portal operators in Table~\ref{tab:gaps}. We find that each portal operator can provide a phenomenologically viable dark matter candidate. The key features of these candidates are summarized in Table~\ref{tab:summary}. Further, Figures~\ref{fig:DMHiggs} - 
\ref{fig:DMglu} and \ref{fig:DMweak} below illustrate the parameter space consistent with the observed dark matter density for each portal. Phenomenological and theoretical constraints on the model will be discussed in detail in Section~\ref{sec:pheno}.  

\subsection{Cosmological History of the Dark Sector}
\label{sec:history}

We consider the regime where the coupling between the SM and the dark sector is sufficiently small that the two sectors are not in thermal equilibrium at any time. At the end of inflation, the Standard Model sector is reheated to temperature $T_R$. We assume that the inflaton does not couple to the dark sector, so that the energy in the dark sector is zero at that time. (Without this assumption, the dark matter density receives a contribution depending on the details of the inflaton couplings and dynamics, and the model loses predictivity.) After reheating, SM collisions and decays can populate dark sector states via the interaction~\leqn{eqn:interaction}. We consider the ``Conformal Freeze-In" (COFI) scenario where
\beq
\gap < T_R < \Lambda_{\rm CFT}\,,
\eeq{CFTreheat}
so that the dark sector is described by a CFT in this epoch. This allows us to calculate energy transfer rates using the ``unparticle" approach of Georgi~\cite{Georgi:2007ek, Grinstein:2008qk}. The energy transferred to CFT quickly thermalizes due to strong coupling among the CFT states, but the CFT temperature $\Tc$ always remains below the SM plasma temperature $\Ts$. The transfer of energy from the SM plasma to the conformal dark sector continues until either the SM states coupled to the CFT become non-relativistic and drop out of equilibrium, or the SM temperature drops below the gap scale $\gap$. In either case, the dark sector eventually undergoes a confining phase transition at $\Tc\sim \gap$. The energy stored in the CFT degrees of freedom is transferred to the particle-like bound states of the dark sector, which then rapidly (compared to Hubble timescale) decay down to stable dark matter states. Given the small coupling of the dark sector to the SM, such decays would typically not involve SM states, so that essentially all of the energy stored in the CFT at the time of the phase transition ends up in dark matter.       

Quantitative predictions of dark matter relic density in the COFI scenario are obtained as follows. Energy transfer between the SM and CFT degrees of freedom is described by a Boltzmann equation,
\beqa
\frac{d\rhos}{dt} + 3H (\rhos + \Ps) =   -\Gamma_E({\rm SM}\rightarrow {\rm CFT}),
\eeqa{BeqSM}
where $H$ is the Hubble expansion rate, $\rhos$ and $\Ps$ are the energy density and pressure of the SM plasma, respectively, and $\Gamma_E$ is the energy transfer rate per unit volume given by
\beqa
\Gamma_E({\rm SM}\rightarrow {\rm CFT}) &=& \sum_{i,j} n_i n_j \langle \sigma(i+j\to {\rm CFT}) v_{\rm rel} E\rangle + \sum_i n_i \langle \Gamma(i\to {\rm CFT}) E\rangle\,.
\eeqa{GammaE}
Here the sums run over all SM degrees of freedom coupled to the CFT. The cross-sections and decay rates can be evaluated using the ``unparticle" technique of Georgi~\cite{Georgi:2007ek, Grinstein:2008qk}; 
an explicit example of such a calculation is given in Appendix~\ref{subapp:rd}. In the COFI scenario, the dark sector temperature $T_D$ remains well below the SM temperature, $T_D\ll \Ts$, throughout the cosmological history.  
For this reason, we have neglected the reverse energy transfer, from the CFT back to the SM sector, in Eq.~\leqn{BeqSM}. Conformal symmetry of the dark sector guarantees that its energy-momentum is traceless, $\Pc=\frac{1}{3}\rhoc$, and thus its energy density redshifts as radiation, $\rhoc \propto a^{-4}$, as the universe expands. At the time when the CFT sector is populated,  the energy density in the SM sector is dominated by relativistic matter, so that SM and CFT energy densities redshift in the same way. The total energy of the two sectors can only change due to work done against the expansion of the universe:
\beq
\frac{d}{dt}\left(\rhoc+\rhos\right) + 4H \left(\rhoc+\rhos\right)  = 0.
\eeq{total}
Subtracting Eq.~\leqn{BeqSM}, we find that the CFT energy density evolves according to 
\beq
\frac{d\rhoc}{dt} + 4H \rhoc  =  \Gamma_E({\rm SM}\rightarrow {\rm CFT})\,.
\eeq{BeqCFT}
Solving this equation, with the initial condition $\rhoc=0$ at $\Ts=T_R$, yields the CFT energy density as a function of the SM temperature $T$. 

\begin{figure}[t!]
	\center
	\includegraphics[width=12cm]{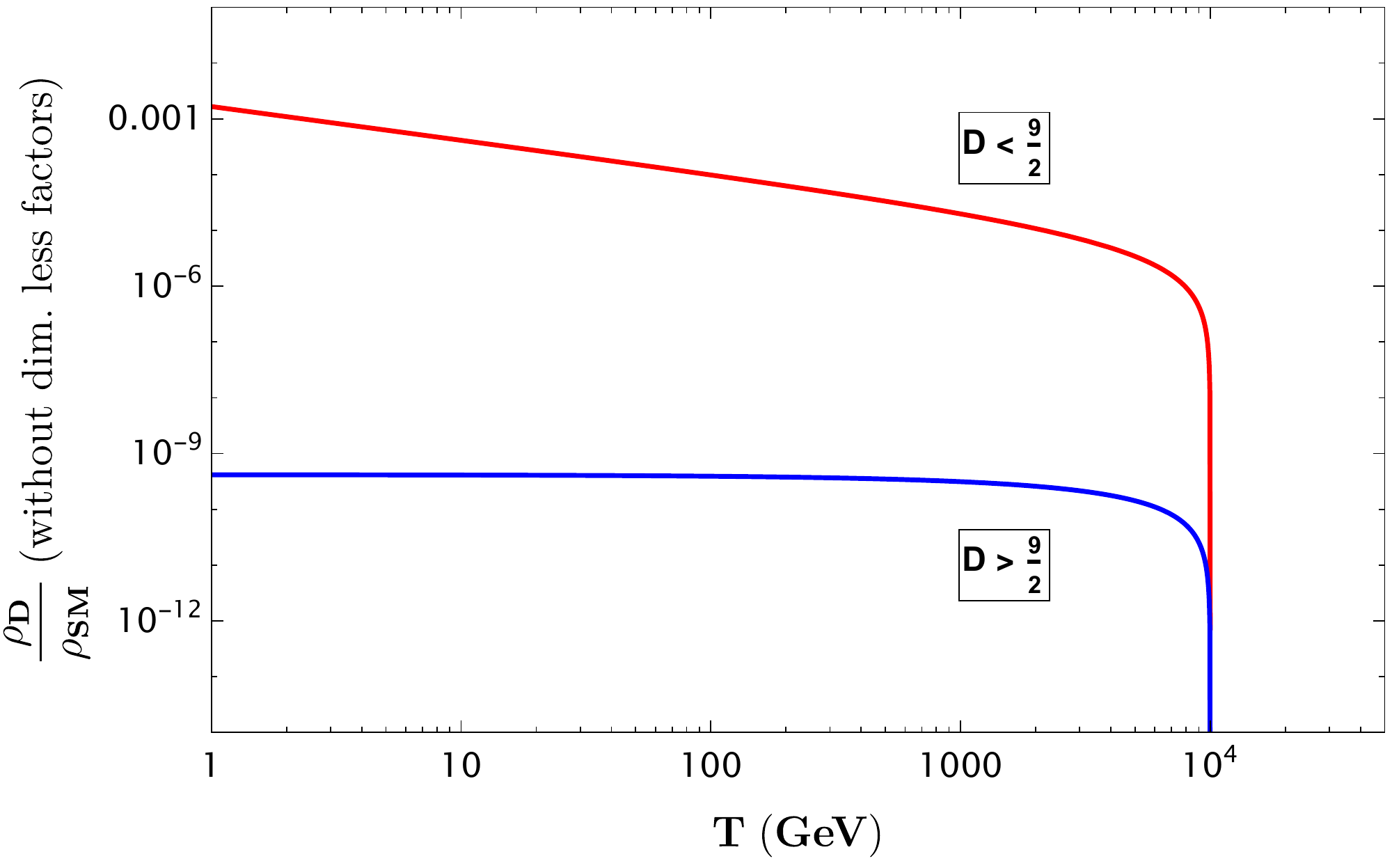}
	\caption{Dark sector energy density (normalized by SM energy density) vs.~temperature of the Standard Model plasma, for two different values of $D$. The red curve ($D < 9/2$) shows IR-dominant production, while the blue curve ($D > 9/2$) shows UV-dominant production.}
	\label{fig:critD}
\end{figure}

It is instructive to discuss an analytic solution of~\leqn{BeqCFT} for the simple case when the energy transfer rate is given by 
\beq
\Gamma_E({\rm SM}\rightarrow {\rm CFT}) \sim \frac{\l^2}{\L^{2(D-4)}}\,\Ts^{2D-3}.
\eeq{examp}
This scaling occurs when the SM temperature $\Ts$ is well above all relevant SM energy scales (such as masses) and the mass gap of the dark sector.\footnote{This regime, where the SM itself is approximately conformal, was also considered in Ref.~\cite{Redi:2021ipn}.} This can be easily shown via simple dimensional analysis, keeping in mind that the SM temperature is the only relevant dimensionful scale besides (the square of) the coupling to the dark sector. Integrating~\leqn{BeqCFT}, the energy density of the dark sector grows as
\beq
\rhoc \sim \frac{M_{\rm pl}}{\L^{2D-8}}\; \left[ \; T^4 \; \left( \frac{T_R^{2D-9}-T^{2D-9}}{2D-9} \right)\right],
\eeq{critD}
where $M_{\rm pl}$ is the Planck mass.

For values of $D$ below the critical dimension $D = \ds + d = 4.5$, most of the dark sector energy density is produced at low temperature (``in the infrared") and the dark matter relic density can be predicted without knowledge of UV physics and the reheating temperature. (See Fig.~\ref{fig:critD}.) This is similar to the original freeze-in scenario of Hall {\it et.al.}~\cite{Hall:2009bx}. For $D> 4.5$, most of the dark sector energy density is produced soon after the reheating. In this case, the predicted dark matter relic density does depend on $T_R$. However, in practice this dependence is weak, due to the low powers in the exponent for $T_R$ compared to the dependence on the mass gap, as will be shown later in this section.  

\begin{table}[t!]
	\begin{center}
		\begin{tabular}{ |c||c|c|c|c| }
			\hline
			&&&&\\[-0.7em]
			$\Os$  & DM Mass           & DM Mass            & Dominant CFT & Dominant  \\
			& (Scalar Mediator) & (Vector Mediator)  & Deformation  & Production Mode
			\\[0.4em]\hline &&&&\\[-0.9em] 
			$H^\dag H$ & 0.4 - 1.2 MeV & 40 - 400 keV &  Tree-level & $h \rightarrow \mathrm{CFT}$
			\\[0.4em]\hline &&&&\\[-0.9em] 
			$H Q^\dag q$ & \makecell{1st: \cancel{SN} \\ All: 0.1 - 1 MeV \\ MFV: 0.5 - 5 MeV} & 
			\makecell{1st: \cancel{SN} \\ All: 50 - 200  keV \\ MFV: 0.1 - 1 MeV} & Radiative mixing & $q\bar{q} \rightarrow \mathrm{CFT} $
			\\[0.4em]\hline &&&&\\[-0.9em]
			$H L^\dagger \ell_R$ & \makecell{1st: \cancel{WDM} \\ All: 3 - 10 keV \\ MFV: 10 - 100 keV } & 
			\makecell{1st: \cancel{WDM} \\ All: \cancel{WDM} \\ MFV:  \cancel{WDM}} & Radiative mixing & $\ell \bar{\ell} \rightarrow \mathrm{CFT} $
			\\[0.4em]\hline &&&&\\[-0.9em] 
			$G^{\mu\nu}G_{\mu\nu}$ & 0.2 - 2 MeV  & 50 - 400 keV & Radiative direct & $g g \rightarrow \mathrm{CFT} $ 
			\\[0.4em]\hline &&&&\\[-0.9em] 
			$B^{\mu\nu}B_{\mu\nu}$ & 0.1 - 10 MeV & 0.05 - 1 MeV & Radiative direct & $ \gamma \gamma \rightarrow \mathrm{CFT} $ 
			\\[0.4em]\hline
		\end{tabular}
		\caption{Summary table for each SM operator portal considered. In this table, \cancel{SN} stands for models that are ruled out by supernova cooling constraints, and \cancel{WDM} stands for models that are ruled out by warm dark matter constraints.}
		\label{tab:summary}
	\end{center}
\end{table}

The Boltzmann equation~\leqn{BeqCFT}, with energy transfer rates calculated within the `unparticle' approach, is valid as long as $\Ts>\gap$ (required for the validity of the collision term) and $T_D>m_{\rm DM}$ (required for radiation-like Hubble term). As the universe expands and cools, both conditions may become invalid, requiring modifications to the Boltzmann equation. For $T_D<m_{\rm DM}$, we simply replace $4H\to 3H$ in the Hubble term, since at these temperatures the dark sector is populated by non-relativistic dark matter particles. For $\gap>\Ts>m_{\rm DM}$, we consider dark matter production in the ``hadronic phase". The corresponding collision term is calculated within the low-energy effective theory discussed in Section~\ref{subsec:IR}. Note that production in the hadronic phase only occurs if the SM particles interacting with the CFT are light (electrons or photons); in all other cases, the relevant SM particles drop out of the thermal bath at $\Ts>\gap$ and all production is in the CFT regime. Moreover, we find that for all portal interactions considered here, dark matter production in the hadronic phase is subdominant to production in the CFT regime, with the exception of a small region in the parameter space of the lepton-portal model. 

We note that in the COFI scenario, it is possible that at some time in the cosmological history $\Ts>\gap > \Tc$. In this regime, the thermal bath of the dark sector is described by particle-like bound-state excitations. However, the energy transfer from the SM to the dark sector can still be described within the unparticle approach, since the energy transferred in a single collision is above $\gap$. This is analogous to using the parton model to calculate (inclusive) rates of hadron production at the LHC, even though no quark-gluon plasma is produced.   

With the low-temperature modifications outlined above, Eq.~\leqn{BeqCFT} remains valid to present day.  Integrating this equation, with energy transfer rates evaluated separately for each portal, provides predictions for current dark matter relic density which can be compared with the observed value, $\Omega \, h^2=0.1$. These predictions will be discussed in the rest of this section. 

\subsection{Higgs Portal: $\Os = H^\dag H$}
\label{sec:HH}

There are multiple mechanisms of SM $\to$ dark sector energy transfer in the $H^\dag H$ portal model. For $\Ts$ between the reheating temperature ($T_R$) and the weak scale, the leading mechanism is the scattering process $HH \rightarrow $ CFT. After the electroweak phase transition, one Higgs in the interaction term can be replaced with its VEV and dark energy density will be produced through Higgs decay. Additionally, there is production from quark and gluon fusion through a Higgs portal. Quark fusion continues until the quarks fall out of the thermal bath. Other contributing processes include heavy quark to light quark + CFT decay and pion annihilation below $\Lambda_{\scriptstyle_{\rm QCD}}$. These are subdominant due to phase space factors and can be neglected. It can be shown that the Higgs decay process is the dominant production mechanism, provided that production is IR dominated with $D < 4.5$ (or equivalently the CFT operator dimension $d < d_* = 2.5$).

\begin{figure}[t!]
	\center
	\subfloat{%
		\includegraphics[width=7.3cm]{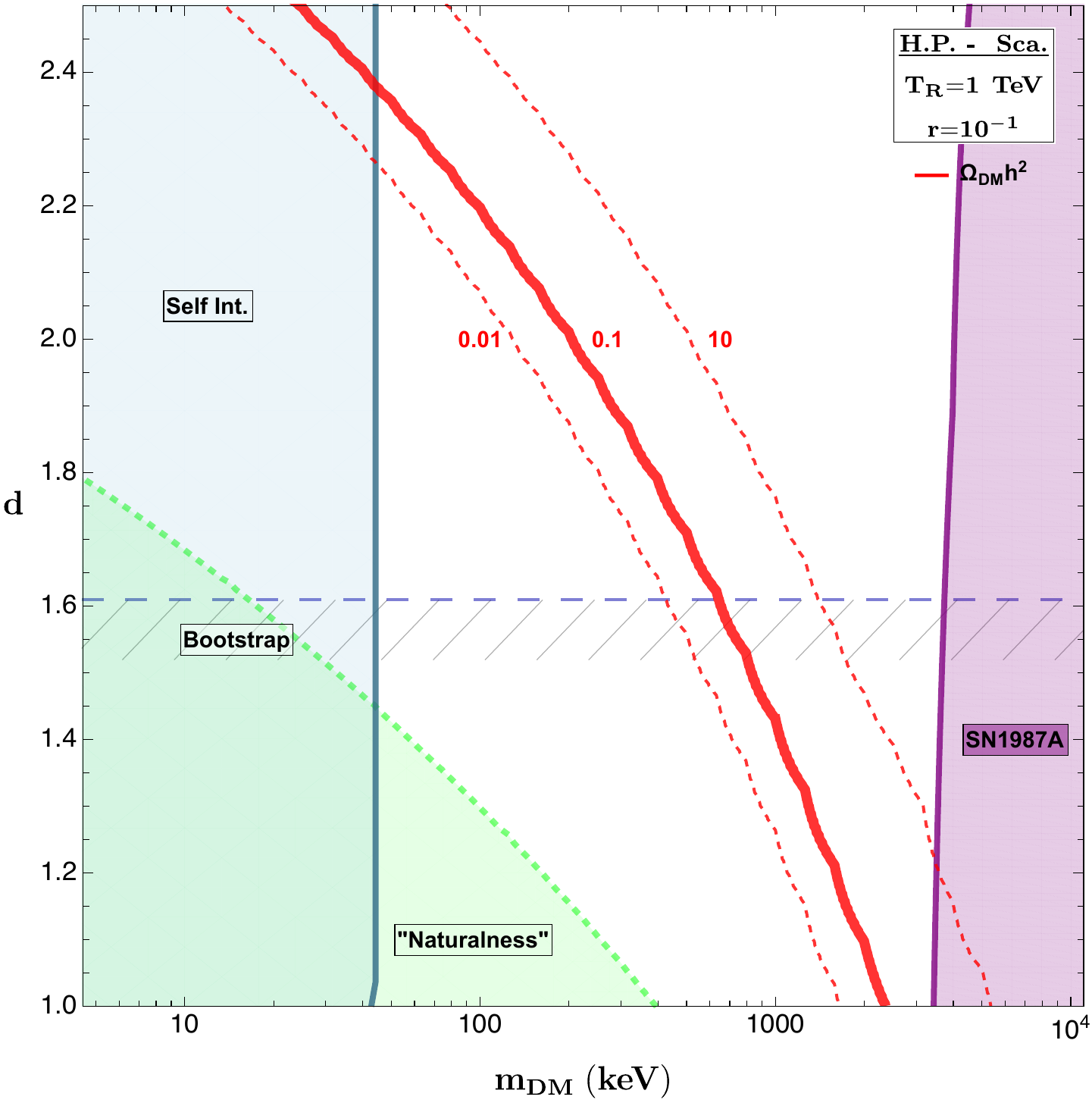}
		\includegraphics[width=7.3cm]{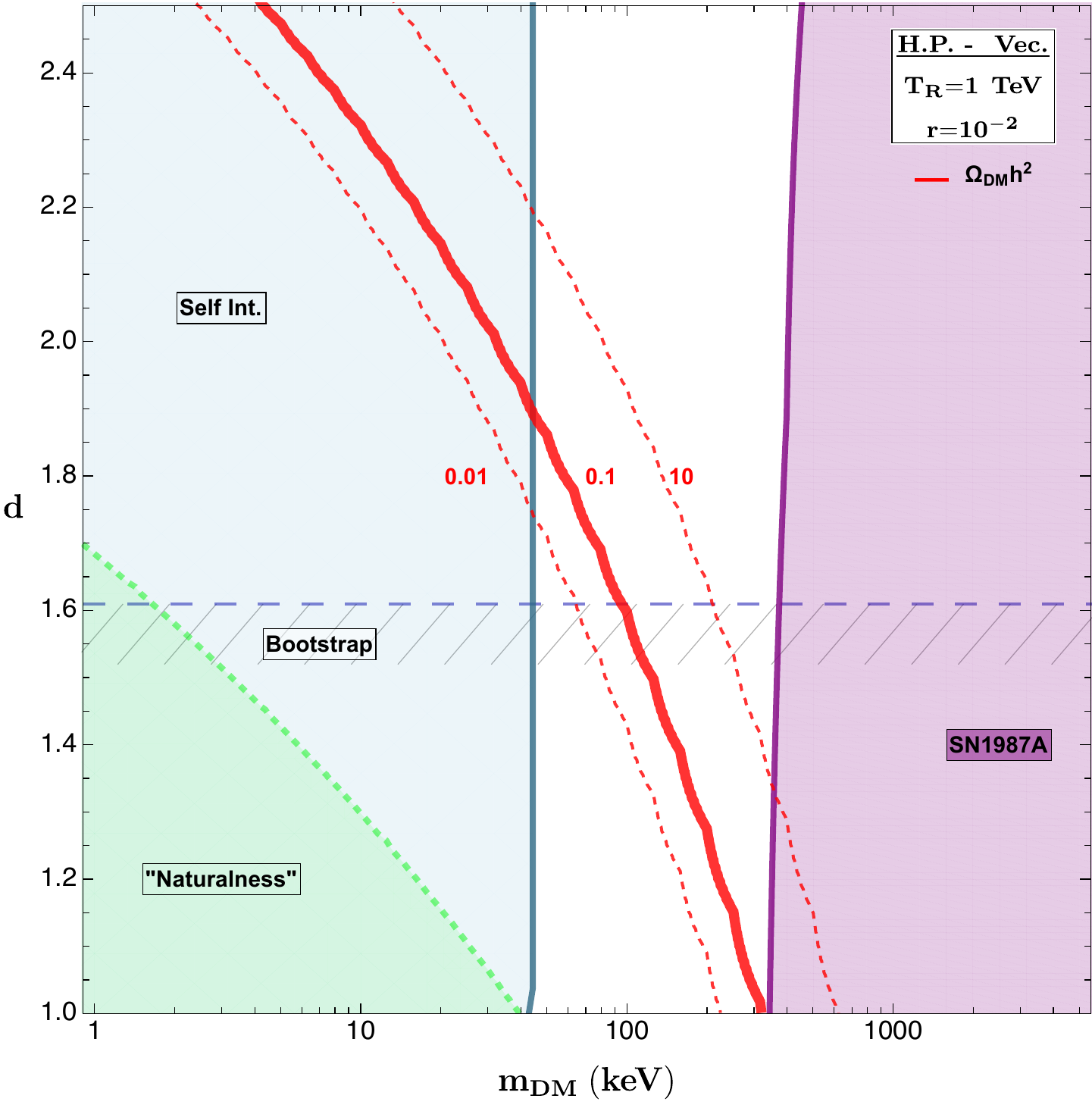}}
	\caption{Dark matter relic density contours (red) and observational/theoretical constraints, in the Higgs portal model, with a scalar mediator (left) and a vector mediator (right). The solid red line indicates parameters where the observed dark matter abundance is reproduced.}
	\label{fig:DMHiggs}
\end{figure}

An analytic approximation for the relic density can be obtained by considering only the dominant mode of production: Higgs decay. The collision term in the Boltzmann equation is given by,
\beq
\Gamma_E({\rm SM}\rightarrow {\rm CFT}) = n_h \langle \Gamma(h\to {\rm CFT}) E\rangle \,=\, \frac{f_d \lambda_{\scriptstyle_{\rm CFT}}^2 v^2 m_H^{2(d-1)} T}{\Lambda_{\scriptstyle_{\rm CFT}}^{2d-4}}\,K_2(m_H/T),
\eeq{hh_decay}
where $f_d=2^{-2d}\pi^{1/2-2d}\Gamma(d+1/2)/(\Gamma(d-1)\Gamma(2d))$, $v$ is the Higgs VEV, $m_H$ is the Higgs boson mass and $K_2(x)$ is the modified Bessel function of the second kind.

Using Eqs.~\leqn{hh_decay} and~\leqn{BeqCFT}, the current relic density of dark matter can be calculated. This yields
\beqa
\frac{\Omega_{\scriptstyle_{\rm DM}} h^2}{0.1} = \left[ \frac{\mdm}{1\mev} \right]\left[ \frac{\left(A \, f_d^3 \, g_*^{-9/2}\right)^{1/4}}{ 10^{-5}} \right]\left[ \frac{\left( \frac{\gap}{m_h} \right)^{(6-\frac{3d}{2})}}{10^{-12}} \right]. 
\eeqa{anal_HP}
Here, $g_* \equiv g_* (m_H)$ is the effective number of SM degrees of freedom when $\Ts = m_H$ and $A$ is a model-dependent constant that represents the number of degrees of freedom of the dark sector as $\rhoc \equiv A \, \mdm^4$. We have used the mass gap formula from Table \ref{tab:gaps} to convert the interaction coupling dependence to mass gap dependence as, $$\gap = \left( \frac{\l}{\L^{d-2}} v^2 \right)^\frac{1}{4-d}.$$
The ratios in each bracket are $\mathcal{O}(1)$ for $1 < d < 2.5$ and $A \sim \mathcal{O}(1)$. Thus, we expect a mass-gap for the Higgs portal model at the MeV scale. For details of this calculation, see Appendix \ref{subapp:rd}. This result is in good agreement with the numerical integration of the Boltzmann equation. 

The dark matter mass $\mdm$ and the dimension $d$ of the CFT operator that produce the correct observed relic density are shown in Fig.~\ref{fig:DMHiggs}. Since the dark sector is mostly populated through Higgs decays which occur at temperatures below the weak scale, the relic density is independent of the reheating temperature or any other UV-scale parameters. Fig.~\ref{fig:DMHiggs} also shows phenomenological and theoretical constraints on the model, which will be discussed in detail in Section~\ref{sec:pheno}. We observe that the model produces a viable DM candidate with masses $\mdm\sim 0.1-1$ MeV. In these figures, we have fixed the value of $r=\mdm/\gap$ (see Section~\ref{subsec:IR} for the discussion of this parameter). The ratio $r$ is tightly constrained by the combination of bounds from large-scale structure (warm dark matter) and dark matter self-interactions. Given these bounds, $r$ can only be varied by a factor of at most a few relative to the values shown. Such variation does not have a strong effect on the predicted dark matter mass range. 

\subsection{Quark \& Lepton Portals: $\Os = H Q^\dag q, \ H L^\dagger \ell_R$ }
\label{subsec:QP_LP_three_scenarios}

Above the weak scale, energy transfer from the SM to the dark sector occurs via scattering processes $ H f \bar{f} \rightarrow \CFT $ and $H f \rightarrow f + \CFT $, where $f$ refers to quarks or leptons depending on the SM operator used. The energy transfer rate in these channels peaks at high temperatures, introducing dependence on the reheat temperature $T_R$. Below the weak scale, $\Os$ is matched onto a dimension-3 bilinear fermion operator. The dominant process contributing to production of CFT energy density is fermion annihilation $f \bar{f} \rightarrow \CFT $. We find that for $T_R\lsim$~few TeV, production below the weak scale is dominant and the resulting DM relic density is independent of $T_R$. For $D < 4.5 \Rightarrow d < 1.5$, the energy transfer through fermion annihilation peaks at low temperatures, while for $d > 1.5$, temperatures of order the weak scale dominate.  

\begin{figure}[t!]
	\center
	\subfloat{%
		\includegraphics[width=7.3cm]{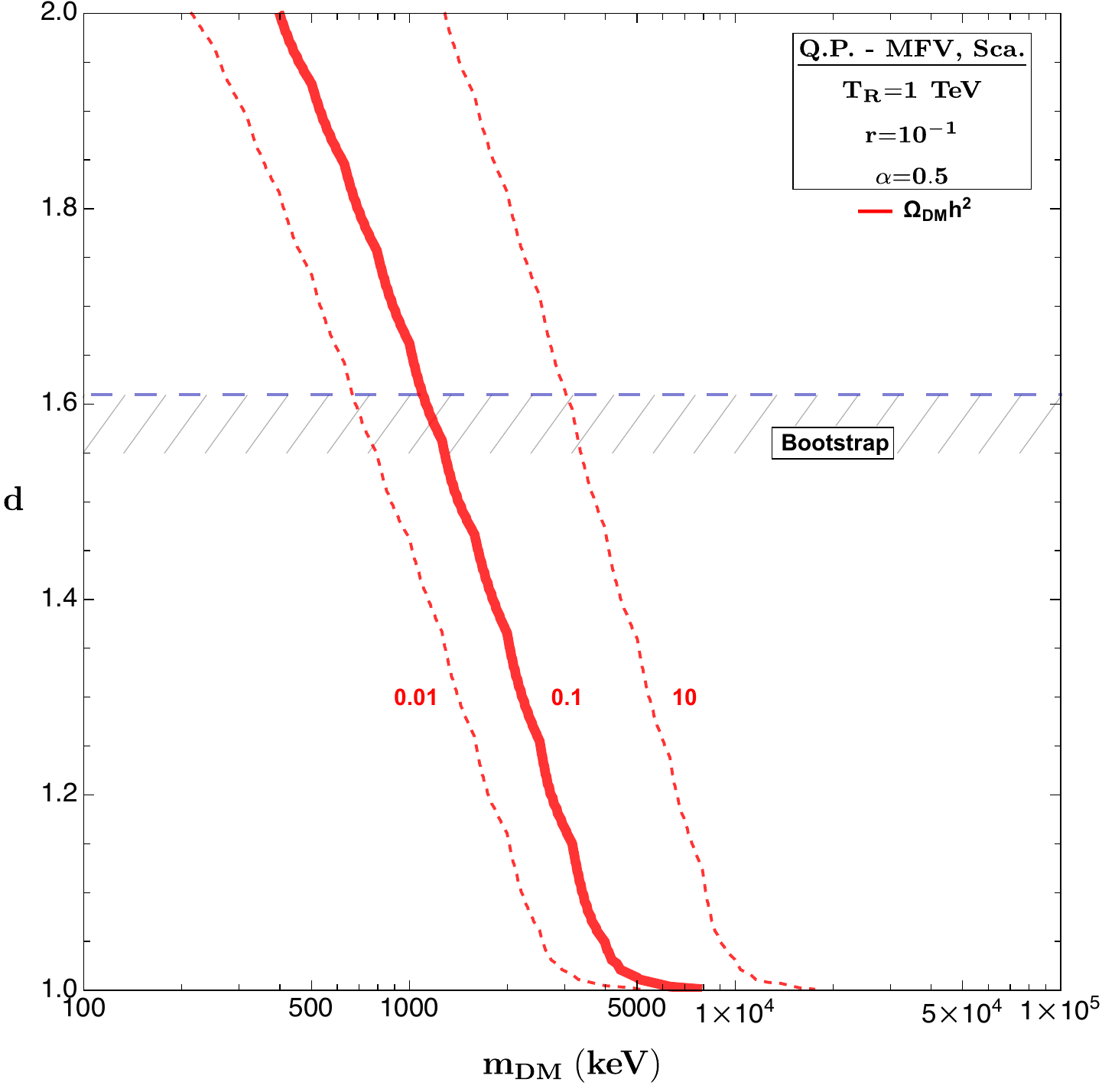}
		\includegraphics[width=7.3cm]{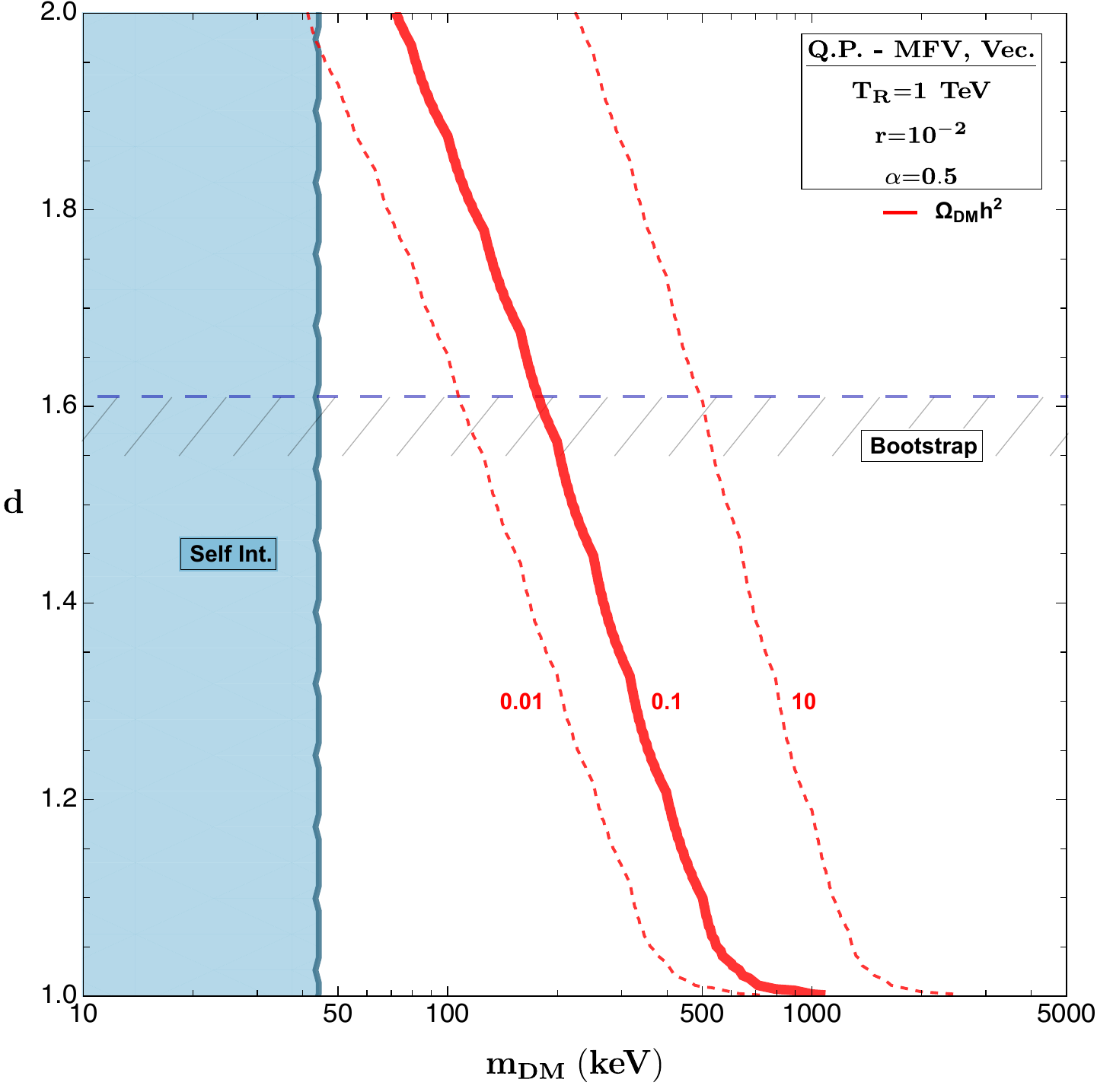}}
	\caption{Dark matter relic density contours (red) and observational/theoretical constraints, in the quark portal model with minimal flavor violation couplings, with a scalar (vector) mediator on the left (right). }
	\label{fig:DMquark1}
\end{figure}

\begin{figure}[t!]
	\center
	\subfloat{%
		\includegraphics[width=7.3cm]{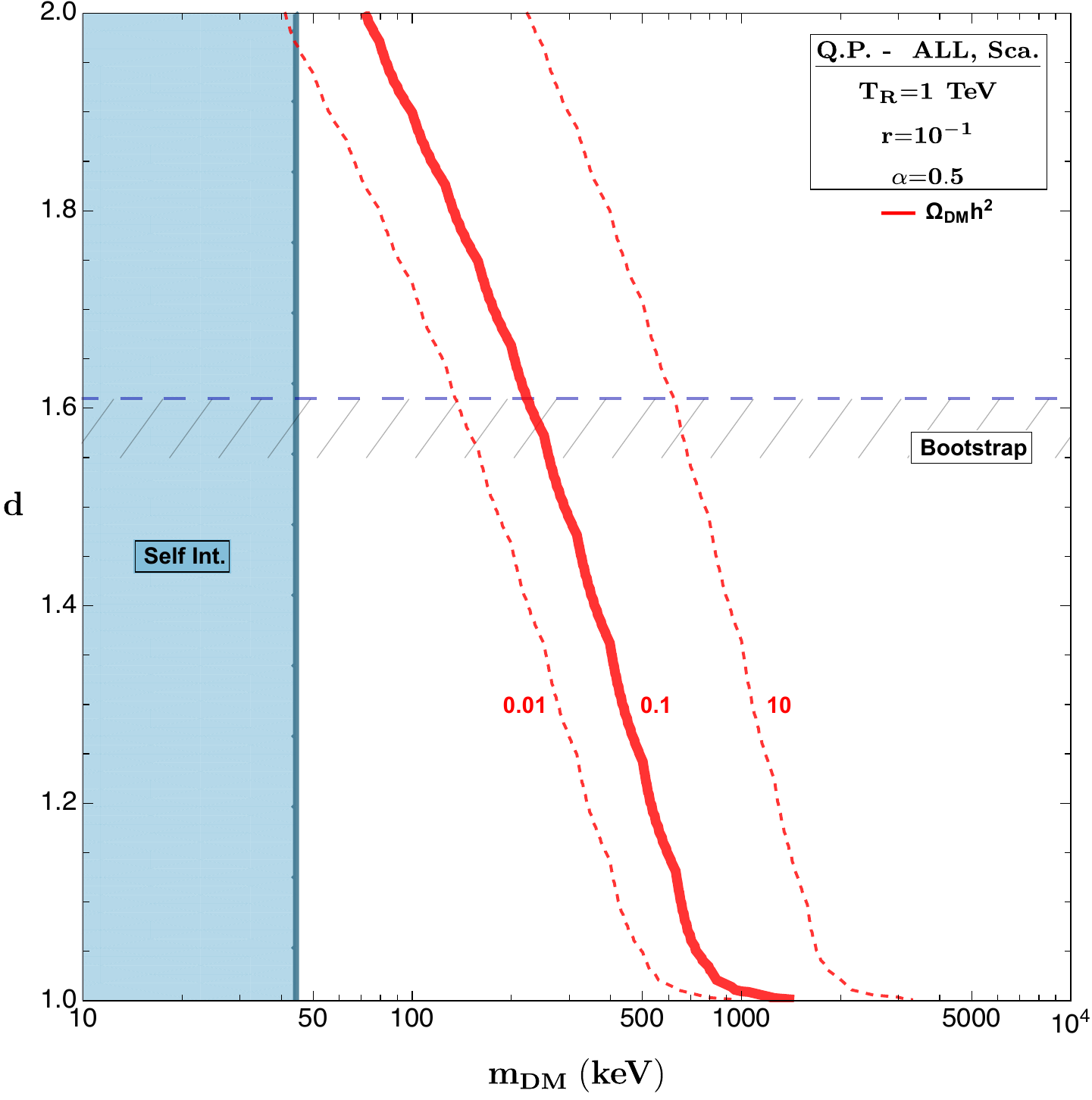}
		\includegraphics[width=7.3cm]{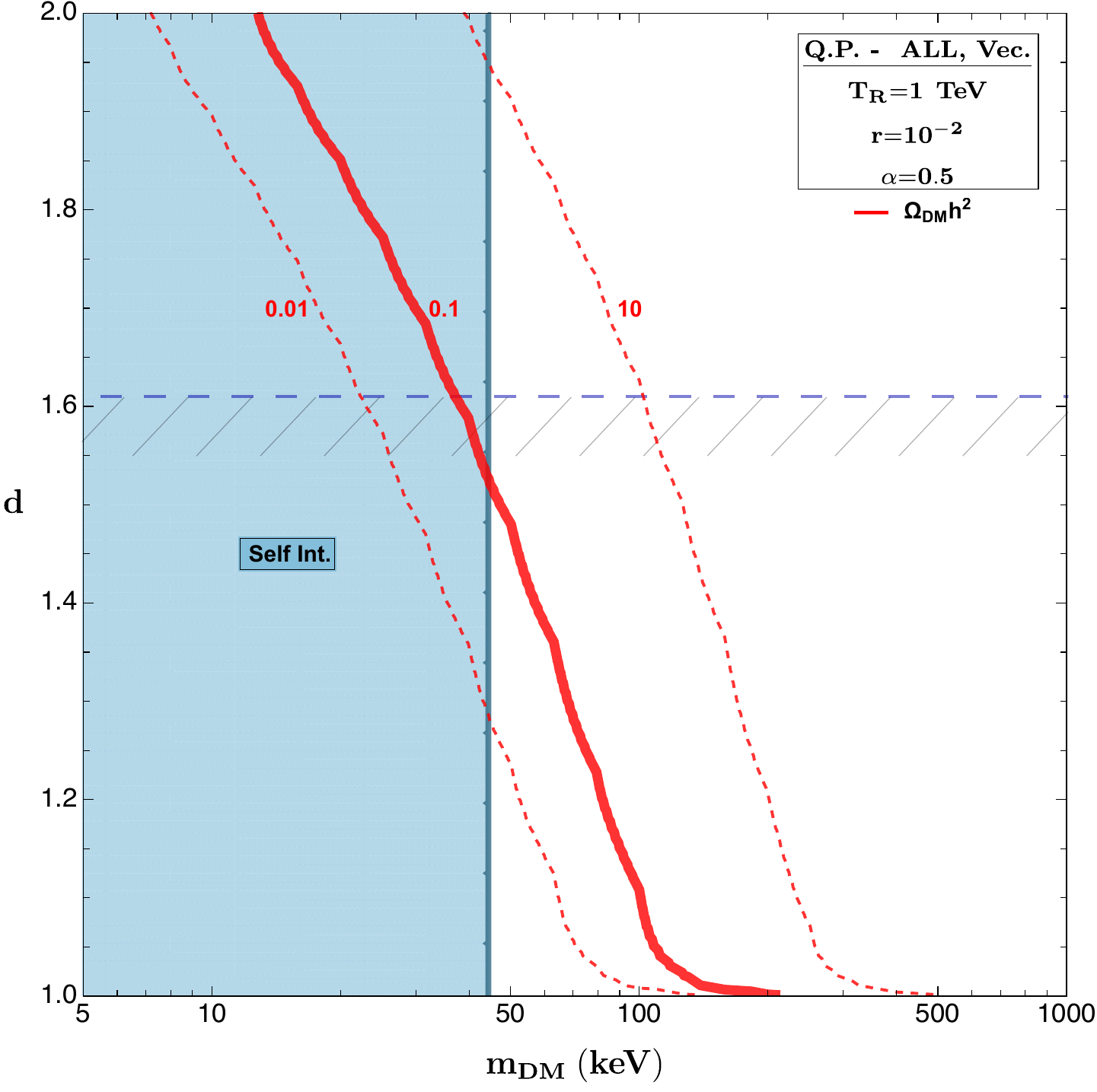}}
	\caption{Dark matter relic density contours (red) and observational/theoretical constraints, in the quark portal model with democratic couplings, with a scalar (vector) mediator on the left (right). }
	\label{fig:DMquark2}
\end{figure}

For the quark portal, conformal freeze-in continues until $T = \Lambda_{\scriptstyle_{\rm QCD}}$ or $T = \gap$, whichever happens first. For the lepton portal, it continues until $T = m_e$ or $T = \gap$. Again, we assume that there are dark pions that form the dark matter relic density we observe today, that are a factor $r \sim 0.01$ (with scalar mediator) or $r \sim 0.001$ (with vector mediator) lighter than the mass gap induced by the Standard Model deformation. The dark sector energy density redshifts as radiation until $\Tc$ hits $\mdm = m_\chi$, and redshifts as matter afterwards, until today. 

Notably, in the lepton portal, it is possible for the SM temperature at which $\Tc$ hits $\mdm$ to be higher than the stopping temperature. In the short period when the universe cools from the former temperature to the latter, the DM energy density is produced in the CFT phase, but hadronizes quickly to matter and redshifts as matter. Additionally, in parts of the parameter space of the lepton portal, production can also be dominated by hadronic processes, where most of the energy density is produced below $T=\gap$ through the processes involving the IR composite states. This is the case for the grey shaded regions in Fig.~\ref{fig:DMlep2}. See Appendix~\ref{subapp:had} for details of thermally averaged hadronic cross-sections and production rates.

As discussed in Section~\ref{sec:Ops}, we consider three scenarios for flavor structure of the quark/lepton portal couplings: Minimal Flavor Violation, Democratic, and First-Generation Only. The three scenarios give different mass gap scales for which the correct relic abundance is produced. 

The energy density ($\rhoc$) produced through the dominant process of fermionic scattering scales as follows for each structure: 
\begin{itemize}
	\item First Generation Only: $\rhoc \sim M_{\rm pl} \left( \frac{m^{4-d}}{\alpha^2 v^2 m_1} \right)^2 \, T^4 \, (v^{2d-3} - T^{2d-3})$
	\hfill\refstepcounter{equation}\textup{(\theequation)}%
	\item Democratic: $\rhoc \sim M_{\rm pl} \left( \frac{m^{4-d}}{\alpha^2 v^2 \sum_i m_i} \right)^2 \,  T^4 \, (v^{2d-3} - T^{2d-3})$ 
	\hfill\refstepcounter{equation}\textup{(\theequation)}%
	\item Minimal Flavor Violation: $\rhoc \sim M_{\rm pl} \; m_j^2 \left( \frac{m^{4-d}}{\alpha^2 v^2 \sum_i m_i^2} \right)^2 \,  T^4 \, (v^{2d-3} - T^{2d-3})$ \hfill\refstepcounter{equation}\textup{(\theequation)}%
\end{itemize}
where $\Lambda_{\scriptstyle \rm SM} \equiv \alpha \, v$ and $m_i$ stands for the relevant fermion masses. At the end of the freeze-in process for each interacting fermion, $T = {\rm Max}[m_i, \, \gap]$ for the lepton portal and $ T = {\rm Max}[m_i, \, \Lambda_{\scriptstyle \rm QCD}, \, \gap]$ for the quark portal. Each of these contributions is summed and appropriately redshifted to obtain the relic density. See Appendix~\ref{subapp:rd} for the relic density equations for each flavor structure and portal.

Of the three scenarios, the MFV model is the least constrained, due to suppressed couplings to the first generation of fermions. In the quark portal, the First-Generation Only scheme is ruled out by supernova cooling constraints from SN1987A data (for both scalar and vector mediators). The other four models are viable and the plots are shown in Figs.~\ref{fig:DMquark1} and ~\ref{fig:DMquark2}.

\begin{figure}[t!]
	\center
	\subfloat{%
		\includegraphics[width=7.3cm]{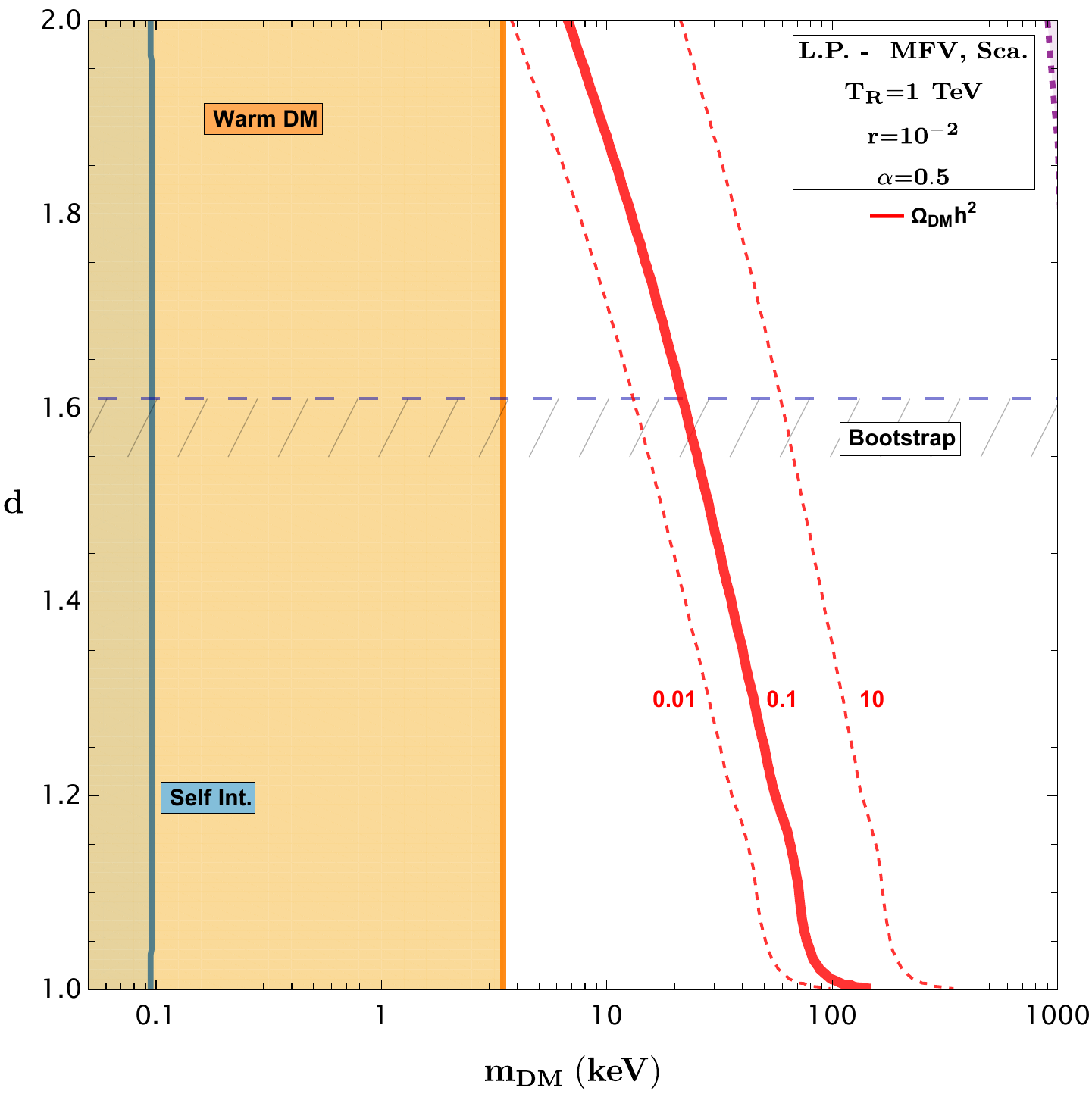}
		\includegraphics[width=7.3cm]{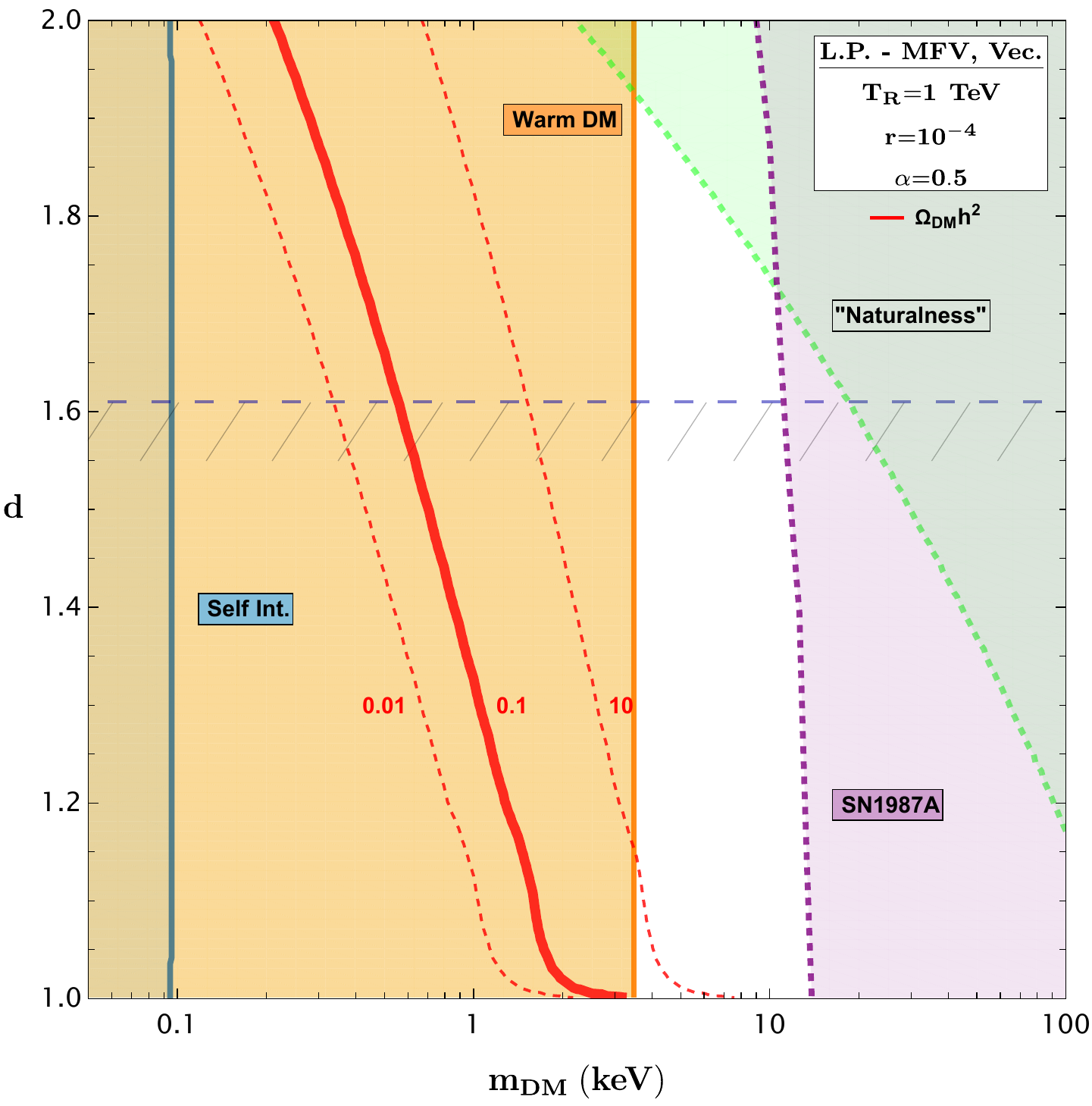}}
	\caption{Dark matter relic density contours (red) and observational/theoretical constraints, in the lepton portal model with minimal flavor violation couplings, with a scalar (vector) mediator on the left (right). }
	\label{fig:DMlep1}
\end{figure}

In the lepton portal, the mass of the DM candidate with correct relic abundance tends to be lower than in other models, and the bound on dark matter free-streaming length from the Lyman-$\alpha$ forest data~\cite{Irsic:2017ixq} plays a major role in constraining the models. This is illustrated in Figs.~\ref{fig:DMlep1} and ~\ref{fig:DMlep2}. The viability of COFI dark matter in this case depends on the details of the model: for example, MFV and democratic models with a scalar mediator predict $m_{\rm DM} \gsim 10$~keV and are consistent with observations, while in other cases $m_{\rm DM} \sim 1$~keV and the models are ruled out.


\begin{figure}[h]
	\center
	\subfloat{%
		\includegraphics[width=7.3cm]{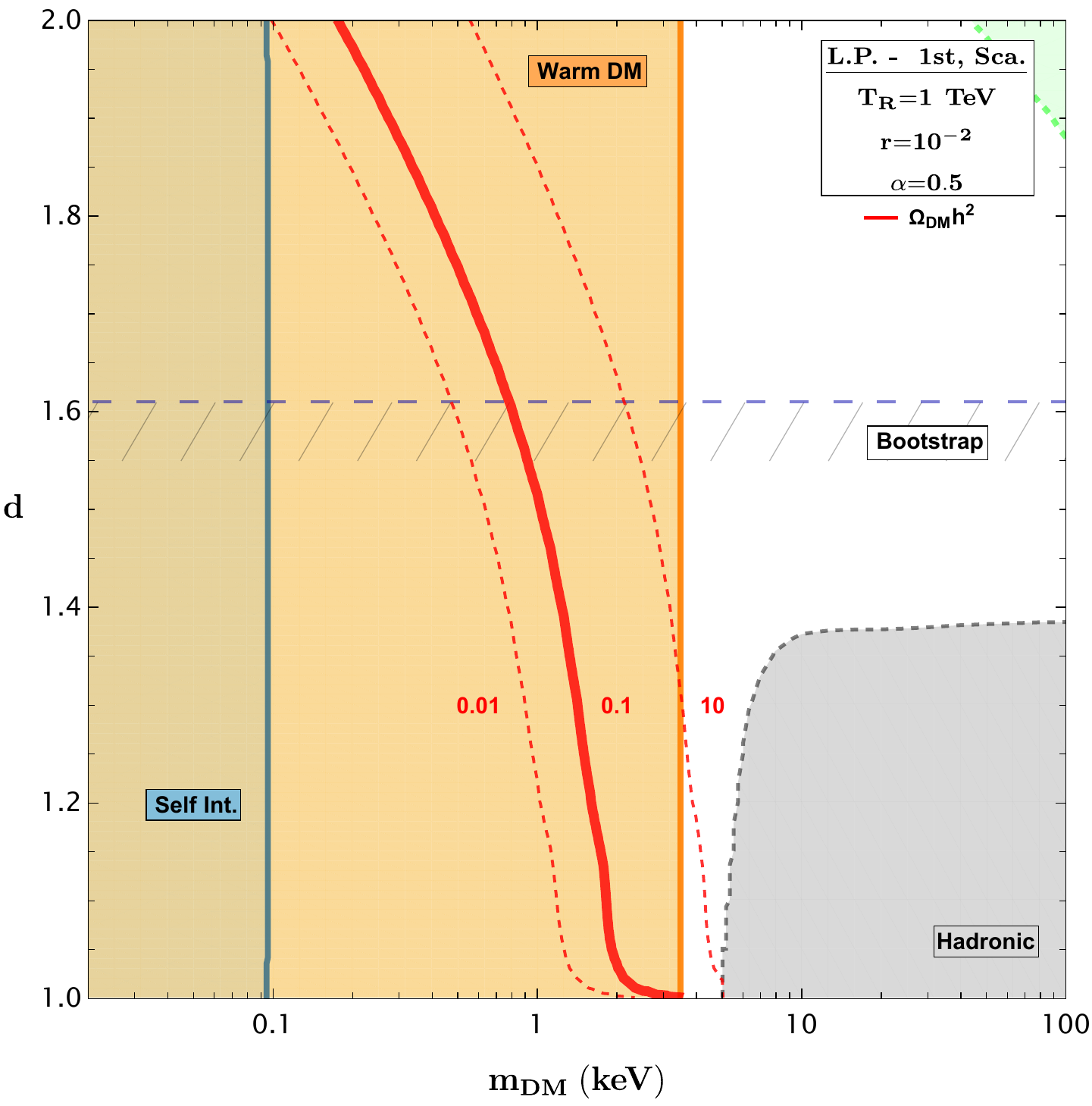}
		\includegraphics[width=7.3cm]{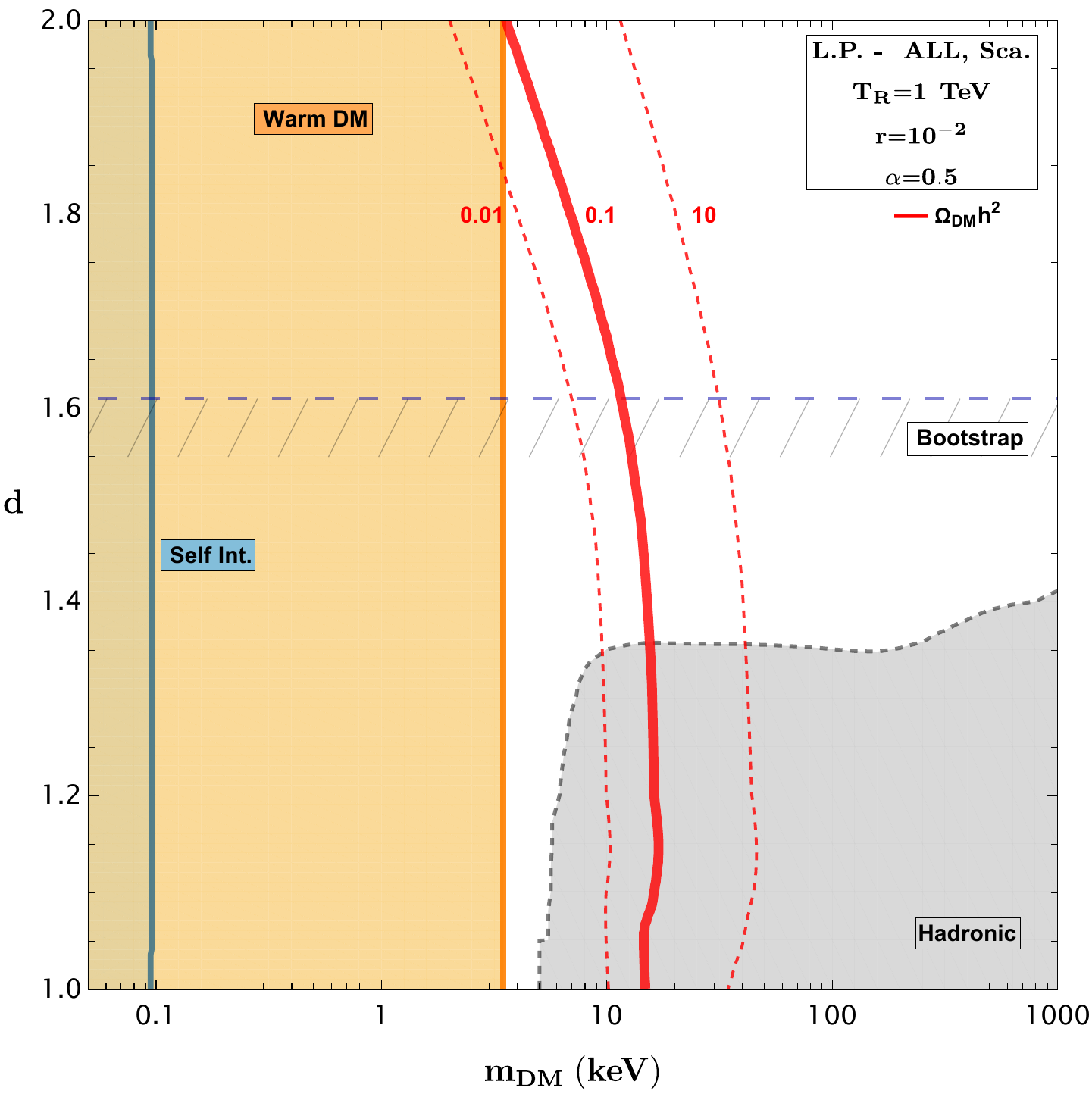}}
	\caption{Dark matter relic density contours (red) and observational/theoretical constraints, in the lepton portal model with only the first generation of leptons on the left and all generations of leptons on the right, with a scalar mediator.}
	\label{fig:DMlep2}
\end{figure}

In summary, we find six models with allowed parameter space that reproduces the relic density: quark portal with MFV or democratic coupling (both scalar and vector mediators), and the scalar mediator lepton portal with MFV or democratic couplings.

\subsection{Gluon Portal: $\Os = G^{\mu\nu}G_{\mu\nu}$}
\label{sec:GG}

\begin{figure}[h]
	\center
	\subfloat{%
		\includegraphics[width=7.3cm]{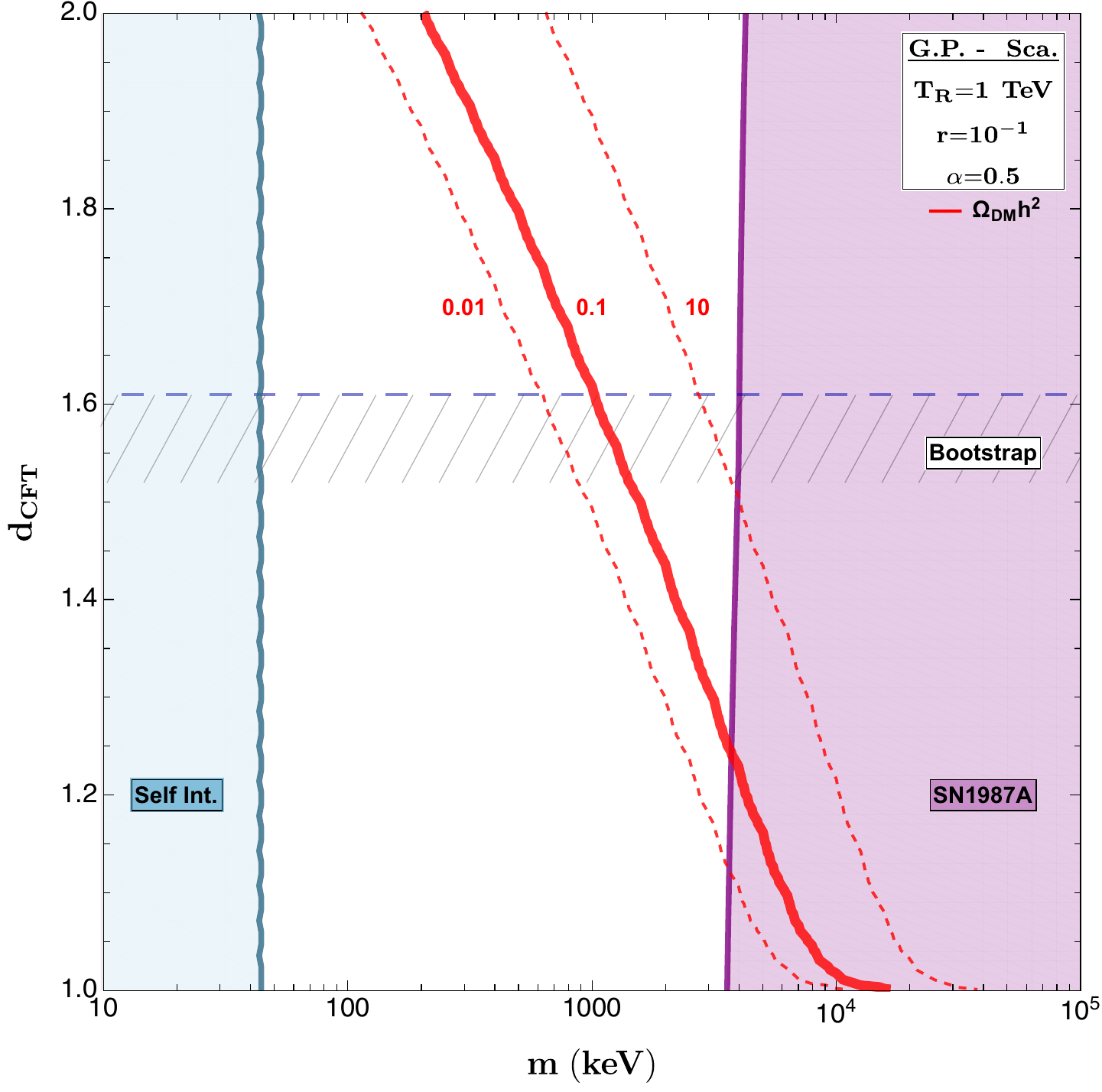}
		\includegraphics[width=7.3cm]{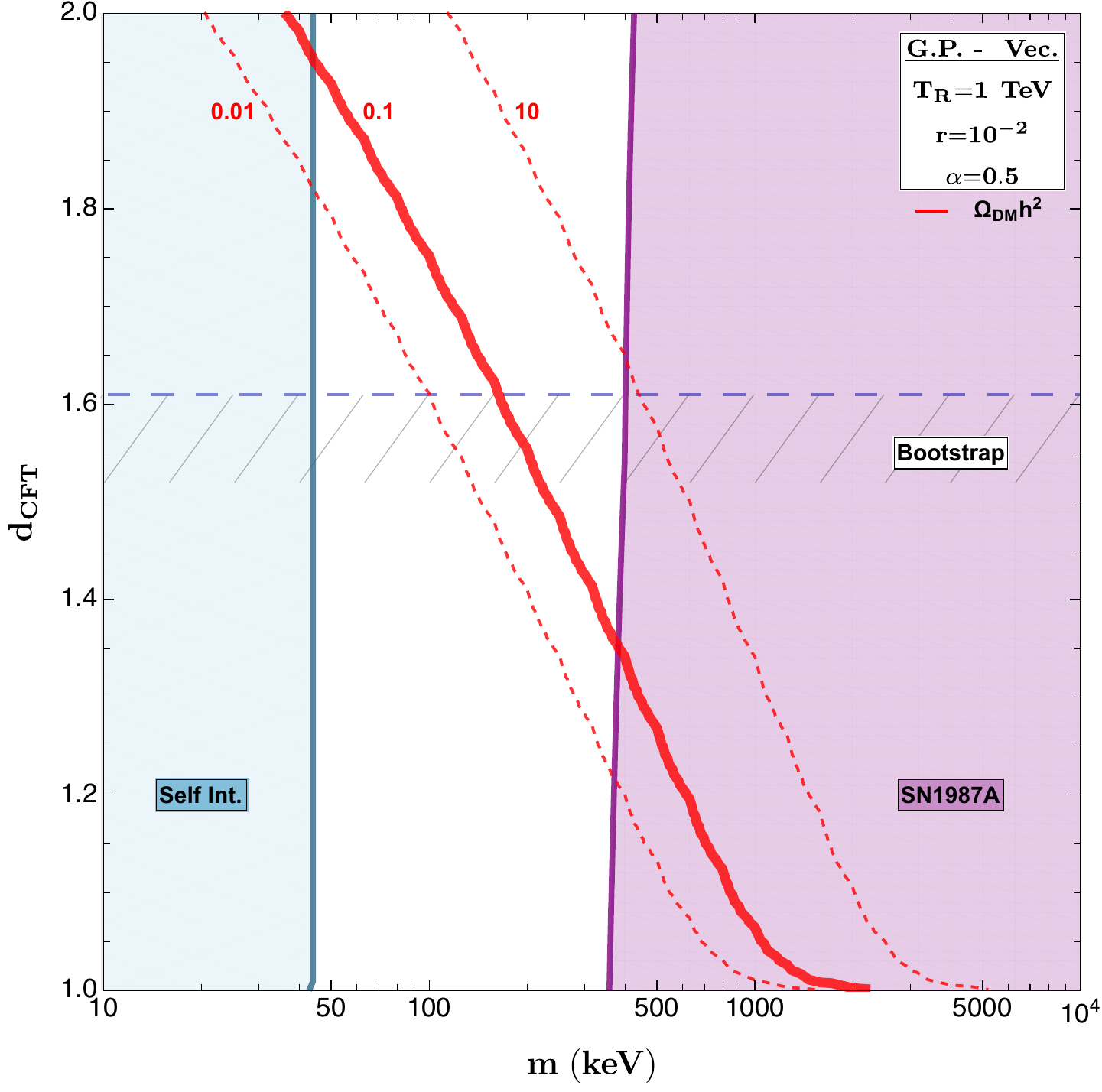}}
	\caption{Dark matter relic density contours (red) and observational/theoretical constraints, in the gluon portal model, with a scalar (vector) mediator on the left (right).}
	\label{fig:DMglu}
\end{figure}

The dominant mode of populating the dark sector is through gluon annihilation, $g g \rightarrow \CFT $. Additionally, there are subdominant processes of production, through loop-induced quark annihilation. The dark sector energy density produced via gluon annihilation scales as,
\beq
\rhoc \sim M_{\rm pl} \left( \frac{m^{4-d}}{16\pi^2 \, \alpha^4 v^4} \right)^2 \, T^4 \, (T_R^{2d-1} - T^{2d-1})
\eeqn

As in the quark portal, production continues until $T = \Lambda_{\scriptstyle_{\rm QCD}}$ or $\Tc = \gap$, whichever happens first. The constraints on the model parameter space are shown in Fig.~\ref{fig:DMglu}. For analytic estimates of the relic density, see Appendix~\ref{subapp:rd}.

\begin{figure}[h]
	\begin{center}
		\subfloat{%
			\includegraphics[width=7.3cm]{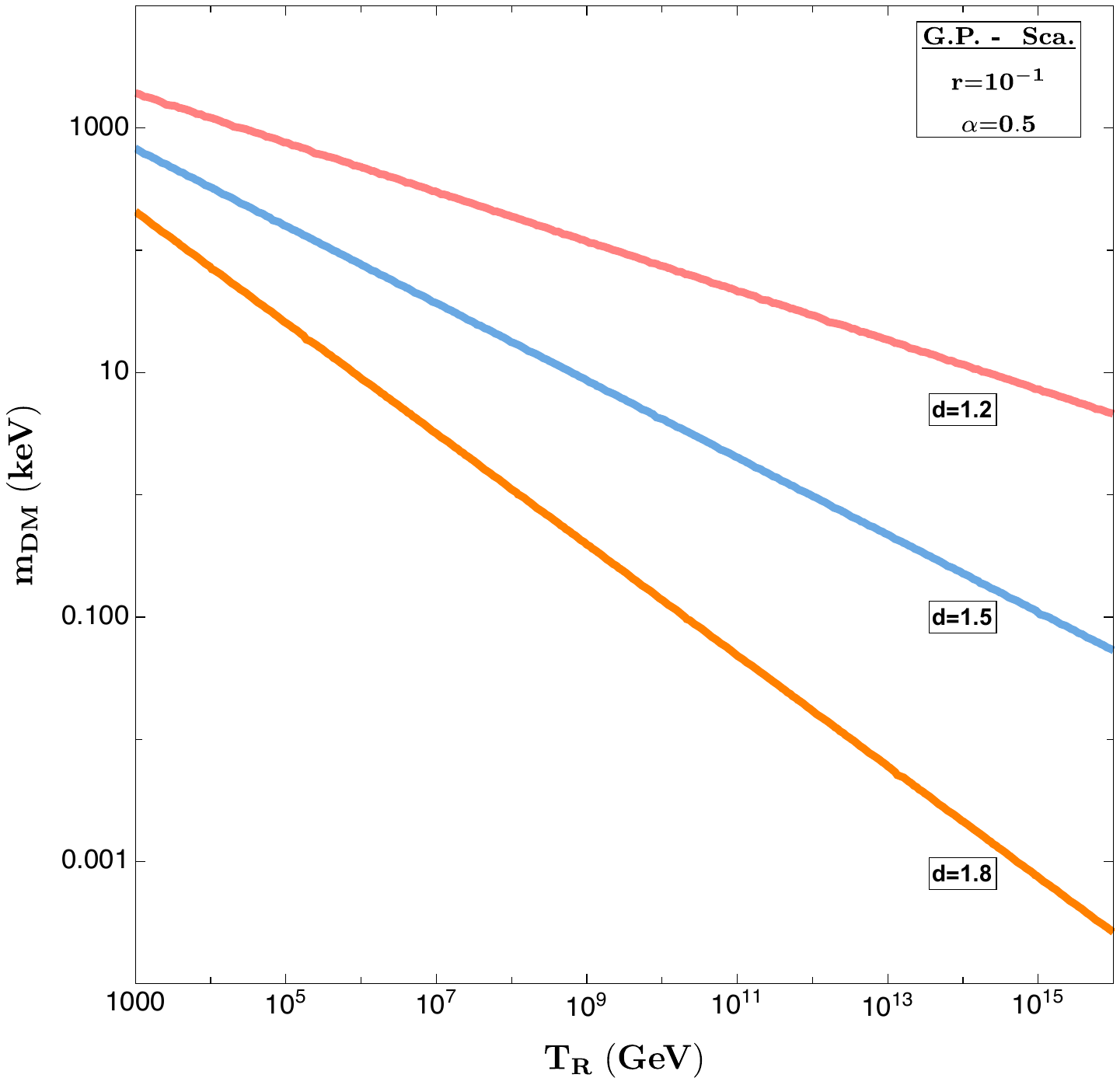}
			\includegraphics[width=7.3cm]{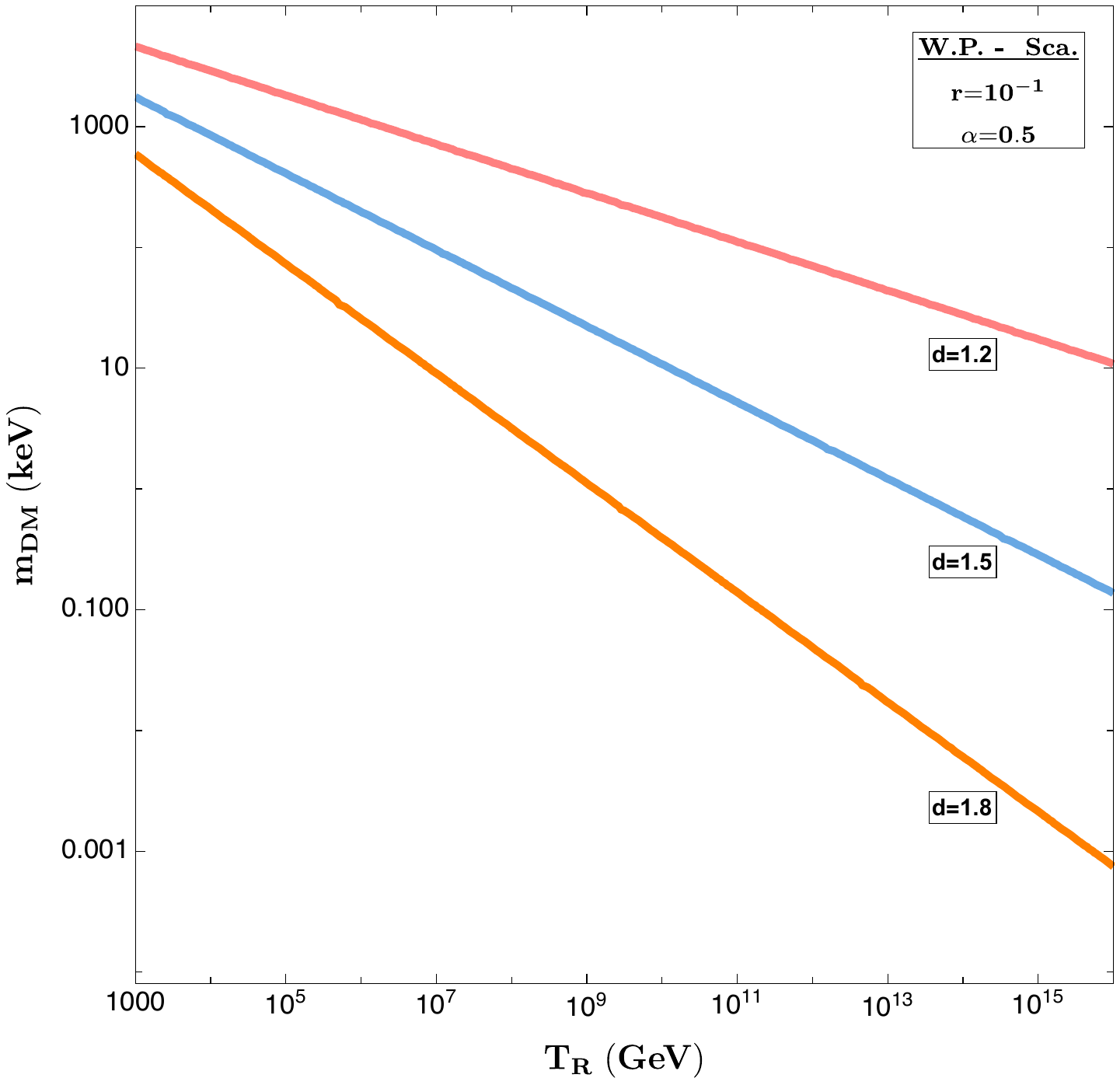}}
		\caption{Dark matter mass that produces the observed relic density, as a function of the reheating temperature, for various values of $d$, for the gluon  (left) and hypercharge (right) portals.}
		\label{fig:otherTR}
	\end{center}
\end{figure}

Since the operator $G^{\mu\nu}G_{\mu\nu}$ is of dimension $\ds = 4$, and the CFT operator dimension $d\geq 1$ by unitarity, the dark sector energy density is always dominated by the production at the highest available temperature, {\it i.e.} the reheating temperature $T_R$ (see Section~\ref{sec:history}). The predictions of these models thus depend on an additional parameter, $T_R$, making it less predictive. However, in practice, the dependence of the predicted dark matter candidate mass on $T_R$ is rather weak. As shown in the derivation in Appendix~\ref{subapp:rd}, the relic density of dark matter today scales as 
\beq
\Omega \, h^2 \propto \left(\gap\right)^{7 - \frac{3}{2} d} \, \left(T_R\right)^{\frac{3}{4}(2 d - 1)}.
\eeq{GPscaling1}
For relic density fixed to the observed value, the dependence of the inferred mass gap on $T_R$ is given by
\beq
\frac{\partial \log \gap}{\partial \log T_R} = \left(\frac{3}{8} \left( \frac{2d-1}{3d-14} \right)\right). 
\eeq{GPscaling2}
The logarithmic derivative is small throughout the range of $d$ considered here. The relationship between the dark matter mass and the reheating temperature is shown in Fig. \ref{fig:otherTR}, for various values of the CFT operator dimension $d$. 

\subsection{Electroweak Boson Portal: $\Os = W^{\mu\nu}W_{\mu\nu},\ B^{\mu\nu}B_{\mu\nu}$}

\begin{figure}[t!]
	\center
	\subfloat{%
		\includegraphics[width=7.3cm]{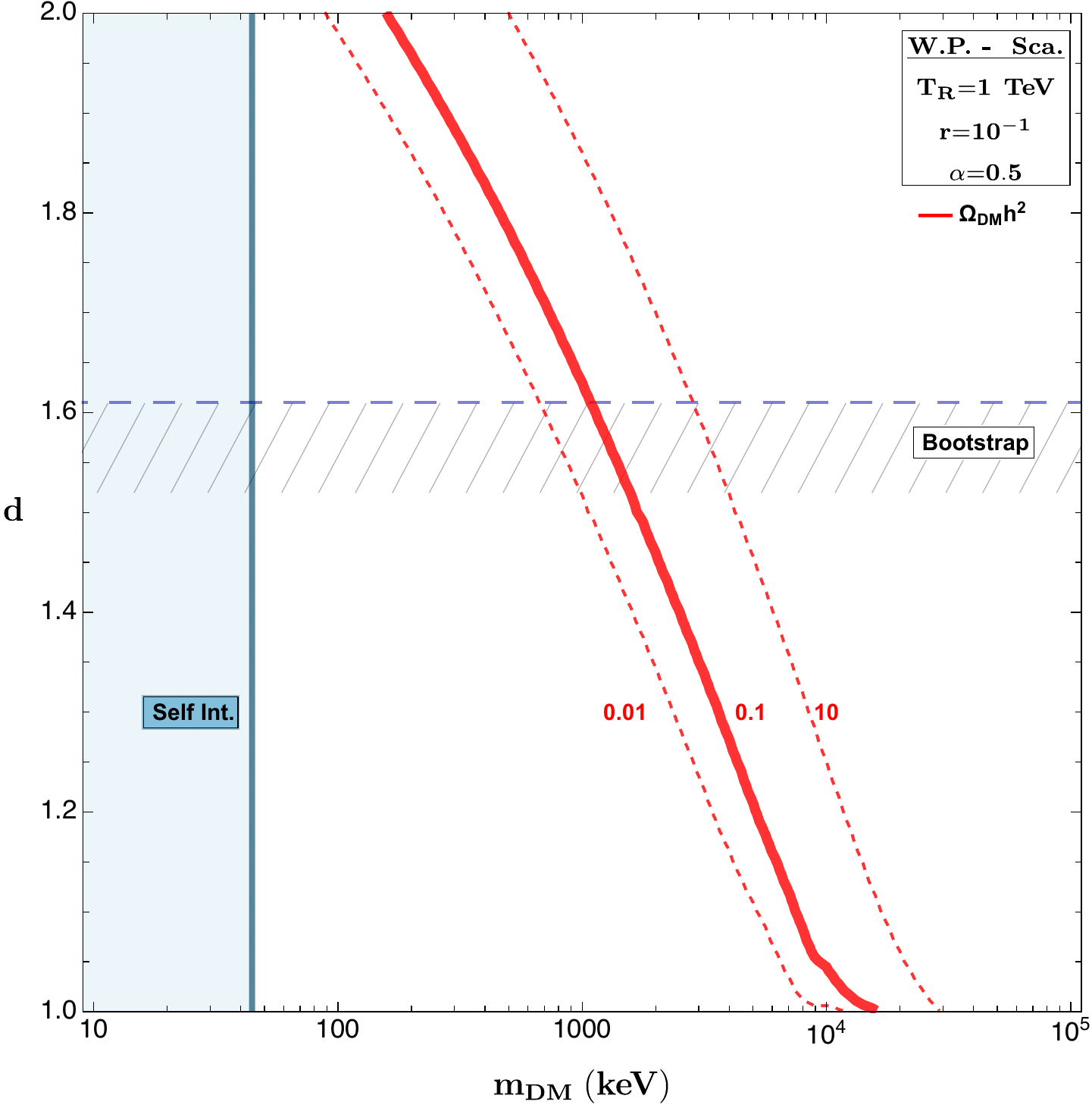}
		\includegraphics[width=7.3cm]{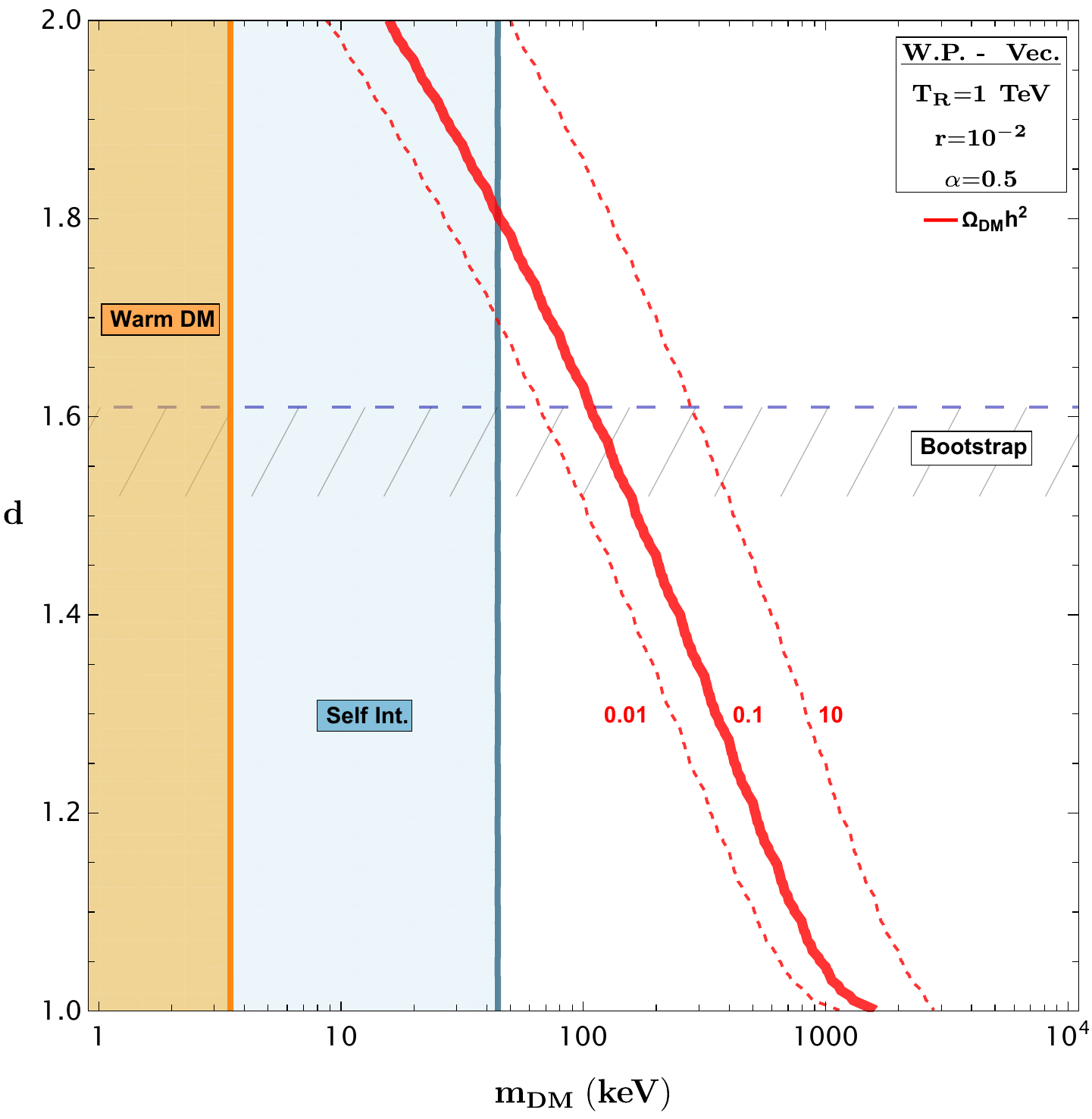}}
	\caption{Dark matter relic density contours (red) and observational/theoretical constraints, in the hypercharge portal model, with a scalar (vector) mediator on the left (right).}
	\label{fig:DMweak}
\end{figure}

Since the phenomenology of both the weak SU(2)$_L$ ($\Os = W^{\mu\nu}W_{\mu\nu}$) and the hypercharge ($\Os = B^{\mu\nu}B_{\mu\nu}$) portals are similar, we consider only the case of $\Os = B^{\mu\nu}B_{\mu\nu}$ to illustrate the salient features of the electroweak boson portal. The dominant production process is that of vector boson annihilation, with the initial dynamical degrees of freedom being hypercharge gauge bosons above the electroweak phase transition (EWPT) and photons below EWPT. Subdominant processes include $Z$ boson decay below the weak scale, and fermion annihilation through the electroweak portal. Photon annihilation continues till $\Ts \sim \gap$, and the dark matter redshifts as matter below $\Tc \sim \mdm$. Photon annihilation to CFT states produces dark sector energy density that scales similarly to the gluon portal model;
\beq
\rhoc \propto M_{\rm pl} \, (1 - \sin^2 \theta_w)\left( \frac{m^{4-d}}{16\pi^2 \, \alpha^4 v^4} \right)^2 \, T^4 \, (T_R^{2d-1} - T^{2d-1})
\eeqn
where $\theta_w$ is the Weinberg angle. For analytic estimates of the relic density, see Appendix~\ref{subapp:rd}.

The viable parameter space and constraints on this model are shown in Fig.~\ref{fig:DMweak}. The value of $r$ is 0.1 and 0.01 respectively for scalar and vector mediators. As in the gluon portal, the interaction term dimension $D$ is always $> 5$ and production is dominant at the reheating temperature $T_R$, making the relic density dependent on an extra parameter. Due to the similarities with the gluon portal, where vector boson annihilation in the UV determines the relic density, equations~\leqn{GPscaling1} and \leqn{GPscaling2} apply in this case as well. Fig.~\ref{fig:otherTR} demonstrates this scaling.

\section{Dark Matter Phenomenology and Constraints}
\label{sec:pheno}

\begin{figure}[h]
	\center
	\subfloat{%
		\includegraphics[width=7.3cm]{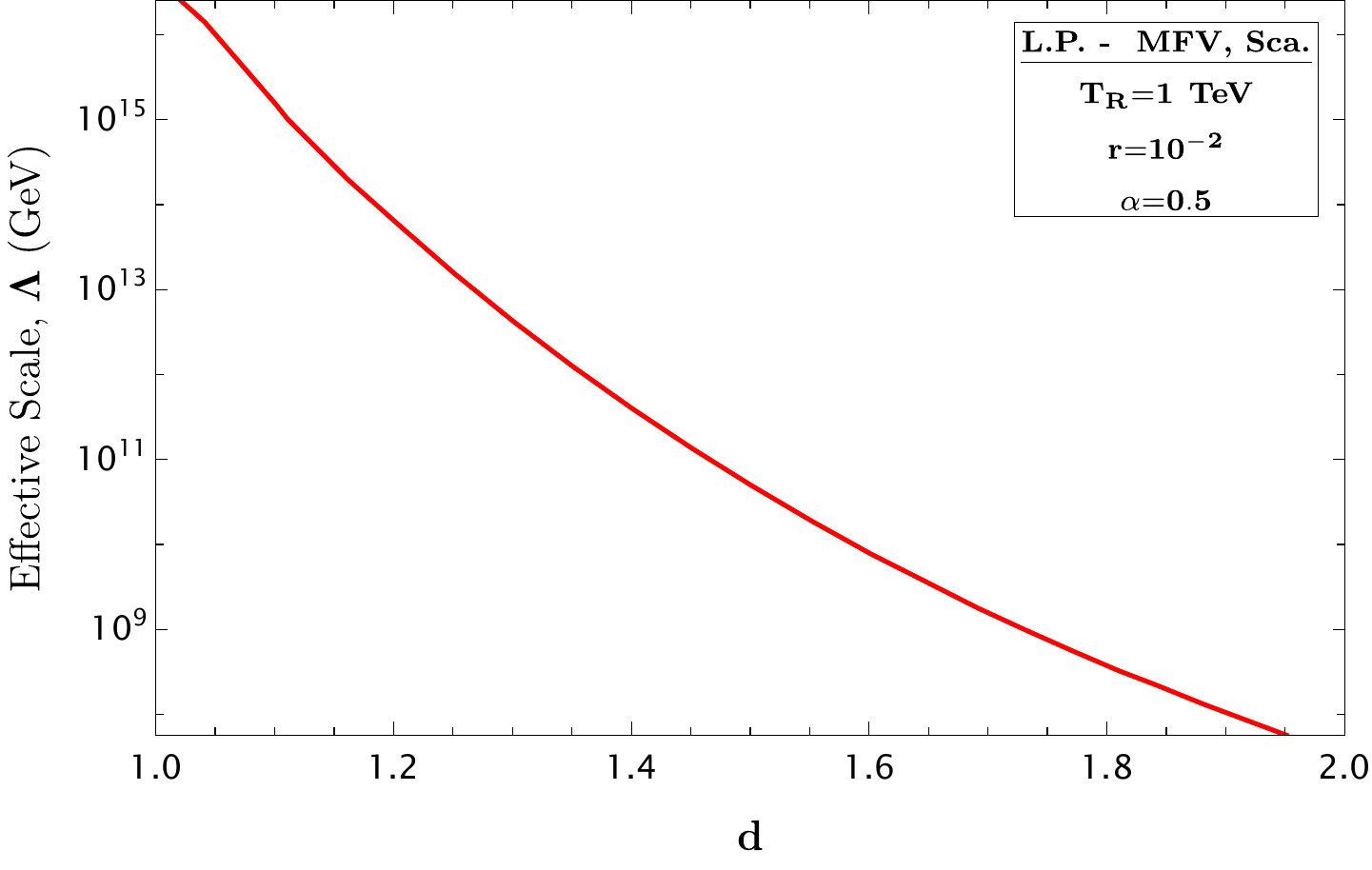}
		\includegraphics[width=7.3cm]{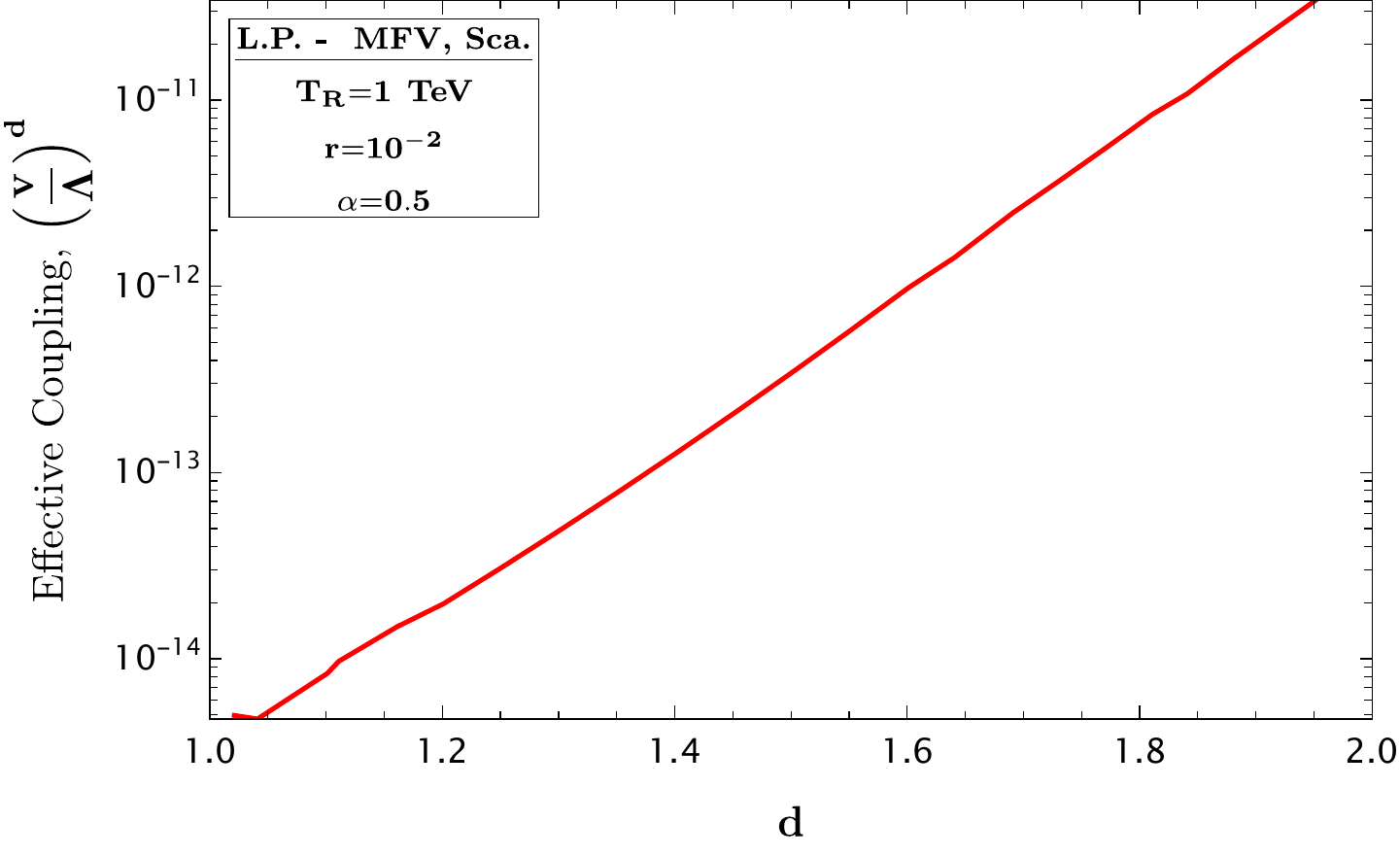}}
	\caption{Left panel: Effective energy scale of the SM-CFT interaction in the MFV lepton portal with a scalar mediator. Right panel: Effective dimensionless strength of the SM-CFT coupling for the same portal, for SM collision energies of order 100 GeV.}
	\label{fig:EffCoup}
\end{figure}

The interactions of the COFI dark matter candidate with the Standard Model particles are extremely weak. The effective energy scale suppressing the SM-CFT interaction is well above the weak scale 
\beq
\Lambda=(\l)^{-\frac{1}{D-4}}\,\cdot \Lambda_{\rm CFT}\,\sim 10^{10}-10^{15}~{\rm GeV}, 
\eeq{EffCoup}
leading to tiny couplings of the DM particles to SM at energies of order the weak scale and below. This is illustrated in Fig.~\ref{fig:EffCoup}, in the case of the lepton portal in the MFV flavor scheme;  other portals produce similar results. As a result, no relevant constraints arise from direct, indirect, and collider searches for DM. However, there are important phenomenological constraints on the model from dark matter self-interaction and large-scale structure (which are independent of the DM-SM coupling), as well as stellar cooling rates (where the small coupling is compensated by large amount of SM particles in the stellar bodies). These constraints will be considered in this section. We will also outline theoretical constraints on the model parameter space related to naturalness and CFT bootstrap bound.    

\subsection{Dark Matter Self-Interaction Bound}

\begin{figure}[h]
	\center
	\includegraphics[width=10cm]{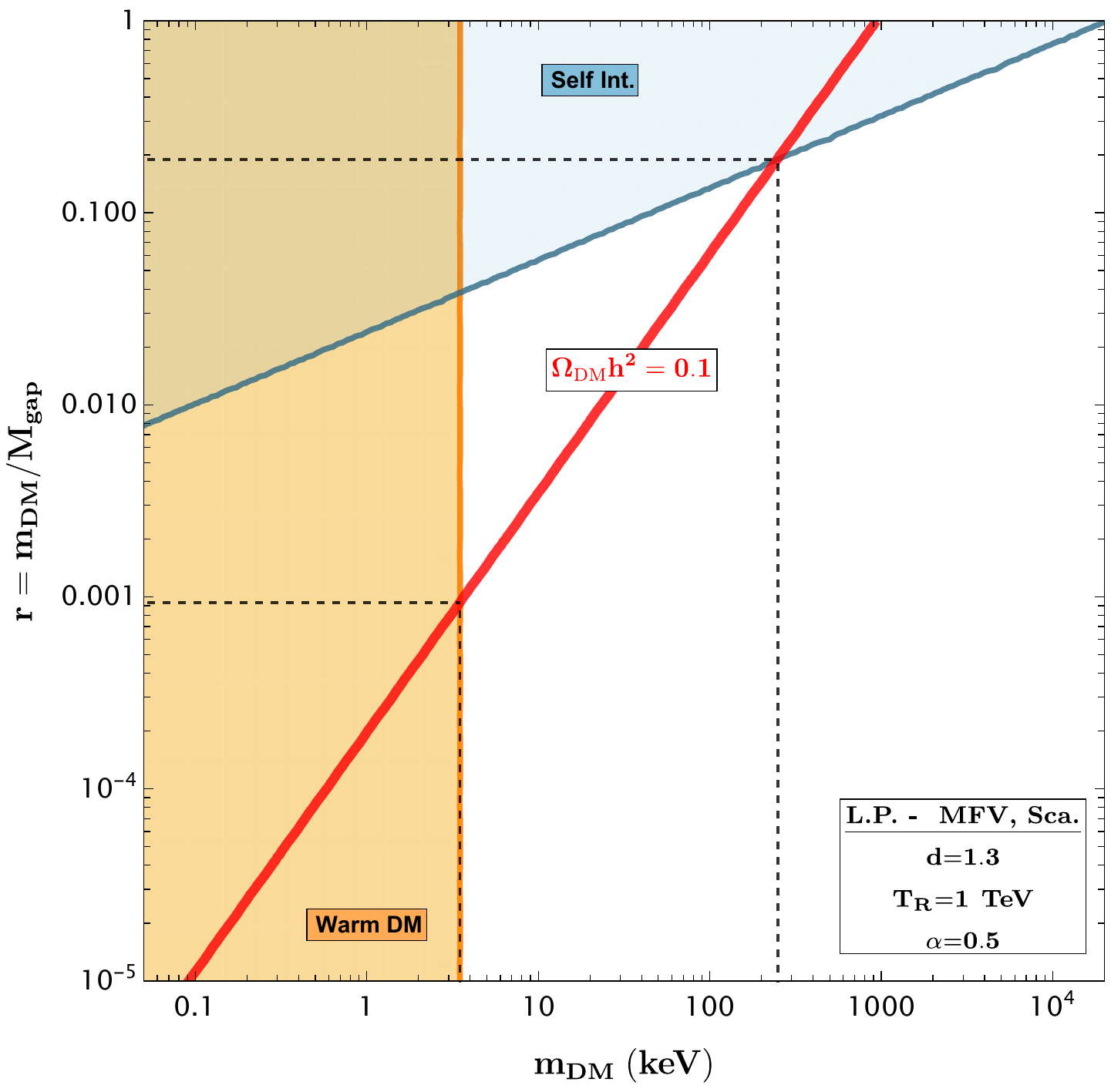}
	\caption{$r$ vs.~$\mdm$ dependence for the lepton portal with the MFV flavor structure, assuming a scalar mediator.} 
	\label{fig:rplot}
\end{figure}

Observations of galactic clusters, such as the Bullet cluster, place an upper bound on the cross-section of elastic scattering of non-relativistic DM particles,
$\sigma_{\rm SI}/m_\chi \lesssim 4500 \; \mathrm{GeV}^{-3}$ \cite{Markevitch:2003at, Lin:2019uvt}. One generally expects that the hadronic phase of our dark sector has characteristic coupling $g_\star \sim \frac{4\pi}{\sqrt{N}}$. If the dark matter is a generic composite state, the elastic scattering cross-section is of the order 
\beq
\sigma_{\rm SI} \sim \frac{g_\star^4}{8\pi \gap^2}.
\eeq{SIgeneric}
For the values of $\gap$ that produce the observed DM relic density, $N\sim 10^4$ would be required for consistency with the observational bound. Such large values of $N$ are possible, but theoretically unattractive. This leads us to consider an alternative possibility that $g_\star\sim 1$ but the DM state is not a (or a collection of) generic composite particle(s), but rather is a derivatively-coupled PNGB. DM elastic scattering is mediated by exchanges of a scalar or vector resonance with mass of order $\gap$. Using the effective theory (\ref{phiDM}), the cross-section for the case of a scalar mediator is estimated to be 
\beq
\sigma_{\rm SI} \sim \frac{r^6}{8\pi \gap^2} , \;\; r = m_\chi / \gap
\eeqn
while for a vector mediator (using (\ref{rhoDM})), 
\beq
\sigma_{\rm SI} \sim \frac{r^2}{8\pi \gap^2}.
\eeqn
Here $r=\mdm/\gap \ll 1$ is a model-dependent parameter. If both vector and scalar mediators are present with similar masses, the vector exchange will dominate. This is the case in QCD where $\rho$ exchange is the main contribution to pion elastic scattering. However, for completeness, we consider both scalar and vector mediator-dominated cases in our phenomenological analysis. We find that in the scalar case, $r \sim 0.01 \text{--} 0.1$ is sufficient for consistency with observational bounds, while in the vector case $r \lsim 10^{-2}$ is required. See Fig.~\ref{fig:rplot} for an illustration of allowed values of $r$ and its effect on the value of $\mdm$ that produces the observed relic density, in one particular model. 

\subsection{Warm Dark Matter Bound}

Since dark matter in COFI models is light as well as relativistic in the early universe, they can free-stream, leading to  suppression of structure/inhomogeneity below a certain length scale $\lambda_{\rm FS}$. Observation of the existence of a DM halo of a certain size then puts an upper bound on $\lambda_{\rm FS}$, and hence on the mass of DM. Typically, observations of the Lyman-$\alpha$ flux spectra, which probes DM halos from redshifts of $z \sim (3 - 5.5)$, are used to set such bounds. Depending on the particular data set used and the systematics of the analysis, the current bound is~\cite{Irsic:2017ixq}\footnote{As discussed above, our DM generically has self-interactions, while the analysis of~\cite{Irsic:2017ixq} is based on an assumption of  collisionless DM. The bound for self-interacting dark matter is somewhat weaker~\cite{Egana-Ugrinovic:2021gnu}, but the difference is not large enough to affect the present discussion.},
\bea
\mdm \gtrsim (3.5 - 5.3) \; {\rm keV}.
\eea
This bound places a non-trivial constraint on certain COFI models with a lepton portal, where the DM mass consistent with relic density is in the $1-10$~keV range. For other portals, COFI DM candidates are much heavier and this bound is irrelevant.

\subsection{Stellar Cooling Bounds}

In this section, we discuss constraints imposed on COFI dark matter models from the evolution of stars. Dark matter candidates in the keV-MeV mass range can be produced in collisions of SM states (nucleons and electrons) in stars. In spite of weak DM-SM coupling, the production can be significant due to large amount of matter in stars. Once a DM state is produced, the weak coupling may allow it to escape the star without interacting, carrying away energy. Systems supported by degenerate pressure, e.g.~supernovae, have a positive heat capacity and production of DM results in an extra cooling mechanism. Systems supported mainly by thermal pressure, such as Main Sequence (MS) stars, have a negative heat capacity. Energy carried by DM states produced in the core does not necessarily lead to extra cooling, but still affects the dynamics of the star, changing the time scale for each stage in its evolution. In either case, existing observations provide constraints on the rate of extra energy loss, which can be translated into bounds on new physics.   

In this work, we consider constraints from the following classes of stars: Main Sequence (MS, e.g.~Sun), Red Giant Branch (RGB), Horizontal Branch (HB), and Supernova (SN).\footnote{It has been pointed out in \cite{Hardy:2016kme} that white dwarfs (WD) and neutron stars (NS) give either comparable or weaker bounds and we do not further consider them.} For the purpose of computing bounds, each system may be characterized by the core temperature $T$, mass density $\rho$ (which is dominated by nuclear mass density), electron number density $n_e$, the degree of electron (and nucleon) degeneracy $E_F$ and $p_F$ (since the Fermi energy and momentum are higher than the temperature only when electrons/nucleons are degenerate), and composition of nuclear matter. From these data and the form of interaction between new physics state and SM states, one then computes energy loss rate per mass $\epsilon$, and compares it to existing bounds. In our estimates, we will adopt the following benchmark parameters for each class of stars~\cite{Hardy:2016kme}:
\begin{center}
	\begin{itemize}
		\item MS: $T \approx 1.3 \; {\rm keV}$, $\rho \approx 156 \; {\rm g \; cm^{-3}}$, $n_e = 6.3 \times 10^{25} \; {\rm cm^{-3}}$, $\epsilon \lesssim 0.2 \; {\rm erg \; g^{-1} \; s^{-1}}$.
		
		\hspace{0.6cm} electrons not degenerate, nucleons not degenerate.  \\
		\item HB: $T \approx 10 \; {\rm keV}$, $\rho \approx 10^4 \; {\rm g \; cm^{-3}}$, $n_e = 3 \times 10^{27} \; {\rm cm^{-3}}$, $\epsilon \lesssim 10 \; {\rm erg \; g^{-1} \; s^{-1}}$
		
		\hspace{0.6cm} electrons not degenerate, nucleons not degenerate. \\
		\item RGB: $T \approx 10 \; {\rm keV}$, $\rho \approx 10^6 \; {\rm g \; cm^{-3}}$, $n_e = 3 \times 10^{29} \; {\rm cm^{-3}}$, $\epsilon \lesssim 10 \; {\rm erg \; g^{-1} \; s^{-1}}$
		
		\hspace{0.6cm} electrons degenerate ($E_F \approx 144 \; {\rm keV}$, $p_F \approx 409 \; {\rm keV}$), nucleons not degenerate. \\
		\item SN: $T \approx 30 \; {\rm MeV}$, $\rho \approx 3 \times 10^{14} \; {\rm g \; cm^{-3}}$, $n_e = 1.8 \times 10^{38} \; {\rm cm^{-3}}$, $\epsilon \lesssim 10^{19} \; {\rm erg \; g^{-1} \; s^{-1}}$
		
		\hspace{0.6cm} electrons degenerate ($E_F \approx p_F \approx 344 \; {\rm MeV}$), nucleons nearly degenerate. \\
	\end{itemize}
\end{center}

The production of DM in stars may occur in one of the two regimes. If energy transferred from the SM into the dark sector in a single collision is above $\gap$, the final state consists of CFT states and the cross-section can be calculated using the unparticle approach. Following the collision, the produced dark sector states quickly hadronize and decay, resulting in multiple DM particles that share the transferred energy. If, on the other hand, energy transferred from the SM into the dark sector in a single collision is between $\mdm$ and $\gap$, the production occurs in the hadronic phase and is estimated using the low-energy effective theory in the dark sector, see Section~\ref{subsec:IR}. For each COFI portal and class of stars, we start by determining which of the two regimes is appropriate, and proceed to estimate the DM production cross-section and the resulting energy loss rate. In cases where the energy loss argument imposes a relevant bound on the COFI scenario, we also estimated the mean-free path $\ell_{\rm MFP}$ of the produced DM particle in the star. If $\ell_{\rm MFP}$ is smaller than the star radius, the DM will typically become trapped in the star, depositing its energy back into the stellar material. In this case, the energy-loss bounds do not apply.

\subsubsection{Quark and Gluon Portals}

In this case, dark sector states are produced in stars primarily through Bremsstrahlung in nucleon collisions. 
For $T < \gap$ (MS, HB, RGB), the final state is the hadronic state of the confined CFT, while for $T > \gap$ (SN), the process is energetic enough to produce CFT state directly.

\subsubsection*{$T < \gap$ (MS, HB, RGB)}

We first consider the case with $T < \gap$. 
The matrix element for the quark scalar operator in nucleons is given by (see for example \cite{Bishara:2017pfq,Hisano:2017jmz})
\beq
\langle N \vert \, \bar{q} q  \, \vert N \rangle \equiv f_{T_q}^N  \, \frac{m_N}{m_q} =: C_q^{(N)}
\label{eq:quark_nucleon_form_factor}
\eeqn
where $f_{T_q}^N$ is the mass fraction parameter of the quark $q$. Values of $f_{T_q}^N$ can be found in \cite{Bishara:2017pfq,Hisano:2017jmz}. The matrix element for the gluon scalar operator can be obtained from the trace anomaly in QCD \cite{Bishara:2017pfq,Hisano:2017jmz}:
\beq
\langle N \vert  \,  G_{\mu\nu}^a G^{\mu\nu a }  \, \vert N \rangle = - \frac{8}{9} \frac{\pi}{\alpha_s} m_N \left( 1 - \sum_q f_{T_q}^N \right) + \mathcal{O} (\alpha_s) =: C_{G}^{(N)}.
\label{eq:Gluon_nucleon_form_factor}
\eeqn
Together with the matching of CFT to its low-energy effective theory described in Section~\ref{subsec:IR}, this provides the effective theories of nucleon-dark hadron interactions. For quark portal, we have
\beq
\mathcal{L} \sim \frac{\l}{\L^d}  \, H Q^\dagger q  \, \Oc \to \mathcal{L} \sim \left( \frac{\l  \, v  \, \gap^{d-1}}{\sqrt{2} g_*  \, \L^d} \right) \left( f_{T_q}^N  \, \frac{m_N}{m_q} \right) \bar{N} N \phi + \frac{g_*}{\gap} \phi \left( \partial \chi \right)^2
\eeqn 
Below the gap scale, $\phi$ can be integrated out, yielding the nucleon-DM coupling:
\beq
\mathcal{L} \sim G_{N\chi}^{(q)} \,\,\bar{N} N \left( \partial \chi \right)^2\,,
\label{eq:quark-portal_EFT_SC}
\eeqn
where 
\beq
G_{N\chi}^{(q)}\,=\, \frac{16\pi^2 \sum_q \kappa_q C_q^{(N)}}{ \left( \sum_q \kappa_q m_q \right) \Lambda_{\rm SM}^2 }\,. 
\eeq{GNchi_q}
Here the sums run over quark flavors, and we have used the mass gap formula for the quark portal. 
%
%

For gluon portal, the effective theory of nucleon-dark hadron interactions
\beq
\mathcal{L} \sim \frac{\l}{\L^d}  \, G_{\mu\nu}^2  \, \Oc \to \mathcal{L} \sim \left( \frac{\l  \, \gap^{d-1}}{g_*  \, \L^d} \right) \left( - \frac{\pi}{\alpha_s} \frac{8}{9} m_N \left( 1 - \sum_q f_{T_q}^N \right) \right) \bar{N} N \phi + \frac{g_*}{\gap} \phi \left( \partial \chi \right)^2.
\label{eq:GP_sc_eeffective_Lagrangian}
\eeqn
%
Below $m_\phi \approx \gap$ we again integrate out the $\phi$. In addition, since $f_{T_q}^{N} \sim 10^{-2}$, we neglect its contribution. Using $\alpha_s (\Lambda_{\rm QCD}) \approx 4 \pi$ and thereby approximating $C_{G}^{(N)} \approx - m_N / 4$, we obtain the nucleon-DM coupling 
\beq
\mathcal{L} \sim G_{N\chi}^{(g)} \,\,\bar{N} N \left( \partial \chi \right)^2\,
\label{eq:gluon-portal_EFT_SC}
\eeqn 
where
\beq
G_{N\chi}^{(g)} = C_{G}^{(N)} \left( \frac{16 \pi^2}{ \Lambda_{\rm SM}^4} \right).
\eeq{GNchi_g}
Using these effective couplings, energy loss rate due to DM emission can be calculated in analogy with the well-known calculation for energy loss through standard model neutrinos~\cite{Raffelt:1996wa}. The energy loss rate per unit volume is given by 
\bea
&& \text{Quark-Portal :} \; Q (\chi\chi) = Q (\bar{\nu} \nu) \, \left( \frac{T^2}{2G_F^2} \right)\, \left( \frac{G_{N\chi}^{(q)}}{\sum_q C_q^{(N)}} \right)^2  \frac{C_{\chi\chi}}{C_{\bar{\nu}\nu}} \nonumber \\
&& \text{Gluon-Portal :} \; Q (\chi\chi) = Q (\bar{\nu} \nu)\, \left( \frac{T^2}{2G_F^2} \right)\, \left( \frac{G_{N\chi}^{(g)}}{\sum_q C_q^{(N)}} \right)^2  \frac{C_{\chi\chi}}{C_{\bar{\nu}\nu}} 
\label{eq:lossrates}
\eea
Here $G_F$ is the Fermi constant. Explicit expressions for $Q (\bar{\nu}\nu)$ (the rate for neutrino-pair production in the standard model) and the constants   
$C_{\bar{\nu}\nu}$ and $C_{\chi\chi}$ are given in the 
Appendix~\ref{subapp:Brem_nucleon}, along with the details of the calculation. The energy loss rate per unit mass ($\epsilon$) that we compare with the observed bounds is calculated as $\epsilon = Q/\rho$. 

\subsubsection*{$T > \gap$ (SN)}

When $T > \gap$ such as in SN, the dark-sector states produced in nucleon collisions are described by a CFT, and the production rate can be calculated using the unparticle approach.\footnote{See~\cite{Davoudiasl:2007jr} for a more heuristic approach to calculating cosmological and astrophysical bounds with unparticles.} It is useful to normalize the energy loss rate using a simple benchmark model of a light scalar particle $\phi$ coupled to nucleons through a Yukawa interaction $\sim g \, \phi \, \bar{\psi}_N \psi_N$. In this case, the energy loss per unit volume $Q(\phi)$ 
is well-known~\cite{Raffelt:1996wa}; see Appendix~\ref{app:energy loss rate} for an analytic formula. The ratio $\epsilon (\text{\tiny CFT}) / \epsilon (\phi)$ can be reliably estimated by a procedure explained in Appendix~\ref{app:energy loss rate}. We obtain
\beq
\epsilon (\text{\tiny CFT}) \sim \frac{G_{\rm eff}^2 (m_N T)^{d-1}}{g^2} \frac{1}{(2\pi)^{2d-2}} \frac{\text{dof}_{\rm \scriptscriptstyle CFT}}{\text{dof}_\phi} \frac{\langle \omega \rangle_{\rm \scriptscriptstyle CFT}}{\langle \omega \rangle_{\phi}} \frac{Q (\phi)}{\rho}\,,
\eeqn
where $d$ is the dimension of the CFT operator, while $\rho$ and $T$ are the nucleon mass density and temperature in the SN core. Note that the dependence on the coupling $g$ in the benchmark scalar model cancels out since $Q(\phi) \propto g^2$.  Here, $\frac{\text{dof}_{\rm \scriptscriptstyle CFT}}{\text{dof}_\phi}$ denotes the ratio of the internal degrees of freedom of the final state produced in the CFT and the benchmark scalar model, while $\frac{\langle \omega \rangle_{\rm \scriptscriptstyle CFT}}{\langle \omega \rangle_{\phi}}$ is the ratio of the average energy carried by the corresponding final states. 
Explicit expressions for the effective coupling $G_{\rm eff}$ in quark and gluon portals are given in the Appendix~\ref{subapp:Brem_nucleon} (see Eqs.~\leqn{eq:G_ff_quark-portal_Brem_nucleon_CFT} and \leqn{eq:G_ff_gluon-portal_Brem_nucleon_CFT}). The factors $\frac{\text{dof}_{\rm \scriptscriptstyle CFT}}{\text{dof}_\phi}$ and $\frac{\langle \omega \rangle_{\rm \scriptscriptstyle CFT}}{\langle \omega \rangle_{\phi}}$ can be determined only if CFT is fully specified, but we expect that they will take values within the range $1 \sim d$. We use $1$ in the constraint plots of Section~\ref{sec:Ops}. 

\subsubsection{Higgs Portal}

For Higgs portal, the COFI dark matter candidate has mass of order MeV, and can only be produced in supernovae. Comparing $\gap$ in the Higgs portal model to $T_{\rm SN}$, we learn that the production is in the CFT regime. Again, the dominant production mechanism for dark states is Bremsstrahlung in nucleon collisions. The relevant part of the Lagrangian is
\beq
\mathcal{L} \sim \frac{\l \, v}{\sqrt{2} \L^{d-2}} \, h \, \Oc + \frac{\alpha_s}{12 \pi v} \, h \, G_{\mu\nu}^a G^{\mu\nu a} 
\eeqn
where the second term is the top-loop induced coupling between the Higgs and gluon (see for example e.g.~\cite{Gunion:1989we,Harlander:2013oja}). 
Integrating out the Higgs and using Eq.~\leqn{eq:Gluon_nucleon_form_factor} yields the effective coupling 
\beq
\mathcal{L} \sim C_G^{(N)} \left( \frac{\alpha_s}{6 \sqrt{2}\pi} \right) \left( \frac{\gap^{4-d}}{v^2 \, m_h^2} \right) \bar{N} N \,  \Oc.
\eeqn
To get this form, we used the mass gap formula for the Higgs portal model. The energy loss rate in the SN is calculated as in the gluon portal (see Appendix~\ref{subapp:Brem_nucleon}) and is given by
\beq
\epsilon (\text{\tiny CFT}) \sim \frac{G_{\rm eff}^2 (m_N T)^d}{g^2} \frac{1}{(2\pi)^{2d-2}} \frac{\text{dof}_{\rm \scriptscriptstyle CFT}}{\text{dof}_\phi} \frac{\langle \omega \rangle_{\rm \scriptscriptstyle CFT}}{\langle \omega \rangle_{\phi}} \frac{Q (\phi)}{\rho}
\eeqn
where
\beq
G_{\rm eff} =  C_G^{(N)} \left( \frac{\alpha_s}{6 \sqrt{2}\pi} \right) \left( \frac{\gap^{4-d}}{v^2 m_h^2} \right).
\eeqn
Numerically, emission from the SN core in the region of the model parameter space relevant for COFI dark matter is well below the observational bound, so that the Higgs portal scenario is unconstrained by stellar cooling considerations.

\subsubsection{Lepton Portal}

Dark sector states are produced through their interactions with electrons in the stellar medium. In all star systems other than the supernova, the electron temperature is below $\gap$, so that the production is in the hadronic phase of the dark sector. The effective theory of electron-dark hadron interactions has the form %
\beq
\mathcal{L} \sim \frac{\l}{\L^d} \left( H L^\dagger \ell_R \right) \Oc \to \frac{\l \, v \, \gap^{d-1}}{\sqrt{2} g_* \, \L^d} \, (\bar{e} e) \, \phi + \frac{g_*}{\gap} \phi \left( \partial \chi \right)^2. \\  
\eeqn
Integrating out the scalar meson $\phi$ yields the electron-DM coupling: 
\beq
\mathcal{L} \sim \frac{16\pi^2 \kappa_e}{\left( \sum \kappa_\ell m_\ell \right) \Lambda_{\rm SM}^2} \left( \bar{e} e \right) \left( \partial \chi \right)^2
\label{eq:Leff_LP_SC}
\eeqn
where the sum runs over all charged lepton flavors. 

The calculation of the energy loss rate due to DM emission is again similar to the case of standard model neutrinos~\cite{Raffelt:1996wa}. The relevant process in MS and HB stars is Compton scattering, $e^- \gamma \to e^- \chi \chi$. The energy loss rate per unit mass is given by
\beq
\epsilon = \frac{Q}{\rho} \sim \frac{9! \, 2 \alpha}{\left( \sum \kappa_\ell m_\ell \right)^2 \Lambda_{\rm SM}^4} \frac{Y_e}{m_u \, m_e^2} \, T^{10}.
\eeqn
Here, $\alpha$ is the fine-structure constant, $Y_e$ is the electron number fraction per baryon and $m_u = 1.661 \times 10^{-24} \; {\rm g}$ is the atomic mass unit. The calculation of this rate is outlined in Appendix~\ref{subapp:compton}. 

In red giants, electrons are degenerate, and Compton scattering receives a strong suppression by the final state Pauli-blocking effect (see Section~3.2 of \cite{Raffelt:1996wa} and footnote~8 in Appendix~\ref{subapp:compton}). Instead, production by a Bremsstrahlung process $e^-N \to e^-N\chi\chi$ is more efficient. The energy loss rate per mass is given by
\beq
\epsilon (\chi \chi) \sim \frac{\pi \, \alpha^2}{189} \left( \frac{Z^2}{A m_u} \right) \left( \frac{16 \pi^2 \, \kappa_e}{\left( \sum \kappa_\ell m_\ell \right) \Lambda_{\rm SM}^2} \right)^2 T^8, \nonumber
\eeqn
where $Z$ is the charge of nuclei and $A$ is the atomic mass. The calculation of this rate is outlined in Appendix~\ref{subapp:Brem_e}.

In the core of the supernova, temperature is sufficiently high for a thermal population of positrons to exist. In this case, $e^+ e^-$ annihilation  becomes the dominant production channel. Moreover, since $T>\gap$, the produced dark-sector states are described by a CFT, and their production rate is estimated using the unparticle approach. 

The energy loss rate is given by
\beq
Q (\text{\tiny CFT}) \sim n_{e^-} n_{e^+} \langle \sigma v E \rangle\,,
\eeqn
where the energy transfer rate $\langle \sigma v E \rangle$ is given by 
\beq
\langle \sigma v E \rangle \sim \left( \frac{\l \, v}{\sqrt{2} \, \L^d} \right)^2 \left( \frac{4\pi^4 \, d(d^2-1)}{(2\pi)^{2d+1}} \right) E_F^{2d-3},
\label{eq:ee_ann_SN_lepton_portal_sigma v E_sec4}
\eeqn
where $E_F \approx 344 {\rm MeV}$ is the electron Fermi energy. This is very similar to the expression that was used in the calculation of relic density produced during freeze-in, with the main difference being that the typical collision energy is now of order $2E_F$ rather than $T$. The positron number density is given by  
\beq
n_{e^+} = 2 \int \frac{d^3 p}{(2\pi)^3} \frac{1}{e^{(E+\mu_{e^-})/T}+1} \approx e^{- \beta \mu_{e^-}} \times n_{\rm th} 
\label{eq:positron_number_density_SN}
\eeqn
with $n_{\rm th}$ being the equilibrium number density at $T=T_{\rm SN}$ with Boltzmann distribution, and the chemical potential $\mu_{e^-}\approx E_F$. 

Unsurprisingly, the MFV flavor scheme lepton portal models are not constrained by supernovae due to the suppressed couplings to electrons and positrons. In both the first generation and democratic flavor schemes however, the dark matter particles end up trapped in the core of the supernova due to significant interactions with the electrons in the plasma. Details of SN trapping calculations can be found in Appendix~\ref{subapp:SN_trapping}. As a result, there is no relevant constraint from supernovae in any of the viable lepton portal models.

\subsubsection{Hypercharge Portal}

In this model, the dark matter candidate has a mass of order MeV, and only the SN has a high enough temperature to produce dark-sector states. There are three possible processes to consider: photon annihilation, $e^+ e^-$ annihilation through a photon loop, and nucleon Bremsstrahlung. Quantitatively, the photon annihilation turns out to be the most important channel as explained in Appendix~\ref{subapp:photon_annihilation}. This is due to the loop- and electromagnetic coupling-suppression for the $e^+ e^-$ annihilation, and phase space- and loop-suppression for the nucleon Bremsstrahlung.
The energy loss rate per volume from the photon annihilation is given by (see Appendix~\ref{subapp:photon_annihilation} for details)
\beq
Q (\text{\tiny CFT}) \sim n_\gamma^2 \, \langle \sigma v E \rangle \sim n_\gamma^2 \; \left( \frac{\l \cos^2 \theta_w}{\L^d} \right)^2 \left( \frac{16 \, d^2 (d^2-1) (d+2)}{(2 d-1) (2\pi)^{2d+1}} \right) T_{\rm SN}^{2d-1}.
\eeqn
where $\cos \theta_w$ is the Weinberg angle. 

\subsection{Naturalness bound}

In addition to the observational bounds discussed so far, we consider two constraints on the model parameter space motivated by theoretical considerations, the naturalness and ``CFT bootstrap" bounds. 

The effective coupling of the SM to the dark sector required to reproduce the observed DM relic density is tiny, ${\cal O}(10^{-14}-10^{-11})$. The naturalness bound is the requirement that such a coupling can be obtained in the effective theory without invoking trans-Planckian mass scales or unexplained small dimensionless parameters. As a concrete example, consider the UV completion of the CFT in terms of a gauge theory with a BZ fixed point, see Section~\ref{subsec:UV}. In addition to consistency requirements of the COFI scenario, $T_R\lsim \L \lsim M_{\rm BZ}$, naturalness requires
\beq
M_{\rm Bz} \lsim M_{\rm Pl},~~~~\lambda_{\rm BZ}\sim 1.
\eeq{nat_req}
In some of the COFI dark matter scenarios, there are parts of the parameter space where these requirements cannot be satisfied; those regions are shaded in green in the plots of Section~\ref{sec:Ops}. However, it is worth keeping in mind that these bounds are model-dependent. Any amount of tuning, or alternative UV completions, may lead to modifications of the naturalness bound.

\subsection{Numerical CFT bootstrap bound }

One of the attractive features of the COFI theories is that the small mass scale in the CFT sector is generated dynamically. This occurs through a combination of cosmological phase transitions in the SM sector followed by a slow RG running of the CFT sector, and finally dimensional transmutation within the CFT sector triggered by the $\mathcal{O} (1)$ CFT breaking effect. Our analysis so far has been based on an assumption that the largest breaking of the conformal invariance is from the interaction between the SM and the CFT sector and associated operator mixing effects. In particular, we assumed that the CFT scalar operator $\Oc$ appearing in the interaction does not show up on its own in the UV Lagrangian. If it did, it would make the CFT RG run from the onset and may result in a larger value of $\gap$ than what we have been taking.

As explained in Section~\ref{subsec:IR}, we may assume a $\mathbb{Z}_2$ discrete symmetry in the CFT sector under which the particular $\Oc$ is odd, hence can not be added to the UV Lagrangian. However, the CFT may contain another $\mathbb{Z}_2$-even scalar operator of dimension $<4$, which may not necessarily couple to the SM but would potentially generate a large $\gap$ on its own. Such an operator would generically appear in the OPE of two of $\Oc$ operators. A useful bound on this indeed does exist in the numerical CFT bootstrap literature \cite{Poland:2018epd, Poland:2011ey}. The idea is that given a scalar operator with scaling dimension $d$, the numerical CFT bootstrap provides an upper bound on the dimension of scalar operators that enter the OPE $\Oc \times \Oc$. This latter dimension turns out to be $\leq 4$ if $d \lesssim 1.6$. We indicate this bound by a dashed line on the plots of the COFI parameter space in Section~\ref{sec:Ops}; the parameter space below the line is potentially problematic. This bound is, however, somewhat model dependent and can be evaded, for example, by assuming a larger global symmetry, e.g.~$\mathbb{Z}_4$, in the CFT sector.

\section{Conclusions}
\label{sec:out}

In this paper, we have considered a dark sector that is invariant under conformal symmetry, broken only by a weak coupling to the Standard Model. This coupling leads to breaking of the conformal invariance in the infrared, at a scale $\gap$. Below this scale, the dark sector is described by a hadronic phase, with the lightest meson (dark pion) playing the role of dark matter. Within a broad range of model and cosmological parameters, the dark matter relic density is dominated by the energy transfer from the SM plasma to the dark sector in the conformal regime. We have labeled this scenario ``Conformal Freeze-In" (COFI). We showed that the COFI scenario provides a viable dark matter candidate, consistent with all phenomenological constraints, for several choices of the SM portal operator primarily interacting with the dark sector. We conclude that a conformal dark sector minimally coupled to the SM can naturally produce the observed dark matter.

\section*{Acknowledgements}

We are grateful to Damon Binder, Jae Hyeok Chang, Thomas Hartman, Luca Iliesiu, Hyung Do Kim, Eric Kuflik, Ofri Telem, Eliott Rosenberg, and Yiming Zhong for useful discussions. SH thanks Wen Han Chiu and Liantao Wang for useful discussions and collaboration on a related subject.
SH also would like to thank Nima Afkhami-Jeddi for helpful discussions about CFTs.

M.P. is supported by the U.S. National Science Foundation through grant PHY-2014071. 
S.H. was supported by the DOE grants DE-SC-0013642 and  DE-AC02-06CH11357, as well as a Hans Bethe Post-doctoral fellowship at Cornell.
G.K. is supported by the Science and Technology Facilities Council with Grant No. ST/T000864/1.

\appendix

\section{Details of Calculations in Cosmology}
\label{app:calc}

\subsection{Analytical Estimates of Relic Densities}
\label{subapp:rd}

\noindent \textbf{Higgs Portal}:

In this section, we show a brief derivation of Eq.~\leqn{anal_HP}, that relates observed dark matter relic density to parameters in the theory in the Higgs portal. In addition, the computation for Eq.~\leqn{hh_decay} is shown in more detail. Using the same procedure, analytical results for relic density can be computed for all portals considered in this paper, and the results for other portals are summarised at the end without going into technical details.

In the Higgs portal case, as mentioned before, below the critical dimension $d_*\,=\, 5/2$, dark matter production is dominated by the Higgs decay process. At temperatures below the electroweak phase transition, the effective interaction between the dark sector and the SM becomes, 
\beq
{\cal L}_\mathrm{int} = \frac{\l}{\L^D} \frac{v}{\sqrt{2}}\ h\ \Oc.
\eeq{L_int_decay}
The energy transfer rate through this process is given by Eq.~\leqn{hh_decay} and can be computed as follows:
\beqa
n_h \langle \; \Gamma(h \rightarrow \CFT) \; E  \rangle =\iint \, \d\Pi_h \, \d\Pi_\CFT \, f_h \, (2\pi)^4 \, \delta^4(p_h - P) \, E_h \, \vert \mathcal{M} \vert^2. \ \ 
\eeqa{hh_1}
Here and below, $P = p_\CFT$ is the momentum carried by the dark sector. The phase space for the CFT sector is chosen to be identical to that of ``unparticles" as prescribed by Georgi in~\cite{Georgi:2007ek}. Using Georgi's notation, we have,
\beqa
&& n_h \langle \; \Gamma(h \rightarrow \CFT) \; E  \rangle \nonumber\\
&& = \iint \frac{\d^3 \vec{p}_h}{(2\pi)^3 2E_h}\ \frac{\d^4 P}{(2\pi)^4} \mathrm{e}^{-\beta E_h} (2\pi)^4\ \delta^4(p_h - P)  \ A_d \ (P^2)^{d-2} \ E_h\ \frac{v^2}{4}\ \frac{\l^2}{\L^{2d-4}} \nonumber\\ 
&&= \frac{A_d \; v^2 \; \l^2}{4\L^{2d-4}} (m_h^2)^{d-2} \int \frac{\d^3 \vec{p}_h}{2(2\pi)^3} \,\exp(-\beta\sqrt{\vert\vec{p}_h\vert^2 + m_h^2}), \\
&&\hspace{12em} \nonumber 
\eeqa{hh_2}
where,
\beq
A_d = \frac{16\pi^{5/2}}{(2\pi)^{2d}}\frac{\Gamma(d+1/2)}{\Gamma(d-1)\Gamma(2d)}.
\eeq{}

Setting $p\equiv\vert\vec{p}_h\vert$ and simplifying gives
\beqa
n_h \langle \; \Gamma(h \rightarrow \CFT) \; E  \rangle &=& \frac{A_d \; v^2 \; \l^2 \; (m_h^2)^{d-2}}{4\L^{2d-4}}  \int 4 \pi p^2 \frac{\d p}{2(2\pi)^3}\, \exp(-\beta \sqrt{p^2 + m_h^2} ) \nonumber\\
&=& \frac{A_d \; v^2 \; \l^2 \; (m_h^2)^{d-2}}{32 \pi^2 \L^{2d-4}}  \int p^2 \,\d p\, \,\exp(-\beta \sqrt{p^2 + m_h^2} ). 
\eeqa{hh_3}
This integral represents a Bessel function of the second kind. Additionally, in our notation, $f_d = A_d/16\pi^2$. Thus, on simplifying, we get,
\beq
n_h \langle \Gamma(h\to {\rm CFT}) E\rangle \,=\, \frac{f_d \, \lambda_{\scriptstyle_{\rm CFT}}^2 v^2 \, m_h^{2(d-1)} T}{\Lambda_{\scriptstyle_{\rm CFT}}^{2d-4}}\,K_2(m_h/T).
\eeq{hh_decay2}
The CFT energy density at any point in time (as a function of the Standard Model bath temperature) can be obtained by integrating the Boltzmann equation given in Eq.~\leqn{BeqCFT}. To get a simple estimate, it suffices to do this calculation in the relativistic approximation where the Higgs is assumed to be massless and is described by a Maxwell-Boltzmann distribution. The process roughly starts around the electroweak scale $\sim v$ and continues till the SM temperature reaches the Higgs mass. 

In the relativistic approximation (i.e., taking the limit $m_h \rightarrow 0$ in the thermal average calculation), the energy transfer rate in this process is given by,
\beq
n_h \langle \Gamma(h\to {\rm CFT}) E\rangle \,=\, 2 f_d \, \l^2 \, v^2 \, \frac{m_h^{2d-4}}{\Lambda_{\scriptstyle_{\rm CFT}}^{2d-4}}\,T^3.
\eeq{hh_decay_rel}
We integrate the Boltzmann equation with this collision term, ignoring the temperature dependence of $g_*$ for now, and enforcing the condition that decays are inactive above the electroweak scale. Thus, we have,
\beq
\rhoc (T) \,=\, \frac{2 M_* f_d \lambda_{\scriptstyle_{\rm CFT}}^2 }{3 \sqrt{g_*(T)} v}\, \left( \frac{m_h}{\Lambda_{\scriptstyle_{\rm CFT}}} \right)^{2d-4} \,T^4\left( \frac{v^3}{T^3}-1 \right),
\eeq{rho_CFT_decay}
where $M_* = 3\sqrt{5}/(2\pi^{3/2}) \, M_{\rm pl}$, comes from the definition of Hubble as $H = \sqrt{g_*}\; T^2/M_*$.

At $T \sim m_h$, as the Higgs falls out of the thermal bath, this process becomes exponentially suppressed, and further production of dark sector energy can be neglected for this analysis. The energy density present in the dark sector then redshifts like radiation ($\rho \propto a^{-4}$) until its temperature $\Tc$ becomes comparable to the mass of the dark matter candidate. After this point, it redshifts like matter ($\rho \propto a^{-3}$) as required.

Thus,
\beq
\rhoc (m_h) \,  =\, \frac{2 M_* f_d \l^2 }{3 \sqrt{g_*(m_h)} v}\, \frac{m_h^{2d}}{\L^{2d-4}}\, \left( \frac{v^3}{m_h^3}-1 \right), 
\eeq{rhomh}
and
\beqa
\rhoc (T_m) \,  =\, \frac{2 M_* f_d \l^2 \, g_*(T_m)}{3 (g_*(m_h))^{3/2} v }\,\left( \frac{m_h}{\L} \right)^{2d-4}  \left( \frac{v^3}{m_h^3}-1 \right) T_m^4,
\eeqa{rho_CFT_end}
where $T_m$ is the \textit{SM temperature} at which the \textit{dark sector temperature} ($\Tc$) drops to the mass of the dark matter candidate. We also define the CFT energy density at this temperature as $\rhoc \equiv A \, \mdm^4$, where A represents a model-dependent measure of the number of degrees of freedom of the CFT (times a constant = $\pi^2/30$). Thus, the relic density is given by
\beq
\rho_{\rm DM} (T_0) \,=\, A \, \mdm^4 \frac{g_*(T_0) T_0^3}{g_*(T_m) T_m^3},
\eeq{relic_1}
where $T_0$ is the current CMB temperature. Additionally, from Eq.~\leqn{rho_CFT_end}, $T_m$ is given by,
\beqa
T_m^4  = A \, \mdm^4 
\left[  \frac{2 M_* f_d \l^2 \, g_*(T_m)}{3 (g_*(m_h))^{3/2} v }  \,\left( \frac{m_h}{\L} \right)^{2d-4}  \left( \frac{v^3}{m_h^3}-1 \right) \right]^{-1}
\eeqa{Tm}
Using Eq.~\leqn{Tm} in Eq.~\leqn{relic_1} gives the relic density of dark matter from the Higgs portal in terms of other parameters in the theory. 

Note that we use $g_*(T_0) \sim g_*(T_m) \sim {\cal O}(1)$. This is a reasonable approximation, as both temperatures are below the QCD scale. $g_*(m_h)$, denoted as just $g_*$ below, is approximately $\mathcal{O}(100)$. We also replace $\left( \frac{v^3}{m_h^3}-1 \right)\to {\cal O}(1)$ for this order-of-magnitude estimate. Additionally, we substitute $\gap$ in the equation instead of $\l$ and $\L$ using the mass gap equations. Taking the ratio of $\rho_{\rm DM} (T_0)$ and the present critical energy density gives Eq.~\leqn{anal_HP}:
\beqa
\frac{\Omega_{\scriptstyle_{\rm DM}} h^2}{0.1} = \left[ \frac{\mdm}{1\mev} \right]\left[ \frac{\left(A \, f_d^3 \, g_*^{-9/2}\right)^{1/4}}{ 10^{-5}} \right]\left[ \frac{\left( \frac{\gap}{m_h} \right)^{(6-\frac{3d}{2})}}{10^{-12}} \right]. 
\eeqa{relic_final}
This simple estimate is in good agreement with the results of numerical integration of Eq.~\leqn{BeqCFT}. 

Following the same procedure, the relic density can be calculated for each of the other three portals. These equations are given below, neglecting derivatives of $g_*$, but keeping all scales intact.

For the quark and lepton portals, the primary production process is that of fermion annihilation below the weak scale, where the Higgs is replaced by its VEV. The thermal averaging process can be repeated for $2 \to $ CFT processes as, 
\beq
n_1 n_2 \langle \sigma(f_1 f_2 \rightarrow \mathrm{CFT}) v_{rel} E \rangle = g_f^2 \, \frac{\l^2}{\L^{2d}} \frac{4 d(d^2-1)}{(2\pi)^{2d+1}}\; v^2 \; T^{2d+3}
\eeqn
where $g_f$ is the number of degrees of freedom of the fermion (considered massless in this limit).

Integrating the Boltzmann equation, we get,
\beq
\rhoc (T) = g_f^2 \frac{\l^2}{\L^{2d}} \frac{4 d(d^2-1)}{(2d-3)(2\pi)^{2d+1}}\;T^4(T_w^{2d-3}-T^{2d-3})
\eeqn
where $T_w$ is the weak scale temperature. 

For the gluon and electroweak portals, the results are similar, since the dominant process is that of vector boson scattering. However, the SM operator is dimension 4, and $T_w$ is replaced by $T_R$ since production starts right away after reheating, and these portals depend on the UV scale of reheating. Thus, in the gluon and electroweak portals, we have, 
\beq
\rhoc (T) = g_V^2 \frac{\l^2}{\L^{2d}} \frac{d^2(d^2-1)(d+2)}{(2d-1)(2\pi)^{2d+1}} \; T^4(T_R^{2d-1}-T^{2d-1})
\eeqn

Following the same procedure as described by Eqs.~\leqn{rhomh}-\leqn{relic_final}, we get the following equations for relic densities in other portals.\\

\noindent \textbf{Quark Portal}:
\begin{enumerate}
	\item First Generation Only:
	\begin{equation*}
		\rho_{\scriptstyle_{\rm DM}}(T_0) = \mdm \gap^{6-3d/2} \, A^{1/4} T_0^3
		\left[ \frac{M_* }{\alpha^4 v^4 (m_u+m_d)^2} \frac{16 \, d(d^2 - 1)}{(2d-3)(2\pi)^{2d+1}}\,  (v^{2d-3} - \Lambda_{\scriptstyle \rm QCD}^{2d-3}) \right]^{3/4} \CR
		\hspace{4.5em}
	\end{equation*}
	\item Democratic:
	\begin{equation*} 
		\rho_{\scriptstyle_{\rm DM}}(T_0) = \mdm \gap^{6-3d/2} \, A^{1/4} T_0^3
		\left[ \frac{M_* }{\alpha^4 v^4 m_{\rm top}^2} \frac{16 \, d(d^2 - 1)}{(2d-3)(2\pi)^{2d+1}}\,  (v^{2d-3} - \Lambda_{\scriptstyle \rm QCD}^{2d-3}) \right]^{3/4} \CR
		\hspace{4.5em}
	\end{equation*}
	\item Minimal Flavor Violation:
	\begin{equation*} 
		\rho_{\scriptstyle_{\rm DM}}(T_0) = \mdm \gap^{6-3d/2} \, A^{1/4} T_0^3
		\left[ \frac{M_* \, m_b^2}{\alpha^4 v^4 m_{\rm top}^4} \frac{16 \, d(d^2 - 1)}{(2d-3)(2\pi)^{2d+1}}\,  (v^{2d-3} - m_b^{2d - 3}) \right]^{3/4} \CR
		\hspace{4.5em}
	\end{equation*}
	Note that in the MFV flavor structure, due to the dependence of the coupling on the fermion mass, the heaviest fermion in the thermal bath below the weak scale contributes more to production than the other flavors. This would be the bottom quark in the quark portal.
\end{enumerate}
\noindent \textbf{Lepton Portal}:
\begin{enumerate}
	\item First Generation Only:
	\begin{equation*}
		\rho_{\scriptstyle_{\rm DM}}(T_0) = \mdm \gap^{6-3d/2} \, A^{1/4} T_0^3
		\left[ \frac{M_*}{\alpha^4 v^4 m_e^2} \frac{ 16 \, d(d^2 - 1)}{(2d-3)(2\pi)^{2d+1}} \, (v^{2d-3} - m_e^{2d-3}) \right]^{3/4} \CR
		\hspace{4.5em}
	\end{equation*}
	\item Democratic:
	\begin{equation*}
		\rho_{\scriptstyle_{\rm DM}}(T_0) = \mdm \gap^{6-3d/2} \, A^{1/4} T_0^3
		\left[ \frac{M_*}{\alpha^4 v^4 m_\tau^2} \frac{ 16 \, d(d^2 - 1)}{(2d-3)(2\pi)^{2d+1}} \, (v^{2d-3} - m_e^{2d-3}) \right]^{3/4} \CR
		\hspace{4.5em}
	\end{equation*}
	\item Minimal Flavor Violation:
	\begin{equation*}
		\rho_{\scriptstyle_{\rm DM}}(T_0) = \mdm \gap^{6-3d/2} \, A^{1/4} T_0^3
		\left[ \frac{M_*}{\alpha^4 v^4 m_\tau^2} \frac{ 16 \, d(d^2 - 1)}{(2d-3)(2\pi)^{2d+1}} \, (v^{2d-3} - m_\tau^{2d-3}) \right]^{3/4} \CR
		\hspace{4.5em}
	\end{equation*}
	Just as in the quark portal, in the lepton MFV case, the $\tau$-lepton contributes most to the dark matter energy density.
\end{enumerate}
\noindent \textbf{Gluon Portal}:
\begin{equation*}
	\rho_{\scriptstyle_{\rm DM}}(T_0) = \mdm \gap^{6-3d/2} \, A^{1/4} T_0^3
	\left[ \frac{M_* }{256 \, \pi^4 \alpha^8 v^8 } \, \frac{36 \, d^2(d^2 - 1)(d+2)}{(2d-1)(2\pi)^{2d+1}} \, T_R^{2d-1} \right]^{3/4} \CR
	\hspace{4.5em}
\end{equation*}
\noindent \textbf{Hypercharge Portal}:
\begin{equation*}
	\rho_{\scriptstyle_{\rm DM}}(T_0) = \mdm \gap^{6-3d/2} \, A^{1/4} T_0^3
	\left[ \frac{M_* \, \cos^4\theta_w}{256 \, \pi^4 \alpha^8 v^8 } \, \frac{16 \, d^2(d^2 - 1)(d+2)}{(2d-1)(2\pi)^{2d+1}} \, T_R^{2d-1} \right]^{3/4} \CR
	\hspace{4.5em}
\end{equation*}

As in the Higgs portal case examined previously in this appendix, these analytical estimates are in good agreement (with DM mass within an order of magnitude) with the numerically integrated results shown in Figs. \ref{fig:DMHiggs}, \ref{fig:DMquark1}, \ref{fig:DMquark2}, \ref{fig:DMlep1}, \ref{fig:DMlep2}, \ref{fig:DMglu}, and \ref{fig:DMweak}. Further, the DM mass dependence on reheating temperature as shown in Figs.~\ref{fig:otherTR} and Eq.~\leqn{GPscaling1} can be shown using these relic density estimates.

\subsection{Estimates for Hadronic Production}
\label{subapp:had}

While the exact nature of IR physics in COFI models depends on the details of confinement and the hadronic spectrum, we can still calculate the thermally averaged cross-sections in this regime up to $\mathcal{O}(1)$ factors, assuming a simple model as described in Section~\ref{subsubsec:IRphysics}. In this section, we show an example calculation for the lepton portal in the democratic flavor scheme, since hadronic production dominates in parts of the parameter space that produces the observed DM relic density in this model. This is one of very few COFI models with this property, and adding the hadronic contribution is important in this case.

As explained earlier, we model the confined/hadronic regime as containing a cosmologically stable pseudo-Nambu-Goldstone Boson, $\chi$, which acts as dark matter and a mediator with mass $\sim \gap$. Note that $\Oc$ is a scalar, and operator matching ensures that the scalar mediator has the dominant coupling to the SM sector. Thus we ignore any contributions from the vector mediator (if it exists). Using Eqns.~\leqn{phiDM} and \leqn{Op_match} the SM-DM interaction can be written as,
\beq
\mathcal{L} \sim \frac{\l}{\L^{d+\ds -4}} \, \frac{\gap^{d-1}}{g_\star} \, \Os \, \phi +  \frac{g_\star}{\gap} \, \phi \, \left( \partial \chi \right)^2,
\eeqn
where $\phi$ is the scalar mediator, $\chi$ is the DM, and $g_\star$ is the coupling ($= 4\pi/\sqrt{N}$ in large-N theories).

Since the hadronic processes are only relevant below the confinement scale, the mediator with mass $\sim \gap$ can be integrated out and we get,
\beq
\mathcal{L} \sim \frac{\l}{\L^{d+\ds -4}} \, \gap^{d-4} \, \Os \left( \partial \chi \right)^2.
\eeqn
In the lepton portal with democratic flavor scheme, using the relation between the coupling and gap-scale, this simplifies to\footnote{Recall that the dominant deformation that leads to confinement and generation of a mass gap in the lepton portal is from radiative mixing with the Higgs operator.}, 
\beq
\mathcal{L} \sim \frac{1}{\alpha^2 v^2 m_{\rm \scriptstyle tot}} \, \bar{e} e \left( \partial \chi \right)^2,
\eeqn
where $m_{\rm \scriptstyle tot} = m_e + m_\mu + m_\tau$ is the sum of masses of all the leptons running in the loop that generates the deformation of the CFT. Since these hadronic processes occur at very low energies ($T < \gap$), we only need to consider electrons as the other leptons are no longer in the bath. The Higgs is also replaced by its VEV below the weak scale.

The energy transfer rate is then given by,
\bea
n_e^2 \; \langle&&\sigma(e^+(p_1) \, e^-(p_2)  \rightarrow \chi(p_3) \, \chi(p_4)) \; E \;  \rangle  \CR
= && \iiiint \, \d\Pi_{e_1} \, \d\Pi_{e_2} \, \d\Pi_{\chi_1} \, \d\Pi_{\chi_2} \, f_{e_1}\, f_{e_2} \, (2\pi)^4 \, \delta^4(\Sigma p) \, (E_1 + E_2) \, \vert \mathcal{M} \vert^2 \CR
= && \prod_i \int \frac{d^3 p_i}{(2\pi)^3 E_i} \, e^{-\beta E_1} \, e^{-\beta E_2} \, (2\pi)^4 \delta^4(\Sigma p) \, \left(\frac{1}{\alpha^2 v^2 m_{\rm \scriptstyle tot}} \right)^2 \, (E_1 + E_2) \, (p_3.p_4)^2 \, (p_1.p_2) \CR
= && \left(\frac{1}{\alpha^2 v^2 m_{\rm \scriptstyle tot}} \right)^2 \prod_i \int \frac{d^3 p_i}{(2\pi)^3 E_i} \, e^{-\beta (E_1 + E_2)} \, (2\pi)^4 \delta^4(\Sigma p) \, (E_1 + E_2) \, (p_1.p_2)^3,
\eea
where $\Sigma p = p_1+p_2-p_3-p_4$, and in the last line, the particles involved are approximated to be relativistic/massless.

Computing the phase space integrals and using the delta function as usual, one gets,
\beq
n_e^2 \; \langle \sigma(e^+(p_1) \, e^-(p_2)  \rightarrow \chi(p_3) \, \chi(p_4)) \; E \;  \rangle  = \left(\frac{1}{\alpha^2 \, v^2 \, m_{\rm \scriptstyle tot}} \right)^2\,\frac{3240}{\pi^8} \, T^{13}.
\eeqn

This `collision term' can be plugged into the Boltzmann equation and integrated to get the energy density of dark matter states ($\chi$) produced in the hadronic phase:
\beq
\rho_{\scriptstyle_{\rm DM}} (T) = \frac{M_*}{\sqrt{g(T)}} \left(\frac{1}{\alpha^2 \, v^2 \, m_{\rm \scriptstyle tot}} \right)^2 \frac{3240}{\pi^8} \, \frac{T^4}{5} \,(\gap^5 - T^5),
\eeqn
where dark matter is assumed to redshift as radiation, and $M_*$ is as defined in the previous subsection. The interaction term in the Lagrangian is very irrelevant, and the power of temperature in this expression is high, as one might expect. Thus, this process is dominant at the temperature it starts, and we can use $(\gap^5 - T^5) \rightarrow \gap^5$ for calculating the final contribution of the hadronic production process. Additionally, we dropped terms of $\mathcal{O}(m_e)$, since $m_e < \gap$.

For most COFI models, hadronic production is very negligible. It is only relevant in lepton portal models with significant couplings to electrons since only electrons are light enough to persist in the thermal bath at such low temperatures (unlike QCD states and Higgs bosons). The same argument could be used in the case of the hypercharge portal, since photons are always present in the SM plasma; however, the CFT-phase energy density production in this case is proportional to positive powers of the reheating temperature which easily overwhelms the hadronic production that is proportional to powers of $\gap$. Thus, regions of parameter space with non-negligible hadronic production exist only in IR-dominated regime ($d<1.5$) in the first-generation and democratic flavor schemes in the lepton portal model.

\section{Derivation of Energy Loss Rates}
\label{app:energy loss rate}

\subsection{Compton scattering in MS and HB} 
\label{subapp:compton}

\begin{figure}[h]
	\center
	\includegraphics[width=0.4\textwidth]{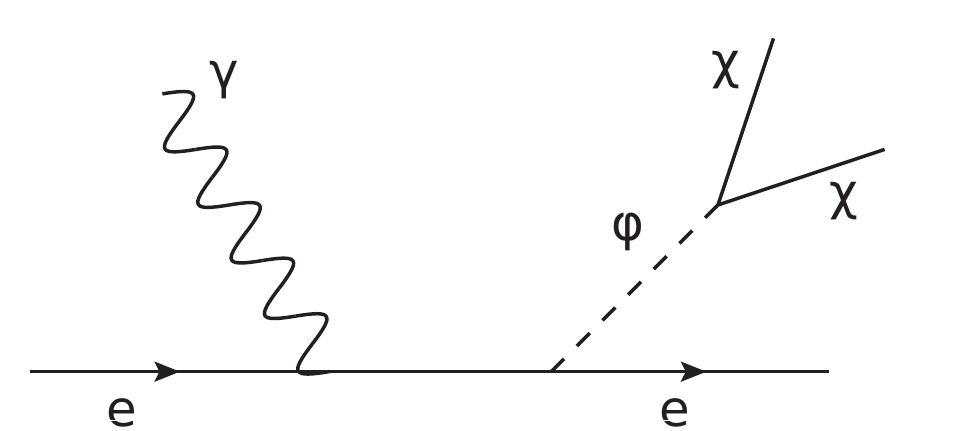}
	\caption{Feynman diagram for Compton scattering in the lepton portal. }
	\label{fig:StellarFeynman_Compton}
\end{figure}

In \cite{Raffelt:1996wa}, the production of a neutrino ($\bar{\nu} \nu$) pair from the Compton scattering of non-relativistic and non-degenerate electrons is studied and the rate is\footnote{If electrons are degenerate, the rate is suppressed by a factor 
	\beq
	F_{\rm deg} \sim \frac{3 E_F T}{p_F^2}. \nonumber
	\eeqn
	However, Compton scattering is important only in MS and HB starts in which electrons are not degenerate.
	Also, throughout the calculation, we use the criterion that if the photon plasma mass $\omega_p$ is less than $3T$ then we can neglect the plasma mass effects. For more details, see \cite{Raffelt:1996wa}
}
\beq
\sigma (\bar{\nu} \nu) \sim \frac{\alpha}{8\pi^2} G_F^2 m_e^4 \left( \frac{\omega}{m_e} \right)^4,
\eeqn
where $\omega$ is the energy carried by the $\bar{\nu} \nu$ pair, which is roughly $\omega \sim T$ (up to $\mathcal{O} (1)$).
Using this result and the effective theory Eq.~(\ref{eq:Leff_LP_SC}), the cross-section for the production of $\chi \chi$-pair is estimated to be
\beq
\sigma (\chi \chi) \sim \frac{2 \pi^2 \alpha}{\left( \sum \kappa_\ell m_\ell \right)^2 \Lambda_{\rm SM}^4} m_e^4 \left( \frac{\omega}{m_e} \right)^6.
\eeqn
For non-relativistic and non-degenerate electrons, they are almost at rest, and the energy loss rate per unit volume $Q$ can be approximated as
\beq
Q \sim n_e \int 2 \frac{d^3 k_\gamma}{(2\pi)^3} \; \frac{\omega \, \sigma (\chi \chi)}{e^{\omega/T} - 1},
\eeqn
where $k_\gamma$ is the photon momentum and the factor 2 is for the two photon polarization degrees of freedom.
The computation proceeds by writing 
\beq
\sigma (\chi \chi) = \sigma_* \left( \frac{\omega}{m_e} \right)^p, \; \sigma_* \equiv \frac{2 \pi^2 \alpha}{\left( \sum \kappa_\ell m_\ell \right)^2 \Lambda_{\rm SM}^4} m_e^4
\eeqn
and $p=6$ for our case. Explicit computation given in \cite{Raffelt:1996wa} shows that
\beq
Q \sim \frac{(p+3) ! \zeta (p+4)}{\pi^2} \frac{\sigma_* n_e T^{p+4}}{m_e^p}
\eeqn 
where $\zeta (n)$ is the Riemann zeta function. We can further use $n_e = Y_e \frac{\rho}{m_u}$ (where $Y_e$ is the electron number fraction per baryon and $m_u = 1.661 \times 10^{-24} \; {\rm g}$ is the atomic mass unit) to finally get the energy loss rate per mass $\epsilon$:
\beq
\epsilon = \frac{Q}{\rho} \sim \frac{9! 2 \alpha}{\left( \sum \kappa_\ell m_\ell \right)^2 \Lambda_{\rm SM}^4} \frac{Y_e}{m_u m_e^2} T^{10}.
\eeqn
Here, we used the expression for $\sigma_*$ and $p=6$.

\subsection{Bremsstrahlung from an electron in RGB}
\label{subapp:Brem_e}

\begin{figure}[h]
	\center
	\includegraphics[width=\textwidth]{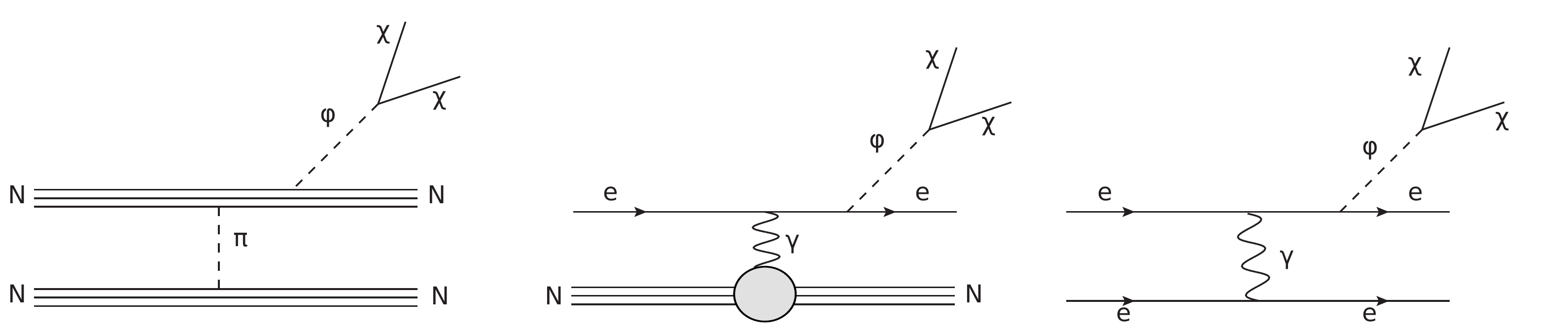}
	\caption{Feynman diagrams for relevant Bremsstrahlung processes. }
	\label{fig:StellarFeynman_Brem}
\end{figure}

The expression for $\epsilon$ for the production of a $\bar{\nu} \nu$ pair from a degenerate electron line is given in \cite{Raffelt:1996wa} and is (with $C_V \approx C_A = 1$ and $F_+ = 1, F_- = 0$ and assuming a single species of nuclei of charge $Z$ and atomic mass $A$)
\beq
\epsilon (\bar{\nu} \nu) \approx \frac{2\pi \alpha^2}{189} \left( \frac{Z^2}{A m_u} \right) G_F^2 T^6.
\eeqn 
One reasonable estimation of the rate for the case of $\chi \chi$ production can be made by comparing the effective coupling between the two cases. The matching condition is 
\beq
\frac{G_F}{\sqrt{2}} \to \frac{16\pi^2 \omega}{\left( \sum \kappa_\ell m_\ell \right) \Lambda_{\rm SM}^2}
\eeqn
where we have included one factor of $\omega$ to make up the right dimension. That $\omega$, rather than $E_F$, is the right factor even in the degenerate situation, is understood as follows. Unlike in $e^- e^+$ annihilation where the final state energy is of the order $E_F$, in Bremsstrahlung, the final states carry only $\omega \sim T$ because of final state Pauli-blocking for the electron: essentially, while the electron has energy $\sim E_F$, the amount of energy change by the momentum transfer, i.e.~displacement in the Fermi surface with radius $E_F$, is limited to $\sim T$. In the end, we get
\beq
\epsilon (\chi \chi) \sim \frac{\pi \alpha^2}{189} \left( \frac{Z^2}{A m_u} \right) \left( \frac{16 \pi^2}{\left( \sum \kappa_\ell m_\ell \right) \Lambda_{\rm SM}^2} \right)^2 T^8.
\label{eq:epsilon_Brem_e}
\eeqn
We emphasize that our estimation is at the level of $\mathcal{O} (1)$ or even an order of magnitude, due to non-trivial combinatoric factors and precise values for $\omega/T$ and so on, that our computation does not take into account.

\subsection{Bremsstrahlung from a nucleon}
\label{subapp:Brem_nucleon}

This process is most relevant for quark-, gluon-, and Higgs-portal. 
For the part of parameter space relevant for the relic density, the production is in the form of hadrons of confined CFT in MS, HB, and RGB, while in SN, the final state is CFT state. 

\subsubsection*{Hadronic final state: MS, HB, RGB}

We first derive rates for the case when the final states are hadrons of confined CFT.
The strategy is the same as before. We take the expressions obtained for $\bar{\nu} \nu$ production and estimate for the $\chi\chi$ production by making necessary modifications. 
In Section~4 of \cite{Raffelt:1996wa}, $Q (\bar{\nu} \nu)$ is shown to be
\beq
Q (\bar{\nu} \nu) = \left( \frac{\sum_q C_q^{(N)} G_F}{\sqrt{2}} \right)^2 \frac{n_B}{20\pi^4} \int_0^{\infty} d \omega \; \omega^6 \; S_\sigma (-\omega)
\eeqn
where in the non-degenerate (for nucleons) limit
\bea
&& S_\sigma (\omega) = \frac{\Gamma_\sigma}{\omega^2} s(\omega/T) \times \left\lbrace
\begin{array}{ll}
	1 \hspace{1cm} {\rm for} \;  \omega > 0 \\
	e^{\omega/T} \;\;\;\; {\rm for} \; \omega < 0
\end{array} \right. \nonumber \\
&& \Gamma_\sigma = 4 \sqrt{\pi} \alpha_\pi^2 n_B T^{1/2} m_N^{-5/2} \label{eq:nucleon_Brem_Had_stuffs} \\
&& s (x) \approx \sqrt{1 + \vert x \vert \pi / 4} \nonumber \\
&& \alpha_\pi = \frac{m_N^2}{\pi m_\pi^2} \approx 15 . \nonumber  
\eea
The convention is that $\omega < 0$ corresponds to the energy taken away from the medium and $n_B = n_p + n_n$ is the nucleon number density. It may be worth clarifying that the factor $C_q^{(N)}$ comes from the matching of quark-neutrino four fermion interaction to nucleon-neutrino four Fermi interaction. Hence, $C_q^{(N)}$ appearing here is literally the same as the one in the effective theory Eq.~(\ref{eq:quark-portal_EFT_SC}). Performing the integration, Raffelt showed that 
\beq
Q (\bar{\nu} \nu) \approx \frac{2048}{385\, \pi^{7/2}} \left(\sum_q C_q^{(N)} \right)^2 G_F^2 \, \alpha_\pi^2 \, \frac{n_B^2}{m_N^{5/2}} T^{11/2}.
\label{eq:Q_nunu_Brem_nucleon_Had}
\eeqn
The rate for $\chi\chi$ production can be obtained from this by (i) matching the effective coupling and (ii) taking into account difference in the integration. For quark- and gluon-portal, the matching of coupling becomes 
\bea
&& \text{Quark-Portal :} \; \left( \frac{\sum_q C_q^{(N)} G_F}{\sqrt{2}} \right)^2 \; \leftrightarrow \; \left( \frac{16 \pi^2 \sum_q \kappa_q C_q^{(N)}}{\left( \sum_q \kappa_q m_q \right) \Lambda_{\rm SM}^2} \right)^2 \left( \frac{\omega}{2} \right)^2 \\
&& \text{Gluon-Portal :} \; \left(  \frac{\sum_q C_q^{(N)} G_F}{\sqrt{2}} \right)^2 \; \leftrightarrow \; \left( C_G^{(N)} \frac{ 16 \pi^2}{ \Lambda_{\rm SM}^4} \right)^2 \left( \frac{\omega}{2} \right)^2
\eea
where $\omega \sim T$ and the form factors $C_q^{(N)}$ and $C_G^{(N)}$ are defined in Eq.~(\ref{eq:quark_nucleon_form_factor}) and (\ref{eq:Gluon_nucleon_form_factor}), respectively. 

Next, in the case of $\bar{\nu} \nu$ production, the integration in the expression of $Q (\bar{\nu} \nu)$ is
\beq
\int_0^\infty d \omega \; e^{-\omega / T} \; \omega ^4 \; s(-\omega/T) \equiv C_{\bar{\nu} \nu} T^5
\label{eq:C_nunu}
\eeqn
In the case of $\chi \chi$ production, on the other hand, it is given by (other than the $\omega$-independent part of effective coupling factors for which we have already shown the matching)
\beq
\frac{1}{4} \int_0^\infty d \omega \; e^{-\omega / T} \; \omega ^6 \; s(-\omega/T) \equiv C_{\chi\chi} T^5 \left( \frac{T}{2} \right)^2.
\label{eq:C_chichi}
\eeqn
Identifying $(T/2)^2 = (\omega / 2)^2$ in the matching of the coupling, we see that a slightly more accurate ratio of the $\epsilon$'s requires a factor of $C_{\chi\chi} / C_{\bar{\nu} \nu}$. More explicitly, we get
\bea
&& \text{Quark-Portal :} \; Q (\chi\chi) = Q (\bar{\nu} \nu) \frac{\left( \frac{16 \pi^2 \sum_q \kappa_q C_q^{(N)}}{\left( \sum_q \kappa_q m_q \right) \Lambda_{\rm SM}^2} \right)^2 \left( \frac{T}{2} \right)^2}{\left( \frac{\sum_q C_q^{(N)} G_F}{\sqrt{2}} \right)^2} \frac{C_{\chi\chi}}{C_{\bar{\nu}\nu}} \label{eq:Quark-Portal_Q_chichi_nucleon_Brem_Had} \\
&& \text{Gluon-Portal :} \; Q (\chi\chi) = Q (\bar{\nu} \nu) \frac{\left( C_G^{(N)} \frac{ 16 \pi^2}{ \Lambda_{\rm SM}^4} \right)^2 \left( \frac{T}{2} \right)^2}{\left( \frac{\sum_q C_q^{(N)} G_F}{\sqrt{2}} \right)^2} \frac{C_{\chi\chi}}{C_{\bar{\nu}\nu}} \label{eq:Glue-Portal_Q_chichi_nucleon_Brem_Had} 
\eea
where $Q (\bar{\nu} \nu)$ is given in Eq.~(\ref{eq:Q_nunu_Brem_nucleon_Had}). 
We make two comments. First, since both $\bar{\nu} \nu$ and $\chi\chi$ productions involve two particles, the phase space factor for the two cases are the same. The internal degrees of freedom, however, can differ. In addition, while the neutrino production is predominantly via axial coupling \cite{Raffelt:1996wa}, the production of $\chi\chi$ pair is likely to be from a vector coupling. We do not think the latter two factors will generate more than an order of magnitude effect. It would be interesting to carry out improved stellar cooling computations for COFI, which we leave for future investigations.

\subsubsection*{CFT final state: SN}

Our strategy is based on the fact that the ratio $\epsilon (\text{\tiny CFT}) / \epsilon (\phi)$, the energy loss rate for COFI to that of real scalars, can be estimated to a reasonable accuracy. To this end, (i) we need to compare the effective couplings, (ii) use powers of $(m_N T)$ to make up the correct dimensions, and (iii) powers of $(2\pi)$ to take into account difference in phase space. In addition, one can improve the estimation by including (iv) ratio of internal degrees of freedom of the energy carrying final states and (v) ratio of thermal averaged energy $\langle \omega \rangle$ of the new physics states. Regarding point (ii), we recall that nucleons are non-relativistic, implying $p^2 \approx 2 m_N T$, and this in turn means that the characteristic size of the energy transfer is $\omega \sim \sqrt{m_N T}$. Next, the correct factors of $(2\pi)$ for the phase space: for each extra particle we have $\frac{1}{(2\pi)^3} \times (2\pi) = \frac{1}{(2\pi)^2}$ where the first factor is the naive one from the phase space integral measure and the second factor of $(2\pi)$ is the result of extra angular integration. 
This seemingly naive argument works even for the case of CFT final state. In this case, we recall that the phase space associated with a dimension $d$ CFT operator can be thought of as $d$ massless particles, and so we can estimate the phase space factor as $\sim \frac{1}{(2\pi)^{3d}} \times (2\pi)^{d-1} = \frac{1}{(2\pi)^{2d+1}}$, again the first factor for $d$ number of naive phase space factor and the second factor for $(d-1)$ extra angular integration. The final answer indeed agrees with explicit computations once we adopt the Georgi's choice for the unparticle phase space density.
To summarize, below we will estimate the ratio $\epsilon (\text{\tiny CFT}) / \epsilon (\phi)$ using
\beq
\frac{\epsilon (\text{\tiny CFT})}{\epsilon (\phi)} = \frac{Q (\text{\tiny CFT})}{Q (\phi)} \sim \frac{G_{\rm eff}^2 (m_N T)^r}{g^2} \frac{1}{(2\pi)^{2d-2}} \frac{\text{dof}_{\rm \scriptscriptstyle CFT}}{\text{dof}_\phi} \frac{\langle \omega \rangle_{\rm \scriptscriptstyle CFT}}{\langle \omega \rangle_{\phi}}
\label{eq:epsilon_ratio_nucleon_Brem_CFT}
\eeqn

\noindent where the exponent $r$ depends on the model and $g$ is the Yukawa coupling, $\sim g \phi \bar{\psi}_N \psi_N$. 
Before we show our results, it is instructive to present the known cases, confirming the validity of our analysis scheme. 
To this end, let us compute the ratio $\epsilon (\bar{\nu}\nu) / \epsilon_a$, i.e.~the ratio of rate of neutrino-pair production to that of axions. Without including the ratio of $\langle \omega \rangle$'s it is given by
\beq
\frac{\epsilon (\bar{\nu}\nu)}{\epsilon_a} =  \frac{\left( \sum_q C_q^{(N)} G_F \right)^2 (m_N T)^2}{g_a^2} \frac{1}{(2\pi)^2} \left( \frac{3 \times 2 \times 2}{1} \right) = \frac{3}{\pi^2}  \frac{\left( \sum_q C_q^{(N)} G_F \right)^2 (m_N T)^2}{g_a^2}
\eeqn

\noindent where 3, 2, 2 are respectively family, spin, and $SU(2)_L$-doublet degrees of freedom of neutrino. The same ratio can also be computed using explicit expressions for $\epsilon (\bar{\nu}\nu)$ and $\epsilon_a$ (Eqns.~(4.8), (4.10), (4.23), and (4.24) in \cite{Raffelt:1996wa}) and the outcome is about twice larger than our estimation. This factor of two, however, can also be explained since $\langle \omega_{\bar{\nu}\nu} \rangle / \langle \omega_a \rangle$ is numerically about 2. 

We now present our results. First of all, $Q (\phi)$ for a degenerate nucleon medium is given by (Eq.~(4.13) in \cite{Raffelt:1996wa})
\beq
Q (\phi) = \frac{g^2 \alpha_\pi^2}{4\pi} \frac{44}{15^3} \left( \frac{T}{m_N} \right)^4 p_F^5 \, G_\phi (m_\pi / p_F) 
\label{eq:Q_scalar_Brem_nucleon}
\eeqn
where $\alpha_\pi$ is defined in Eq.~(\ref{eq:nucleon_Brem_Had_stuffs}) and numerically $G_\phi (m_\pi / p_f ) \approx 0.8$. Also, since nucleons are borderline degenerate in SN, we can use $p_F \approx m_N T$ (which in a sense we have been using so far already).
We can now use Eq.~(\ref{eq:epsilon_ratio_nucleon_Brem_CFT}) with the effective couplings 
\bea
&& \text{Higgs-Portal :} \;
G_{\rm eff} =  C_G^{(N)} \left( \frac{\alpha_s}{6 \sqrt{2}\pi} \right) \left( \frac{\gap^{4-d}}{v^2 m_h^2} \right) \\
&& \text{Quark-Portal :} \; G_{\rm eff} = \sum_q \kappa_q C_q^{(N)} \frac{16 \pi^2 \gap^{4-d}}{\left( \sum_q \kappa_q m_q \right) \Lambda_{\rm SM}^2}
\label{eq:G_ff_quark-portal_Brem_nucleon_CFT} \\
&& \text{Gluon-Portal :} \; G_{\rm eff} = C_G^{(N)} \frac{16 \pi^2 \gap^{4-d}}{ \Lambda_{\rm SM}^4} \label{eq:G_ff_gluon-portal_Brem_nucleon_CFT}
\eea
and $r = d-1$ for all cases since the mass dimension of $G_{\rm eff}$ is $1-d$ for all three cases. The factors $\frac{\text{dof}_{\rm \scriptscriptstyle CFT}}{\text{dof}_\phi}$ and $\frac{\langle \omega \rangle_{\rm \scriptscriptstyle CFT}}{\langle \omega \rangle_{\phi}}$ depend on the details of the CFT and on a general grounds they are expected to be within $1 \sim d$.

\subsection{Electron-positron annihilation}
\label{subapp:ee_annihilation}

\begin{figure}[h]
	\center
	\includegraphics[width=\textwidth]{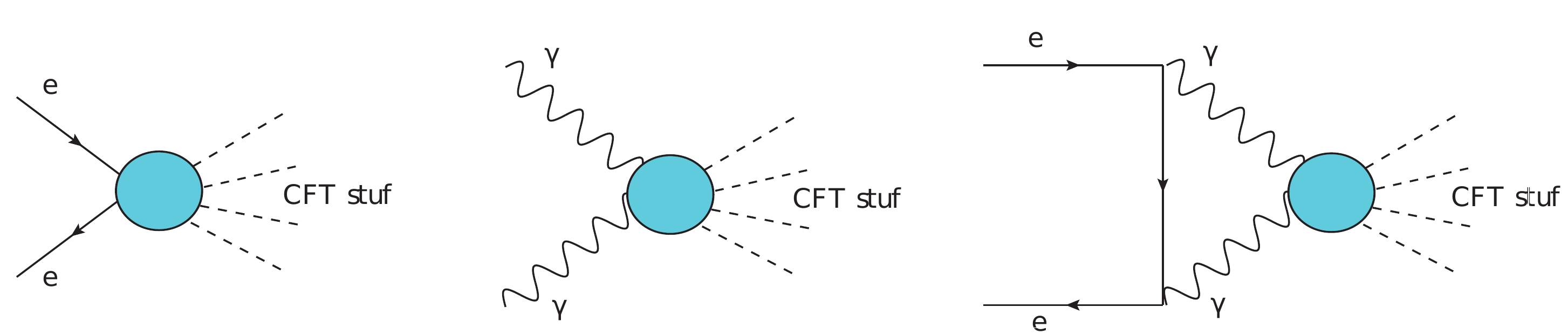}
	\caption{Feynman diagrams for relevant annihilation processes in the lepton and hypercharge portals, in the CFT phase $(T> \gap)$ . }
	\label{fig:StellarFeynman_ee_annih}
\end{figure}

The process of interest is the annihilation of $e^-e^+$ to a CFT final state and this is relevant for the lepton-portal and hypercharge-portal models in the SN. In lepton-portal it is via tree-level coupling, and for the hypercharge-portal it is a one-loop generated coupling.

\subsubsection*{Lepton-portal}

From the Lagrangian
\beq
\mathcal{L} \sim \frac{\l v}{\sqrt{2} \L^d} \, \bar{e} e \, \mathcal{O}_{\rm CFT}
\eeqn
we can estimate the energy transfer rate in the exactly the same way we do for the freeze-in calculation 
\beq
\langle \sigma v E \rangle \sim \left( \frac{\l v}{\sqrt{2} \L^d} \right)^2 \left( \frac{4\pi^4 d (d^2-1)}{(2\pi)^{2d+1}} \right) E_F^{2d-3},
\label{eq:ee_ann_SN_lepton_portal_sigma v E}
\eeqn
where $E_F \approx 344 {\rm MeV}$ is the electron Fermi energy.
From this we get the energy loss rate per volume
\beq
Q (\text{\tiny CFT}) \sim n_{e^-} n_{e^+} \langle \sigma v E \rangle
\eeqn
and hence $\epsilon$ by $\epsilon = Q / \rho$. For the number density of electrons in SN, we use $n_{e^-} \approx 1.8 \times 10^{38} \; {\rm cm^{-3}}$, while for the positrons, we note that the process $e^- e^+ \leftrightarrow \gamma \gamma$ imposes the relationship among chemical potentials, $\mu_{e^+} = - \mu_{e^-}$ and we know $\mu_{e^-} \approx 344 \; {\rm MeV}$. Using this, the number density for the positron can be shown to be
\beq
n_{e^+} = 2 \int \frac{d^3 p}{(2\pi)^3} \frac{1}{e^{(E+\mu_{e^-})/T}+1} \approx e^{- \beta \mu_{e^-}} \times n_{\rm th}
\eeqn
with $n_{\rm th}$ being the equilibrium number density at $T=T_{\rm SN} \, (=1/\beta)$ with Boltzmann distribution. We see that the positron number density is suppressed compared to the thermal density by the factor $ e^{- \beta \mu_{e^-}} $.

\subsubsection*{Hypercharge-portal}

The computation goes through the exact same steps as in the lepton-portal. To this end, we first derive an effective action by computing the loop-diagram shown in Fig.~\ref{fig:StellarFeynman_ee_annih}. The result is estimated to be
\beq
\mathcal{L} \sim \left( \frac{\l}{\L^d} \right) \left( \frac{2 e^2 m_e}{\pi^2} \log \left( \frac{\Lambda_{\rm SM}}{E_F} \right) \right) \bar{e} e \Oc
\eeqn
The appearance of the log is due to the massless particle (i.e.~photon) running in the loop, and we used the external momentum to be $E_F$ appropriate for the degenerate electrons in the SN core. To get the energy loss rate per mass, we now simply need to replace the effective coupling in the lepton-portal computation:
\beq
\frac{\l v}{\sqrt{2} \L^d}  \to\left( \frac{\l}{\L^d} \right) \left( \frac{2 e^2 m_e}{\pi^2} \log \left( \frac{\Lambda_{\rm SM}}{E_F} \right) \right).
\eeqn

\subsection{Photon annihilation}
\label{subapp:photon_annihilation}

This process is relevant for the hypercharge portal at the core of SN. The photon number density is that of a thermal Boltzmann distribution,
\beq
n_\gamma \approx \frac{2 \, \zeta(3)}{\pi^2} \, T_{\rm SN}^3
\eeqn
and since the plasma frequency $\omega_p \sim 19 \; {\rm MeV}$ is less than $3 \, T_{\rm SN} \sim (60 \, \text{--} \, 80) \; {\rm MeV}$, we ignore plasma mass effects. The energy loss rate per volume is estimated to be 
\beq
Q (\text{\tiny CFT}) \sim n_\gamma^2 \langle \sigma v E \rangle \sim n_\gamma^2 \; \left( \frac{\l \cos^2 \theta_w}{\L^d} \right)^2  \left( \frac{16 \, d^2 (d^2-1) (d+2)}{(2 d-1) (2\pi)^{2d+1}} \right) T_{\rm SN}^{2d-1}.
\eeqn
This process is the dominant process for supernova cooling in hypercharge-portal. To illustrate this, we compute the ratio
\beq
\frac{Q (e^-e^+ \to \text{\tiny CFT})}{Q (\gamma\gamma \to \text{\tiny CFT})} \sim e^{-E_F/T_{\rm SN}} \left( \frac{E_F}{T_{\rm SN}} \right)^{2d} \left( \frac{m_e}{T_{\rm SN}} \right)^2 \left( \frac{8 \alpha \log \left( \frac{\Lambda_{\rm SM}}{E_F} \right)}{\cos^2 \theta_w} \right) \ll 1.
\eeqn
Numerically, for $\Lambda_{\rm SM} = 1$ TeV, the above ratio is $\lesssim 10^{-2}$ for $1 \leq d \leq 3$.


\subsection{Trapping in Supernovae}
\label{subapp:SN_trapping}

In this section, we describe how to estimate the cross-section required to evaluate the trapping of hadronic states of the confined CFT at the core of SN. For concreteness sake, we focus on the lepton-portal case, where trapping is important. (See~\cite{Dreiner:2013mua} for discussion of supernova constraints on dark sectors.)

\begin{figure}[h]
	\center
	\includegraphics[width=0.6\textwidth]{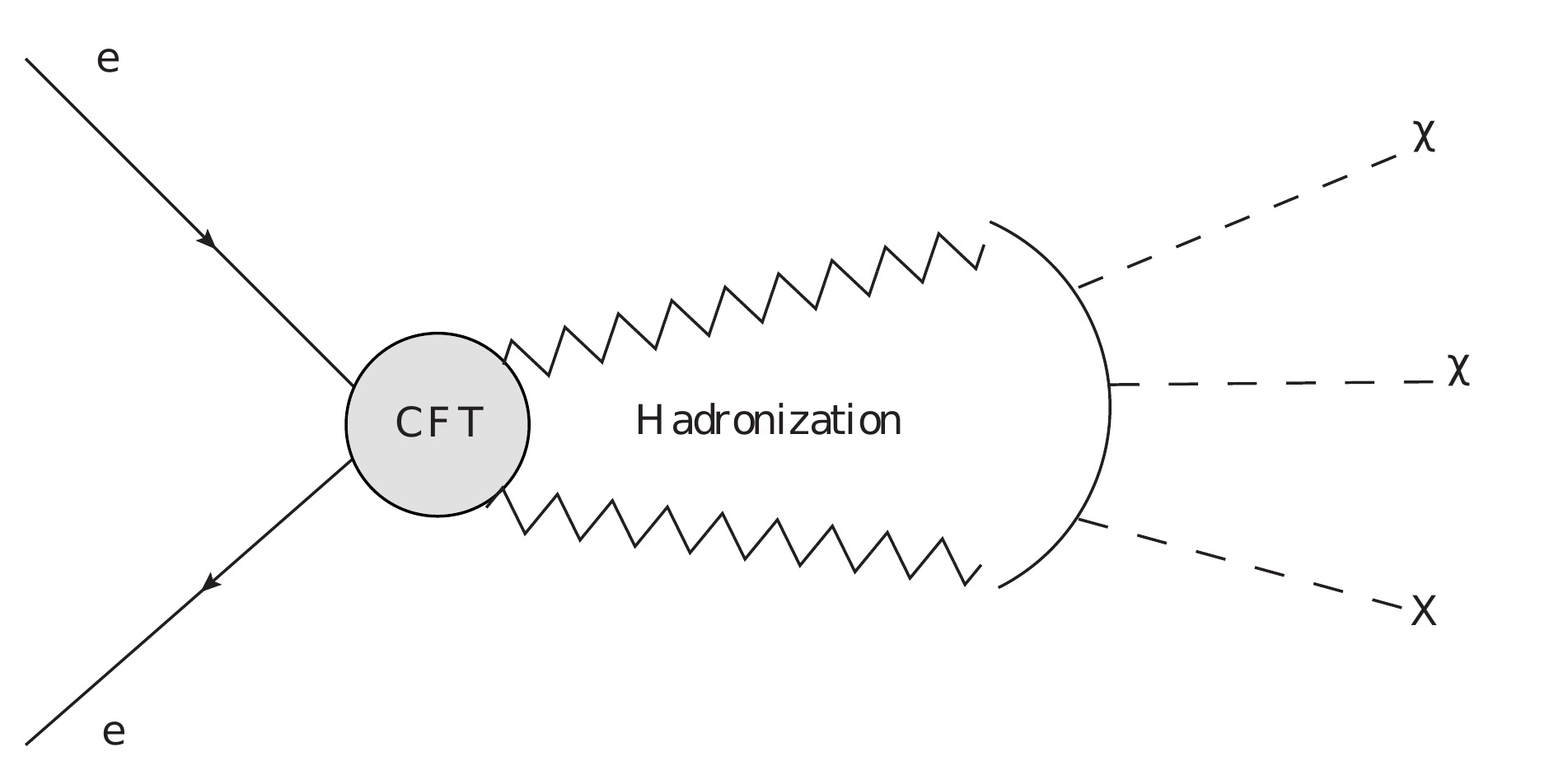}
	\caption{Production of CFT state via $e^- e^+$-annihilation in SN followed by a hadronization into composite dark matter ($\chi$) plus $\mathbb{Z}_2$-even final states (denoted as $X$) in lepton portal model. }
	\label{fig:SN_production_CFT_hadronization}
\end{figure}

Let us first discuss production of a pair of dark matter states $\chi$ near the core of the supernova. We assume that the CFT is described by a large-$N$ gauge theory. When such a theory confines in the IR, we can use large-$N$ analysis which we follow. Schematically, when $\gap < T_{\scriptstyle \mathrm{SN}}$, an annihilation of $e^-e^+$ produces directly the state associated with $\Oc$ which can be thought of as the ``partonic'' state of a confining CFT. Once these ``partonic'' CFT states are produced, they will go through the hadronization process, somewhat similar to the QCD jets. The situation is shown schematically in Fig.~\ref{fig:SN_production_CFT_hadronization}. Some of the hadronic states then can travel out of the supernova, resulting in an extra mechanism for its cooling. Assuming $\mathcal{O} (1)$ fraction of energy is transferred to the dark matter state $\chi$ (which is consistent with our freeze-in calculations), we can make a rough estimation as follows. Here, we assume that the dark matter $\chi$ is a goldstone boson created by a current operator $J_\mu$ of a broken global symmetry. 
We are interested in the rate for $\Oc$ to result in a pair of $J_\mu$'s which in turn ``hadronizes'' into the dark matter $\chi$ and other hadronic states. This information will be used below to estimate the cross-section responsible for trapping in SN. The rate for $\Oc$ to turn into a pair of $J_\mu$'s is encoded in the OPE (Operator Product Expansion) coefficient 
\begin{equation}
	J^\mu J_\mu \sim c \, \Oc + \cdots.
\end{equation}
where $c$ is the OPE coefficient which carries a scaling dimension of $6-d$ (recall that the conserved current has dimension 3 and acquires no anomalous dimensions).
In addition, the probability for a current $J_\mu$ to produce a single $\chi$ can be summarized in an interpolation relation of the form (in large-$N$ limit)
\begin{equation}
	J_\mu \sim \frac{1}{g_\star} \gap \partial_\mu \chi
	\label{eq:J_pi_interpolation}
\end{equation}
where the factor of $g_\star \sim \frac{4\pi}{\sqrt{N}}$ is inserted to be consistent with large-$N$ behavior
\beq
\langle J J \rangle \sim \frac{N}{16\pi^2} \sim \frac{1}{g_\star^2}.
\eeqn
The other factors are fixed by dimensional analysis and the fact that $\chi$ is a goldstone boson associated with a broken current $J_\mu$.

The matrix element for the pair production diagram in Fig.~\ref{fig:SN_production_CFT_hadronization} is then estimated to be
\begin{equation}
	\mathcal{M} \sim \left( \frac{v}{\lambda^{d}} \right) \cdot \frac{1}{c} \cdot \left( \frac{\gap  p_\mu}{g_\star} \right)^2 \cdot \left( \bar{u} u \right)
	\label{eq:star_cooling_production_CFT_picture}
\end{equation}
We wish to determine $g_\star$-dependence of the OPE coefficient $c$ which is needed to figure out correct $g_\star$-counting for the rates. While there is no fully rigorous and systematic means to answer this question, ``matching'' between the above estimation and the fully-hadronic picture may be used to get a reasonable assessment. To this end, we first note that at leading order in $1/N$-expansion the hadronic cubic interaction vertex is $\Gamma_3 \sim g_\star$. This is understood by noting that 
\begin{equation}
	\langle \Oc J J \rangle \sim \frac{N}{16\pi^2} \sim \frac{1}{g_\star^2}, \;\;\;\; \Oc \sim \frac{M_{\rm gap}^{d-1}}{g_\star} \phi
\end{equation}
where $\phi$ is a scalar meson interpolated by $\Oc$ with mass of the order $M_{\rm gap}$. Using these together with Eq.~(\ref{eq:J_pi_interpolation}), we get,
\begin{equation}
	\langle \Oc J J \rangle \sim \frac{1}{g_\star^2} \sim \left( \frac{M_{\rm gap}^{d-1}}{g_\star} \right) \left( \frac{M_{\rm gap}}{g_\star} \right)^2 \cdot \Gamma_3 \;\; \rightarrow \;\; \Gamma_3 \sim g_\star.
\end{equation}
With this information at hand, the matching to the fully-hadronic picture gives
\begin{equation}
	\mathcal{M} \sim \left( \frac{v}{\Lambda^d} \right) \cdot \left( \frac{M_{\rm gap}^{d-1}}{g_\star} \right) \cdot \frac{1}{M_{\rm gap}^2} \left( \frac{M_{\rm gap} p_\mu}{g_\star} \right)^2 \cdot \Gamma_3 \cdot \left( \bar{u} u \right).
	\label{eq:star_cooling_production_hadron_picture}
\end{equation}
Comparing this to Eq.~(\ref{eq:star_cooling_production_CFT_picture}) finally shows that
\begin{equation}
	c \sim M_{\rm gap}^{d-6} \; g_\star^0 .
\end{equation}

\begin{figure}[h]
	\center
	\includegraphics[width=0.8\textwidth]{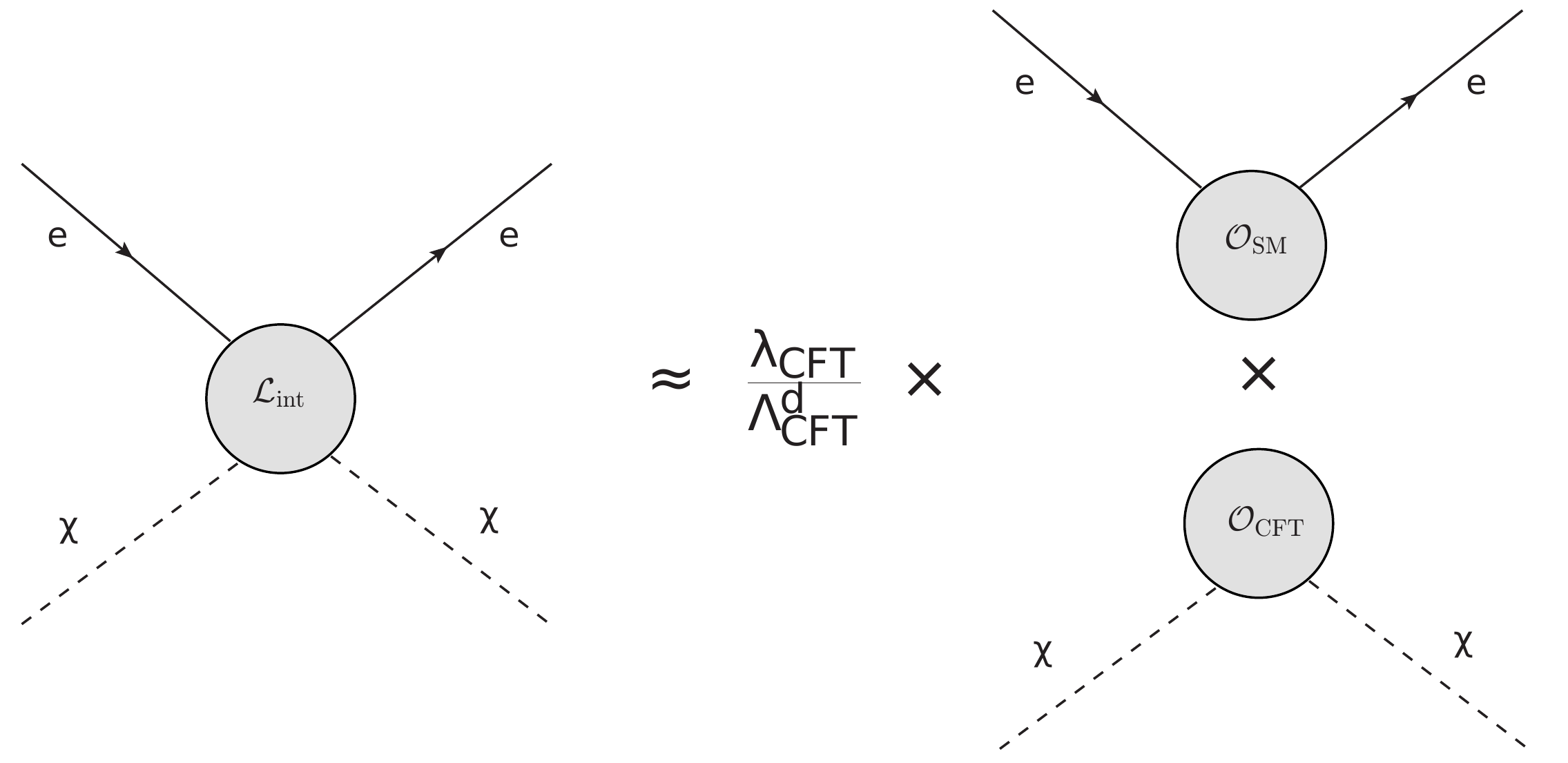}
	\caption{Diagram relevant for trapping in SN, and the assumption of factorization. }
	\label{fig:SNTrapping}
\end{figure}

With these preparations, we now discuss trapping. Again, with a simplifying assumption that most of the CFT energy is processed to the DM state $\chi$, the relevant picture is: we have DM particles produced in the core of SN and we are interested in the cross-section of $\chi + e^- \to \chi + e^- + X$, where $X$ denotes any other collectively $\mathbb{Z}_2$-even states in the final state. To the extent that the leading contribution comes from $\chi + e^- \to \chi + e^-$,\footnote{A simple argument based on phase space suppression seems to support this assumption, although multiplicity of the diagrams and any unknown non-perturbative physics could in principle invalidate the claim. Here, we simply assume, which is certainly enough for the stellar cooling bound, that at least $\mathcal{O} (1)$ contribution comes from the simple $2 \to 2$ process.} a reasonable estimate is possible assuming the factorization shown in Fig.~\ref{fig:SNTrapping}. From the discussions given above about the OPE, the lower part of the diagram is given by
\beq
\langle \chi_2 \vert \Oc \vert \chi_1 \rangle \sim \frac{1}{c} \langle \chi_2 \vert J_\mu J^\mu \vert \chi_1 \rangle \sim \frac{M_{\rm gap}^{d-4}}{\hat{c}\, g_\star^2} \, p_{\chi_1} \cdot p_{\chi_2}
\eeqn 
where we have introduced a dimensionless quantity $\hat{c}$ defined by $c = \hat{c}\, M_{\rm gap}^{6-d}$. The full matrix element is then computed to be
\beq
\mathcal{M} \sim \left( \frac{\lambda v}{\Lambda^d \sqrt{2}} \right) \left( \frac{M_{\rm gap}^{d-4}}{\hat{c}\, g_\star^2} \, p_{\chi_1} \cdot p_{\chi_2}\right) \bar{u} (k_1) u (k_2) 
\eeqn
with $k_{1,2}$ being the four-momentum of the incoming and outgoing electrons, respectively. The cross-section can finally be estimated and one gets,
\beq
\sigma \sim \frac{1}{8\pi \gamma \,\hat{c} \,g_\star^4} \frac{E_F^4}{\left( \sum_\ell \kappa_\ell m_\ell \; \alpha^2 v^2 \right)^2}
\eeqn  
where $E_F$ is the electron Fermi energy and $\gamma$, defined by $E_\chi = \gamma \, E_F$, encodes the fraction of energy  carried by the DM $\chi$ upon CFT-hadronization. $v$ is the VEV of the Higgs and $\alpha$ is defined by $\Lambda_{\rm \scriptscriptstyle SM} = \alpha \, 4\pi v$. Also, we used the formula for the $M_{\rm gap}$ given in Section~\ref{sec:gen}. The mean free path is obtained from
\beq
\lambda_\chi = \frac{1}{n_e \sigma}
\eeqn
and we use the optical depth criterion
\beq
\int_{r_0}^{R_c} \frac{d r }{\lambda_\chi} \gtrsim \frac{2}{3} \;\;\; \Rightarrow \;\;\; \frac{3}{2} \frac{0.1 R_c}{\lambda_\chi} \gtrsim 1 \;\;\;\;\;\; \text{(trapped)}
\eeqn
to assess the possibility of the trapping. To get the final expression, we used $r_0 \approx 0.9 R_c$ and $R_c \approx 13 \; {\rm km}$ is the radius of the core~\cite{Dreiner:2013mua}. 

In the lepton portal, models of both first generation and democratic flavor schemes have trapping cross-sections many orders of magnitude above the optical depth criterion, and the dark matter particles produced are completely trapped. Thus, there is no relevant supernova constraint.

\bibliography{refs}
\bibliographystyle{jhep}

\end{document}